# Untersuchungen zum Anpassungs- und Etablierungspotential der invasiven Asiatischen Tigermücke *Aedes* (*Stegomyia*) *albopictus* (SKUSE)

Dissertation

zur Erlangung des Doktorgrades

der Naturwissenschaften

vorgelegt im Fachbereich Biowissenschaften

der Goethe-Universität Frankfurt am Main

von

Aljoscha Kreß

aus Frankfurt am Main

Frankfurt (2016)

(D30)

Untersuchungen zum Anpassungs- und Etablierungspotential
der invasiven Asiatischen Tigermücke
*Aedes* (*Stegomyia*) *albopictus* (S<small>KUSE</small>)

Aljoscha Kreß

# Inhalt

# Summary

The Asian tiger mosquito, *Aedes albopictus* (synonym: *Stegomyia albopicta*), was scientifically described as a new species by Skuse in 1894. This otherwise black mosquito owes its name to a pattern of distinct silverish stripes on its thorax, abdomen and legs. There is general agreement that the former (breeding) macrohabitat of this species has been described as phytotelmata in the forested areas of Southeast Asia, however, this has changed in the last four decades as it has adapted to more urban areas and antrotelmata. Capable of producing eggs with a certain dry resistance as well as cold hardiness, populations of *Ae. albopictus* became distributed around the globe by the international trade of used tires and other preferred breeding places of this species. The invasion of the *Ae. albopictus* across all continents (with the exception of Antarctica) is thought to be the most rapid spread of any insect species in the last four decades. As a consequence, many countries are now reporting stable established populations of *Ae. albopictus* in tropical, subtropical and temperate climate zones. Models predict further range expansions in the future as global climate change will render additional regions in higher altitudes and longitudes suitable for colonization by this mosquito species. In addition, *Ae. albopictus* is a potential vector for at least 27 viruses as well as for several parasites and plays a major role in the global transmission of dengue virus and chikungunya virus; and its contribution to the rapid spread of zika virus is currently under investigation. Therefore, this species is considered a serious threat to public health worldwide.

This work broaches the issue of three investigations about the potential to adapt and establish stabile populations of the invasive Asian tiger mosquito *Ae. albopictus*.

In a first approach, we focused on the problem of there being only two toxicological test systems for culicidae available to date. Therefore, we developed a dose-response bioassay that paves the way for many biological endpoints and their integrative parameters in order to provide a better understanding of the toxicological mode of actions of insecticides. By doing so, we were able to ask whether there was a difference in the ecotoxicological response of the invasive tropic/subtropic *Ae. albopictus* and the temperate European counterpart – the northern house mosquito *Culex pipiens* – to the insecticide λ-cyhalothrin. We also were able to investigate the impact of higher temperatures on the ecotoxicological response to λ-cyhalothrin and how different nutrition supply regimens alter the effect of this insecticide. Finally, we conducted a risk assessment of the consequences of our findings for a possible mosquito control measurement, which revealed possible fitness advantages for this species, e.g. the fact higher temperatures decrease the insecticide sensitivity of the mosquitoes or that aquatic predator species have a higher sensitivity to the insecticide used than the mosquitoes.

In a second approach, we focused on the mechanism of the cold hardiness of the eggs of *Ae. albopictus* (cold adaptation and diapause), which is thought to be one of the key traits of this species allowing for its successful establishment in temperate habitats all over the world. After a log lasting hypothesis that polyols decrease the super cooling point and thus increase the cold hardiness was falsified, the current state of scientific knowledge suggested an increase in the wax layer that is located in the chorion of the egg. However, no detailed evaluation of the chorion of mosquitoes had been conducted yet. We therefore investigated the chorion layers with an electron scanning microscope in order to reveal the mechanism of wax accumulation induced by diapause and cold acclimation of the egg. We were then able to reveal that the location of the wax layer was not in the outer serosal cuticle, as it had been hypothesized, but more likely in the dark endochorion of the egg shell. In addition, as the wax layer decreased in size due to diapause and we thus hypothesised a compaction of this layer.

In a third approach, we focused on the adaptation potential of the species. In the literature the fast adaptation of the phenotype to temperate climate zones is described as "adaptive phenotype", "rapid adaptation", "rapid adaptive evolution" or "asymmetric evolution". However, these founder populations in freshly invaded areas have a relatively low genetic diversity and it is unlikely that they have a constant influx of alleles in order to adapt on a genetic level. However, the concept of epigenetic adaptation may give an explanation to this phenomenon. We therefore asked, whether it is possible to detect a heritable diversification in low-temperature phenotype after random epigenetic alteration of the DNA. Given this, we investigated the transgenerational effects of two epigenetic agents (and a solvent) on the cold hardiness of the eggs of the phenotypes. We revealed a correlation pattern that links the level of methylation of the agent with the cold hardiness of the egg. As a result, we saw hints of the possible mechanism of epigenetic temperature adaptation of this species.

As a result of these investigations, a high potential for this species to become a threat to public health in many more countries, especially in temperate climatic zones, can be seen. Due to the fact on the one hand that *Ae. albopictus* can gain fitness advantages due to misapplied vector control measures and on the other hand of the high epigenetically adaptation potential, it is in summary recommended to focus further research on the development of vaccinations for viruses and other pathogens. By doing so, citizens will be protected without putting ecosystems and their services in danger and it is the even more economic solution.

# Zusammenfassung

Die Asiatische Tigermücke *Aedes albopictus* (synonym: *Stegomyia albopicta*) wurde erstmals 1894 von Skuse wissenschaftlich beschrieben. Diese von der Grundfarbe her schwarze Stechmücke verdankt ihren Namen dem silbrigen Streifenmuster auf Thorax, Abdomen und den Beinen. Es wird davon ausgegangen, dass das ehemalige Larven-Mikrohabitat die Phytotelmata in den Waldgebieten von Südostasien darstellte. In den letzten vier Jahrzehnten adaptierte sich die Art jedoch an urbanere Regionen und ihre Antrotelmata. Dank ihrer Eigenschaft, Eier mit einer gewissen Trocken- und Kältetoleranz zu produzieren, verbreitete sich die Art zusammen mit den international gehandelten Waren weltweit. Durch diese Invasion in Länder auf allen Kontinenten (mit Ausnahme der Antarktis) wird *Ae. albopictus* als die am schnellsten sich ausbreitende Insektenart in den letzten Jahrzehnten gesehen. Infolgedessen dokumentieren mittlerweile viele Länder in tropischen, subtropischen und gemäßigten Breiten stabile Populationen. Modellrechnungen sagen eine weitere Ausbreitung im Zuge des Klimawandels voraus und stufen weitere Höhenlagen und Längengrade als potentiell geeignete Habitate für die Zukunft ein. Zudem ist *Ae. albopictus* ein theoretischer Vektor für mindestens 27 Viren sowie Parasiten und spielt eine Hauptrolle bei der Übertragung von Dengue-Viren und Chikungunya-Viren; ihre Rolle bei der aktuell grassierenden Zika-Epidemie ist Bestandteil aktuellster Forschung. Daher wird die Art als große Gefahr für die öffentliche Gesundheit betrachtet.

Die vorliegende Arbeit thematisiert drei Untersuchungen zum Anpassungs- und Etablierungspotential der invasiven Asiatischen Tigermücke.

In einem ersten Ansatz wurde das Problem behandelt, dass es lediglich zwei standardisierte toxikologische Testverfahren für Culicidae gab. Daher wurde ein Dosis-Wirkungs-Testsystem entwickelt, das den Weg für weitere biologische Endpunkte und ihre integrativen Parameter freimachte und dadurch ein besseres Verständnis für die Wirkweisen von Insektiziden ermöglicht. Hierdurch konnte nun der Frage nachgegangen werden, ob es Unterschiede in der ökotoxikologischen Reaktion zwischen der invasiven tropisch-subtropischen Asiatischen Tigermücke und der einheimischen nördlichen Hausstechmücke *Culex pipiens* auf das Insektizid λ-Cyhalothrin gibt. Weiter wurde der Einfluss von Temperatur und die Verfügbarkeit von Nahrung auf die Insektizidsensitivitäten der Arten getestet. Schließlich konnte in einer Risikobewertung festgestellt werden, dass bei falsch angewendeten Bekämpfungsmaßnahmen höhere Temperaturen sowie der Ausfall von aquatischen Top-Prädatoren zu Fitnessvorteilen für die Art führen können.

In einer zweiten Untersuchung wurde der Mechanismus der Kältetoleranz der Eier (Kälteakklimatisierung und Diapause) näher untersucht, da dieser für die erfolgreiche Invasion in gemäßigten Breitengraden verantwortlich gemacht wird. Nachdem eine lang vorherrschende Hypothese verworfen wurde, dass die Einlagerung von Polyolen die Frosttoleranz bewirken würde, war der aktuelle Stand der Wissenschaft, dass eine Verdickung der Wachsschicht des Chorions dafür verantwortlich sei. Jedoch lag keine detaillierte Evaluierung von Stechmücken-Eihüllen vor. Mittels einer transmissionselektronenmikroskopischen Studie konnte gezeigt werden, dass nicht nur die Wachsschicht nicht in der Serosa-Cuticula zu verorten ist, sondern im Endochorion und sie zudem im Zuge der Diapause in der Mächtigkeit schrumpft. Daher wird auf Basis der gewonnenen Erkenntnisse auf eine Kompaktierung der Schicht geschlossen.

Die dritte Untersuchung schließlich hatte das hohe Adaptationspotential in gemäßigten Breiten zum Gegenstand. Eine Adaptation auf genetischen Level gilt als unwahrscheinlich, da Gründerpopulationen in den neu besiedelten Gebieten eine niedrige genetische Diversität aufwiesen und ein regelmäßiger Neueintrag von Allelen unwahrscheinlich ist. Jedoch bietet das Konzept der epigenetischen Temperatur-Adaptation einen Erklärungsansatz für dieses Phänomen. Daher wurde die Frage gestellt, ob es möglich ist, eine vererbbare Diversifizierung dieses kältetoleranten Phänotyps nach einer randomisierten epigenetischen Behandlung der DNA zu detektieren. Es wurde eine transgenerationale Untersuchung der Effekte von zwei epigenetischen Agenzien (und einem Lösemittel) auf die Kältetoleranz der Eier durchgeführt. Die Ergebnisse zeigten ein Korrelationsmuster, das den durch die Agenzien veränderten Methylierungsgrad der DNA mit der Forsttoleranz verband, was die gestellte Hypothese unterstützte.

In Folge dieser drei Untersuchungen wurde festgestellt, dass *Ae. albopictus* ein hohes Potential hat, in weiteren Ländern – vor allem in gemäßigten Breiten – ein Gesundheitsrisiko darzustellen. Da die Art einerseits Fitnessvorteile durch falsche Bekämpfungsmaßnahmen und andererseits möglicherweise eine hohes epigenetisches Adaptationspotential besitzt, kann zusammenfassend empfohlen werden, dass der Fokus für weitere Forschung maßgeblich auf der Entwicklung von Impfstoffen für die übertragenen Viren und Pathogene liegen sollte. Dadurch kann die Bevölkerung geschützt werden, ohne Ökosysteme und ihre Dienstleistungen zu gefährden, und dies wäre zudem ökonomisch gesehen die effektivere Lösung.

# 1 Einleitung in das übergeordnete Thema

## 1.1 Erstbeschreibung und Systematik von *Aedes* (*Stegomyia*) *albopictus*

„The banded mosquito of Bengal" nannte der britisch-australische Entomologe Frederick A. Askew Skuse (1863–1896) diese von ihm zum ersten Mal wissenschaftlich beschriebene Spezies (Skuse, 1894). Sie war ihm von E. C. Cortes mit der Bemerkung zugeschickt worden, dass dieses Insekt in Kalkutta ein großes Ärgernis darstelle (Skuse, 1894). In seiner Erstbeschreibung für die „Indian Museum Notes" beschreibt Skuse die Imagínes von *Aedes albopictus* (siehe Abbildung 1) mit einem schwarzen Grundton und einer weißen bis silbrigen Zeichnung der Beschuppung, worauf er mit dem Namen „*Culex albopictus*" hinweist, und verglich sie morphologisch mit der nahe verwandten Art *Aedes aegypti* (Skuse, 1894; Sota & Mogi, 2006; Neal & Gon III, 2007).

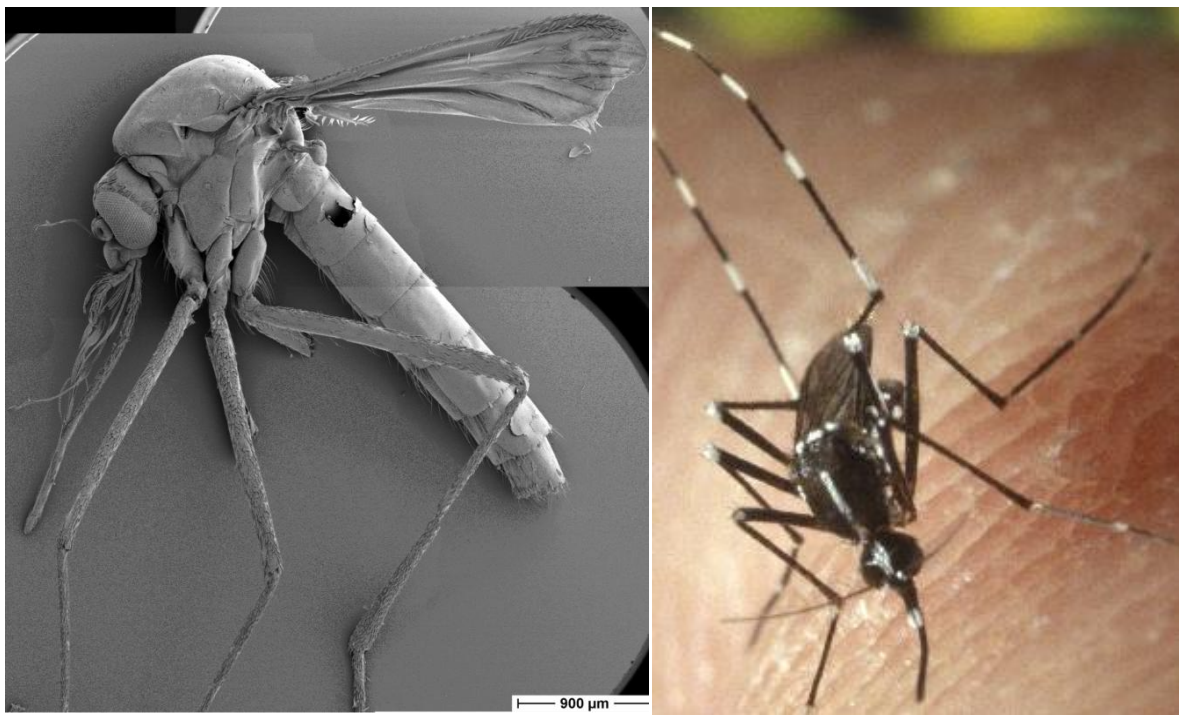

**Abbildung 1a (links, eigenes Werk, unveröffentlicht): Rasterelektronenmikroskopische Aufnahme von *Aedes albopictus*. Kombination aus vier Einzelbildern, verbunden zur Gesamtansicht eines adulten Weibchens. Antennen und große Teile der Thorakalbeschuppung sind durch die Fixierung verloren gegangen. Das Abdomen ist im ersten Segment durch die Trocknung eingebrochen und die drei rechten Beine wurden zur besseren Ausrichtung auf dem Objektträger entfernt. 1b (rechts, Ausschnitt aus CDC/ James Gathany, 2000[1]): Weibchen von *Aedes albopictus* in typischer Haltung mit angewinkelten Hinterbeinen auf menschlichem Wirt mit gut sichtbarer Zeichnung auf Thorax und Beinen.**

---

[1] CDC/ James Gathany (2000), Public Health Image Library unter Public Domain, Bild-ID# 1865, „A female *Aedes albopictus* mosquito feeding on a human host."

Innerhalb der Gattung *Aedes* (respektive Untergattung *Stegomyia*), in welche die Art später gestellt wurde, lässt sich *Ae. albopictus* mit bloßem Auge leicht an dieser Zeichnung auf dem Thorax erkennen: Im Vergleich zu *Ae. aegypti*, die mit 4 diskreten Linien gezeichnet ist, und *Ae. cretinus*, die drei Linien besitzt, überzieht den Thorax von *Ae. albopictus* eine einzige breite und klare Linie, die sich erst kurz vor dem Skutellum aufgabelt (Estrada-Franco & Craig, 1995). Die Larven von *Ae. albopictus* lassen sich z.B. durch das Fehlen ausgeprägter Haken am Thorakalsegment von *Ae. aegypti* unterscheiden (Estrada-Franco & Craig, 1995).

Um Polyphylien zu eliminieren, regte John F. Reinert im Jahr 2000 eine systematische Neubewertung des Genus *Aedes* und später des kompletten Tribus Aedini an, wodurch das ehemalige Subgenus der Asiatischen Tigermücke *Stegomyia* zur Gattungsebene erhoben wurde (Reinert, 2000; Reinert et al., 2004, 2009; Neal & Gon III, 2007). Diese Neuordnung wurde heftig kritisiert und Reinert als „paraphylliophob" bezeichnet (z.B. Savage, 2005); Befürworter der Revision bezeichneten dagegen den vormaligen Zustand des Tribus Aedini als „taxonomische Anarchie" (Hennig et al., 1999; Black, 2004). Bisher wird die Art in der einschlägigen Literatur weiterhin in der Gattung *Aedes* geführt; die Untergattung *Stegomyia* findet regelmäßige Erwähnung (Neal & Gon III, 2007; Pluskota et al., 2008).

Die Diskussionen zur Systematik des Tribus Aedini werden sich erst nach umfänglichen genetischen Charakterisierungen auflösen lassen. Es ist jedoch davon auszugehen, dass dann eine Überarbeitung der Taxonomie und Nomenklatur erfolgen wird (Reidenbach et al., 2009). Zum jetzigen Zeitpunkt sollte jedoch die neu vorgeschlagene Nomenklatur für diese Art zunächst nicht verwendet werden, um möglichen Verwirrungen, Missverständnissen und Fehlinterpretationen bei der Benennung einer medizinisch relevanten Art vorzubeugen. Dieses im Sinne des Internationalen Kodex für Zoologische Nomenklatur (ICZN, 1999) konservative Vorgehen wird vom wissenschaftlichen Mainstream weitgehend befolgt (z.B. Higgs 2005, JME 2005, Weaver 2005): Im „Web of Science" des Institute for Scientific Information ergibt der Suchbegriff „*Stegomyia albopicta*" 47 Suchergebnisse, wohingegen „*Aedes albopictus*" 4098 Suchergebnisse generiert (Suchzeitraum: 2004–2016, Stand: 22.04.2016).

Die vorliegende Dissertation folgt dieser konservativen systematischen Einordnung, jedoch mit expliziter Nennung der Untergattung im Hinblick auf eine zukünftige taxonomische Revision der Aedini, die auf robusten Analysen umfangreicher molekularer Datensätze beruht (Paupy et al., 2009):

| | |
|---|---|
| Klasse: | Insecta (Insekten) |
| Unterklasse: | Pterygota (Fluginsekten) |
| Überordnung: | Neoptera (Neuflügler) |
| Ordnung: | Diptera (Zweiflügler) |
| Familie: | Culicidae (Stechmücken) |
| Unterfamilie: | Culicinae |
| Tribus: | Aedini |
| Genus: | *Aedes* |
| Subgenus: | *Stegomyia* |
| Art: | *Aedes* (*Stegomyia*) *albopictus* (Synonym: *Aedes albopictus*, *Stegomyia albopicta*) |

## 1.2 Makrohabitat

Vor ihrer schlagartigen weltweiten Invasion in den letzten vier Jahrzehnten wird das ursprüngliche Verbreitungsgebiet von *Ae. albopictus* (vgl. Kapitel 1.8) im Bereich der tropischen und subtropischen Wälder Südostasiens sowie auf den Inseln im westlichen Pazifik und dem Indischen Ozean vermutet (Estrada-Franco & Craig, 1995; Bonizzoni et al., 2013). Im Westen lagen die ersten historisch dokumentierten Ausbreitungsgrenzen bei Madagaskar, im Osten bei Neuguinea und im Norden bei Peking, Seoul und Sendai zwischen dem 36sten und 40sten nördlichen Breitengrad (Hawley, 1988; Estrada-Franco & Craig, 1995). Das ursprüngliche Verbreitungsgebiet dieser Art lässt sich jedoch nur schwer rekonstruieren (Hawley, 1988). Einerseits trat *Ae. albopictus* zusammen mit den beiden nahe verwandten Arten *Ae. pseudoalbopictus* und *Ae. seatoi* auf, und diese drei Arten konnten erst durch die taxonomischen Schlüssel von Huang zwischen 1969 und 1979 eindeutig bestimmt werden (Hawley, 1988). Andererseits wurden Bestimmungsarbeiten erschwert, weil früh der Holotyp verloren ging, welcher erst 1968 durch einen Neotyp von Huang ersetzt wurde (Hawley, 1988), obwohl die Terra typica der Indische Bundesstaat Westbengalen ist. Dazu kommt, dass tropische Stechmücken in entomologischen Aufsammlungen erst um die vorletzte Jahrhundertwende mehr Beachtung fanden, als Forschungen die Fähigkeit von Stechmücken zur Krankheitsübertragung aufgedeckt hatten (vgl. Kapitel 1.7.3) (Savage, 2005).

Die Brutgewässer von *Ae. albopictus* waren ursprünglich vor allem Phytotelmata wie Astlöcher, Blattachseln und Bambusstümpfe in ruralen Gebieten (Higa, 2011). In frühen Habitatsstudien werden Begriffe wie z.B. „Wald" jedoch nicht eindeutig definiert, und so bleibt häufig offen, ob damit Waldränder in Siedlungsnähe, Sekundärwälder oder Primärwälder gemeint sind (Hawley, 1988). Allerdings ist *Ae. albopictus* in Ursprungsgebieten häufiger in Waldrändern in Siedlungsnähe als in Primärwäldern zu finden, da sie die Nähe zum Menschen sucht (Hawley, 1988). Zudem sind mittlerweile neben den ruralen Habitaten auch suburbane und urbane Makrohabitate betroffen, da die Art Anthrotelmata wie Blechdosen, Autoreifen oder Regentonnen sehr gut als Ersatzhabitate annimmt (Hawley, 1988). Daher wird eine Adaptation der Art von den ursprünglichen Habitaten in Waldrändern hin zu urbaneren Habitaten angenommen (Estrada-Franco & Craig, 1995). In stark urbanisierten Gebieten ist *Ae. aegypti* konkurrenzstärker (Estrada-Franco & Craig, 1995). In hochurbanen Gebieten ohne jegliche Vegetation ist *Ae. albopictus* nahezu nicht auffindbar (Hawley, 1988; Estrada-Franco & Craig, 1995).

# 1.3 Lebenszyklus und Populationsdynamik

Stechmückeneier werden erst kurz vor der Oviposition befruchtet (Farnesi et al., 2009). Die Entwicklungsdauer des Embryos bis zur dormanten $L_1$-Larve hängt maßgeblich von der Temperatur ab (Farnesi et al., 2009). So liegt bei der physiologisch vergleichbaren *Ae. aegypti* die Entwicklungsdauer unter der optimalen Temperatur von 25 °C bei 77,5 Stunden (Farnesi et al., 2009). Der Schlupfzeitraum der dormanten Larve von *Ae. albopictus* ist von vielen Faktoren, wie z.B. der Diapause oder dem Auslösen des Schlupfreizes, massiv beeinflussbar und daher sehr variabel (Wang, 1966; Estrada-Franco & Craig, 1995). Die weitere Entwicklungsdauer der Lebensstadien von Larve ($L_1$–$L_4$), Puppe und Adultus hängen maßgeblich von den vier Faktoren Temperatur, Ernährung, Populationsdichte und Geschlecht ab (Hawley, 1988; Estrada-Franco & Craig, 1995; Delatte et al., 2009). Das Larvenstadium erreicht die kürzeste Entwicklungsdauer bei Temperaturen um 25 °C, jedoch werden bei geringeren Temperaturen die Individuen größer (Estrada-Franco & Craig, 1995; Delatte et al., 2009). Bei Temperaturen von unter 11 °C stoppt die Entwicklung (Hawley, 1988; Estrada-Franco & Craig, 1995; Delatte et al., 2009). Nährstoffmangel kann die Entwicklungsdauer genauso verlängern wie erhöhte Individuendichte im Brutgewässer und zu einem geringeren Wachstum der Larve führen, wobei die interaktiven Effekte von Futterknappheit durch intraspezifische Konkurrenz nicht klar voneinander zu trennen sind (Hawley, 1988; Estrada-Franco & Craig, 1995). Letztlich entwickeln sich die Larvenstadien der Männchen in der Regel schneller als die der Weibchen und sind auch kleiner als diese, wobei dieser Unterschied sich bei reduzierter Nährstoffsituation verstärkt (Hawley, 1988; Estrada-Franco & Craig, 1995). Beim Puppenstadium spielen die beiden Faktoren Nährstoffsituation und Individuendichte keine Rolle, allerdings sind die Faktoren Geschlecht und Temperatur diejenigen, die einen entscheidenden Einfluss auf die Entwicklungsdauer der Puppen ausüben – mit derselben Tendenz wie beim Larvenstadium (Estrada-Franco & Craig, 1995; Delatte et al., 2009). Insgesamt dauert so die Entwicklung von der $L_1$-Larve bis zum Adultus minimal 8,8 (±0,6) Tage, wobei sich Männchen insgesamt schneller entwickeln (Hawley, 1988; Estrada-Franco & Craig, 1995; Delatte et al., 2009). Als adultes Tier haben die Männchen eine geringere Lebenserwartung als die Weibchen (Delatte et al., 2009). Letztere können nun mehrfach in den gonotrophischen Zyklus eintreten, in dem im Wechsel auf eine Blutmahlzeit eine Eiablage folgt (Vinogradova, 2000; Delatte et al., 2009). Dabei erhöht sich die Lebensdauer abhängig von der Verfügbarkeit von Wirten und vollzogenen Blutmahlzeiten (Estrada-Franco & Craig, 1995). Unter Laborbedingungen liegt die Lebensdauer bei bis zu 117 Tagen, und es wurden bis zu acht gonotrophe Zyklen gemessen (Estrada-Franco & Craig, 1995; Delatte et al., 2009). Mit einer zuckerreichen Ernährung im Labor können Weibchen von *Ae. albopictus* autogen werden und eine erste Eiablage auch ohne vorherige Blutmahlzeit vollziehen (Estrada-Franco & Craig, 1995).

# 1.4 Charakterisierung des Eis von *Aedes (Stegomyia) albopictus*

Die Eier von *Ae. albopictus* sind rund 0,6 mm lang und 0,2 mm breit und zigarrenförmig, wobei das anteriore Ende eher stumpf und das posteriore Ende eher spitz zuläuft (siehe Abbildung 3) (Estrada-Franco & Craig, 1995).

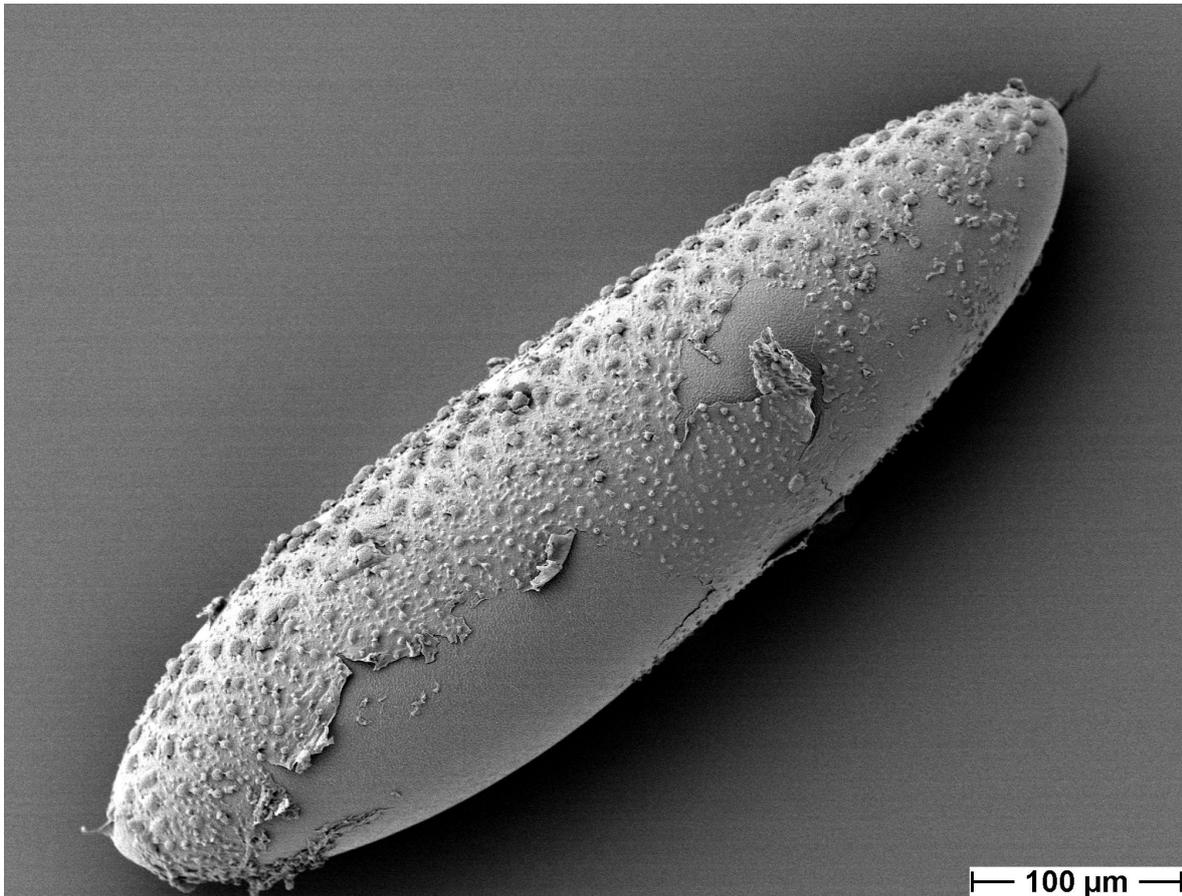

**Abbildung 3 (eigenes Werk, als Teilbild veröffentlicht in Kreß et al 2016a): Rasterelektronenmikroskopische Aufnahme eines Eis von *Aedes albopictus*. Das Exochorium ist durch die Präparation teilweise vom Endochorion abgelöst.**

Die Oberfläche ist durch wachsige Tuberkel strukturiert, die für die Orientierung auf der Wasseroberfläche verantwortlich sein sollen und vor allem ab den 1990er Jahren für die Bestimmungsschlüssel der Culicidae intensiv untersucht wurden (Hinton & Service, 1969; Sahlén, 1996). Stechmückeneier werden in der Regel erst kurz vor der Eiablage befruchtet (Estrada-Franco & Craig, 1995). Durch die dorso-ventrale Rotation des Embryos liegt die dormante $L_1$-Larve mit dem Kopf zum posterioren Ende hin und „blickt" auf die dorsale Seite des Eis (Valle et al., 1999). Eingehende Beschreibungen zum Aufbau sowie der Ontogenese der Schichten der Eihülle wurden in Kreß et al. (2016a) zusammengefasst (vgl. Kapitel 2.2 und Annex A.1.2).

Die Eier sind vor allem drei Risiken ausgesetzt, die im Folgenden behandelt werden: Prädation, Austrocknung und Frost (Hawley, 1988; Estrada-Franco & Craig, 1995).

## 1.4.1 Prädation

Die Prädation der Eier von *Ae. albopictus* geht hauptsächlich von anderen Arthropoden, wie vor allem Ameisen, aus und ist abhängig von der Gelegedichte (Estrada-Franco & Craig, 1995). Sie ist mit 50–90% Mortalität ein entscheidender Faktor für die Populationswachstumsraten dieser Art (Estrada-Franco & Craig, 1995). Die gattungstypische Streuung der Eier durch die Weibchen mittels Einzeleiablage an der Wasserkante selbst in unterschiedlichen Gewässern ist daher nicht verwunderlich, steht jedoch im Kontrast zu der Ablage ganzer Eierschiffchen von Arten z.B. der Gattungen *Culex* oder *Culiseta*, die jedoch auch eine chemische Verteidigungsstrategie durch Vergrämungsstoffe aufweisen (Hinton, 1981; Estrada-Franco & Craig, 1995; Vinogradova, 2000). Arten der Anophelinae legen dagegen mit Schwimmelementen ausgestattete Einzeleier frei treibend auf der Wasseroberfläche ab (Kettle, 1984).

## 1.4.2 Austrocknung

Im Anschluss an eine Entwicklungsphase des Embryos, in der die Serosa-Cuticula (engl. „serosal cuticle") gebildet wird, gefolgt vom Aufbau einer Wachsschicht und der Sklerotisierung durch Chitin, entwickeln die Eier der Arten des Tribus Aedini eine Trockentoleranz, für die sie weithin bekannt geworden sind (Telford, 1957; Beckel, 1958; Harwood & Horsfall, 1959; Furneaux & McFarlane, 1965a; Hinton, 1981; Sota & Mogi, 1992a, 1992b; Sota et al., 1993). Für die Eier von *Ae. albopictus* wurde nach einer zweimonatigen Trockenphase (60–70% relative Luftfeuchte bei 25 °C) eine Überlebensrate von 97,4% dokumentiert sowie eine maximale Überlebensrate von bis zu 243 Tagen (70–75% relative Luftfeuchte bei 25 °C) gemessen (Estrada-Franco & Craig, 1995). Damit ist unter den bisher untersuchten Arten der Familie der Culicidae lediglich *Ae. aegypti* noch trockentoleranter (Sota & Mogi, 1992a). Die volle Trockentoleranz wird jedoch nur bei einer Diapause der Eier von *Ae. albopictus* induziert (Wang, 1966; Sota & Mogi, 1992b). Ausgelöst durch eine reduzierte Photoperiode (≤ 13–14 Stunden Licht pro Tag) in Kombination mit nicht zu hohen Temperaturen (≤ 26 °C) produzieren Weibchen diapausierende Eier, in denen die dormante $L_1$-Larve mit reduziertem Metabolismus und reduzierter Schlupfaktivität suboptimale Jahreszeiten durch Hibernation überbrückt (Mori et al., 1981; Pumpuni et al., 1992; Estrada-Franco & Craig, 1995).

## 1.4.3 Frost

Durch die Diapause der Eier erhöht sich jedoch nicht nur die Trockentoleranz, sondern auch die Toleranz gegenüber Temperaturen unter 0 °C – dieses Phänomen wird „*cold hardiness*" genannt und als einer der entscheidenden Faktoren für die Ausbreitung der Art in gemäßigten Breitengraden angesehen (Hanson, 1991; Hanson & Craig, 1994; Estrada-Franco & Craig, 1995). Durch die Fähigkeit der Eier von *Ae. albopictus* zu einer Kälteakklimatisierung (0–5 °C) kann die Kältetoleranz weiter erhöht werden (Hanson & Craig, 1994). Temperatur-Grenzwerte liegen im Bereich von bis zu 15 Tagen bei -5,0 °C, bis zu 24 Stunden bei -10 °C und bis zu 1 Stunde bei -12 °C (Hanson & Craig, 1995a; Mogi, 2011; Thomas et al., 2012).

Die Fähigkeit zur Diapause und Kälteakklimatisierung mancher Populationen von *Ae. albopictus* ist ein Phänomen, das seit der raschen globalen Ausbreitung dieser Art diskutiert wird. Wurde zunächst angenommen, dass es stabile Populationen sind, da die Frosttoleranz der Populationen z.B. in den USA von einem frosttoleranten Stamm mit nördlicher Verbreitung im asiatischen Ursprungsgebiet herrührt (Hawley, 1988), so geht man mittlerweile von einer schnellen Etablierung und Veränderung vor Ort aus, die als „asymmetrische Evolution", „rapide Evolution", „rapide lokale Selektion" „rapide selektive Kontrolle" oder „rapide adaptive Evolution" bis hin zu „Adaptation" bezeichnet worden ist (Focks et al., 1994; Kobayashi et al., 2002; Lounibos et al., 2003; Urbanski et al., 2012; Bonizzoni et al., 2013). Eigene Untersuchungen beschäftigen sich in Kreß et al. (2016b) näher mit diesem Phänomen (vgl. Annex A.1.3).

In gemäßigten Breiten können frosttolerante Stämme von *Ae. albopictus* sich durch ihre an suboptimale saisonale Witterungen angepassten Eier gegenüber tropischen und subtropischen Stämmen und Arten durchsetzen, da sie nach strengen Wintern mit ihren diapausierenden Eiern eine höhere Schlupfrate haben (Hawley, 1988; Costanzo et al., 2005a). Dies zeigt sich beispielsweise in den USA bei der flächendeckenden Verdrängung von *Wyeomyia*-Arten durch *Ae. albopictus*, die seit der Einschleppung in der Region gemeinsam in Blattachseln von Bromelien brüten (Estrada-Franco & Craig, 1995).

Zurückzuführen ist dieses Phänomen jedoch nicht wie bei vielen frosttoleranten Arthropoden auf die erhöhte Einlagerung von sogenannte Polyolen (niedermolekulare, meist alkoholische Antifrostverbindungen) und die damit verbundene Reduzierung des Punktes der unterkühlten Schmelze („supercooling point") (Hanson & Craig, 1995b). Im Anschluss an die Falsifizierung dieser Hypothese rückte die Idee in den Fokus, dass wie bei der Etablierung der Trockentoleranz der Eier eine vergrößerte Wachsschicht und deren Sklerotisierung die Ursache für die stärkere Kältetoleranz der Eier sein könnte (Furneaux & McFarlane, 1965b; Rezende et al., 2008); jedoch liegen noch keine Daten über die Dicken der Eihüllen von Stechmücken vor (Farnesi et al., 2015). Eigene Untersuchungen beschäftigen sich mit den hieraus entstehenden Fragen in Kreß et al. (2016a) (vgl. Annex A.1.2).

### 1.4.4 Schlupfreiz

Ein wichtiger Faktor für das Auslösen des Schlupfreizes bei *Ae. albopictus* ist – neben Alter, Austrocknungszustand und Temperatur – eine Reduzierung des gelösten Sauerstoffgehalts im umgebenden Wasser (Imai & Maeda, 1976; Hawley, 1988). Es wird davon ausgegangen, dass dies von der dormanten Larve im Ei als Signal für eine Eutrophierung in Zusammenhang mit einem hohen mikrobiellen Abbau wahrgenommen wird und gute Lebensbedingungen signalisiert (Hawley, 1988). Jedoch schlüpfen meist nicht alle Eier auf einmal, sondern es kommt zum Schlupf von gewissen Kohorten (Estrada-Franco & Craig, 1995). Es wird vermutet, dass dieses sogenannte „installment hatching" (Ratenschlupf) für eine zeitliche Ausbreitung sorgt, um so die Risiken von suboptimalen Konditionen wie Austrocknung des Standgewässers oder Prädation für die Larven eines kompletten Eigeleges zu verringern (Estrada-Franco & Craig, 1995).

# 1.5 Charakterisierung der Larve von *Aedes* (*Stegomyia*) *albopictus*

Die Larven von *Ae. albopictus* werden meist in stehenden oder sehr langsam fließenden temporären Kleinstgewässern mit mäßiger Trübung vorgefunden (Estrada-Franco & Craig, 1995). In der Natur reicht die Spanne der pH-Werte in diesen Gewässern von 5,2 bis 8,35 (Estrada-Franco & Craig, 1995). Dabei scheint ein neutraler pH-Wert von 6,8 bis 7,6 optimal (Estrada-Franco & Craig, 1995). Larven kommen auch bei geringen Mengen von gelöstem Sauerstoff im Wasser vor (1,3 mg/L), da es ihnen aufgrund ihres Siphons möglich ist, Luftsauerstoff zu atmen (Estrada-Franco & Craig, 1995). Die höchsten Populationsdichten wurden in Mikrohabitaten mit einer β-Mesosaprobie gefunden (Estrada-Franco & Craig, 1995). Zwar erwiesen sich α-mesosaprobische Mikrohabitate auch als geeignet, jedoch fiel dort die Larvendichte geringer aus (Estrada-Franco & Craig, 1995).

Die Larven von *Ae. albopictus* verfügen über eine gewisse Resistenz gegenüber Austrocknung: Bis zu einem Tag können $L_3$- und $L_4$-Larven auf Filterpapier bei 26 bis 30 °C und 87% rellative Luftfeuchte überleben (Hawley, 1988). Im Labor wurde gezeigt, dass sich die Larven auch komplett ohne externe Nahrung entwickeln können, wobei vermutet wird, dass dies nur durch die Verwertung der eigenen Kadaver eines Geleges möglich ist (Hawley, 1988). Arten der Gattung *Aedes* haben unterschiedliche Strategien zur Nahrungssuche: Sie hängen kopfüber mit ihrem Siphon an der Wasseroberfläche angeheftet und sammeln und filtern Plankton und Kleinstpartikel durch ihren Bürstenapparat (Kettle, 1984). Sie gehen auch aktiv auf Futtersuche in der Wassersäule oder weiden Oberflächen ab (Kettle, 1984). Zum Teil ist es Arten der Gattung *Aedes* auch möglich, Blattwerk zu zerbeißen, weshalb sie mitunter als Vertreter der „Zerkleinerer" geführt werden (Kettle, 1984).

Die Überlebensrate der Larven von *Ae. albopictus* ist hauptsächlich von drei Faktoren begrenzt: Prädation, Konkurrenten und Parasiten sowie Pathogene.

## 1.5.1 Prädation

Wichtige Räuber für Stechmückenlarven in der Throphodynamik von temporären Kleinstgewässern sind zu allererst Strudelwürmer (Turbellaria) der Gattung *Mesostoma*, da sie nach einem erneuten Fluten eines trockengefallenen Systems relativ früh aus ihren Dauereiern schlüpfen (Kumar & Hwang, 2006).

Unter den Crustaceen gibt es nicht nur die Notostraca (Schildkrebse), welche sich nachweislich von Stechmückenlarven ernähren; hier sind vor allem auch cyclopoide Copepoden (Ruderfußkrebse) zu nennen (Kumar & Hwang, 2006). Arten wie *Mesocyclops thermocyclopoides* zeigen sogar ein derart ausgeprägtes Jagdverhalten, dass sie mit 30 bis 40 getöteten Stechmückenarven pro Tag ihren eigenen Nahrungsbedarf übersteigen (Kumar & Hwang, 2006).

Kannibalisch und räuberisch lebende Larven aus Stechmückengattungen wie *Toxorhynchites*, *Psorophora* (Untergattung *Psorophora*), *Aedes* (Untergattung *Mucidus*) sowie der Art *Culex* (*Lutzia*) *tigripes* sind bekannte Stechmückenprädatoren und besiedeln annähernd dieselben Habitate wie *Ae. albopictus*, weswegen von einer natürlichen Prädation durch diese Arten ausgegangen werden kann (Kettle, 1984; Hawley, 1988; Estrada-Franco & Craig, 1995; Basabose, 1996). Weit verbreitete invertebrate Top-Prädatoren von fischarmen Kleinstgewässern, wie aquatische Coleoptera (Käfer) aus der Familie der Dytiscidae (Schwimmkäfer) und Nepomorpha (Wasserwanzen) der Gattung Notonectidae (Rückenschwimmer) oder Belostomatidae (Riesenwanzen) der Gattung *Spaerodema* sind bekannt für ihre Prädation an Stechmückenlarven und werden versuchsweise zur biologischen Schädlingsbekämpfung eingesetzt (Hawley, 1988; Kumar & Hwang, 2006). Bei den Odonata (Libellen) machen nicht nur die aquatischen Entwicklungsstadien Jagd auf Stechmückenlarven, auch die Adulten sind effektive Räuber von schwärmenden Stechmücken (Kumar & Hwang, 2006).

In größeren und weniger austrocknungsbedrohten Habitaten, wie Überschwemmungsgebieten oder Reis-Anbauflächen, werden weltweit auch vertebrate Prädatoren, darunter 253 Fischarten, als biologische Stechmückenbekämpfung eingesetzt (Walton, 2007). Dabei kommt den beiden Fischarten *Gambusia affinis* und *G. holbrooki* (auch als westlicher und östlicher Moskitofisch bezeichnet) eine besondere Bedeutung zu (Walton, 2007). In größeren Gewässertypen stellen auch Kaulquappen von Fröschen eine weitere nennenswerte Prädatorengruppe dar; sie sind aber in tropischen Gebieten z.B. auch in temporären Phytotelmata zu finden (Kumar & Hwang, 2006).

Es wurde gezeigt, dass Stechmücken Vermeidungsstrategien entwickelt haben: Durch Chemosensoren an Tarsen, Antennen und den Spitzen des Proboscis überprüft das Weibchen das potentielle Ablagegewässer auf Kairomone von Prädatoren wie Fischen und legt in deren Gegenwart signifikant weniger Eier ab, um den Bruterfolg zu verbessern (Kumar & Hwang, 2006).

## 1.5.2 Konkurrenten

Die Konkurrenz der Larven scheint einen stärkeren Einfluss auf die Abundanz von *Aedes*-Arten zu haben als ihre Prädation (Meyabeme Elono et al., 2010). In Anbetracht dessen, dass Stechmücken wie *Ae. albopictus* tendenziell meist temporäre Kleinstgewässer besiedeln, deren Biozönose oft aufgrund der Größe und Kurzlebigkeit des Gewässers keine limnischen Prädatoren aufweist, wird dies verständlich (Kumar & Hwang, 2006). Einer der wohl stärksten Konkurrenten ist die ebenfalls invasive, ursprünglich aus Afrika stammende *Ae. aegypti*, die annähernd dieselben Brutgewässer und Makrohabitate nutzt wie *Ae. albopictus* (Hawley, 1988). Auch wenn die Dokumentation in manchen Bereichen lückenhaft ist, wurde klar, dass in großen Städten in Südost-Asien, in denen *Ae. albopictus* natürlicherweise vorherrschte, *Ae. aegypti* einwanderte und nun die dominante Art darstellt (Hawley, 1988). Dies hat auch damit zu tun, dass sich in dieser Zeitspanne der urbane Raum weiter verdichtete und sich sowohl das Makrohabitat als auch die Mikrohabitate und die von *Ae. albopictus* benötigte Restvegetation verloren gingen (Hawley, 1988; Murrell & Juliano, 2008; Higa, 2011; Kraemer et al., 2015). Auch in Laborversuchen wurde gezeigt, dass *Ae.-aegypti*-Larven unter suboptimalen Futterbedingungen kompetitiv überlegen sind (Murrell & Juliano, 2008). Andererseits verdrängte *Ae. albopictus* bei ihrer Expansion im Südosten der USA *Ae. aegypti* in weiten Teilen (Lounibos et al., 2002). Dort wird *Ae. aegypti* nur noch in Brutgewässern mit sehr hohen Temperaturen gefunden (33–37 °C) und scheint bei solchen Temperaturen kompetitiv überlegen zu sein (Lounibos et al., 2002).

Gegen andere Stechmückenarten, etwa die nahe verwandten *Ae.* (*Stegomyia*) *guamensis*, *Ae.* (*Stegomyia*) *polynesiensis* oder auch *Ae.* (*Ochlerotatus*) *triseriatus*, *Ae.* (*Ochlerotatus*) *japonicus* und *Culex pipiens*, scheint sich *Ae. albopictus* unter Laborbedingungen tendenziell besser durchsetzen zu können, auch wenn die Ergebnisse von zwangsvergesellschafteten Laborversuchen und den wirklichen Vergesellschaftungen im Freiland mit hinzukommenden biotischen und abiotischen Faktoren nur bedingt extrapolierbar sind (Hawley, 1988; Livdahl & Willey, 1991; Carrieri et al., 2003; Costanzo et al., 2005b, 2011; Armistead et al., 2008). Daher gibt es mehrere Regionen, in denen *Ae. albopictus* konkurrenzstärker als andere vergesellschaftete Arten ist, was vor allem auf eine bessere Futterverwertung des vorhandenen Detritus zurückzuführen ist (Juliano, 1998; Lounibos et al., 2002; Bevins, 2007; Murrell & Juliano, 2008). Dagegen korreliert die Präsenz von Futterkonkurrenten in temporären Flachgewässern wie Arten der Ordnung Cladocera (*Ceriodaphnia*, *Chydorus*, *Daphnia*, *Simocephalus*), der Ordnung Copepoda (*Calanoida*) sowie Dipteren-Larven (Chironomidae) negativ mit der Abundanz von *Aedes*-Arten (Kumar & Hwang, 2006; Meyabeme Elono et al., 2010).

Intraspezifische Konkurrenz spielt nur bei hohen Dichten eine Rolle und es zeigt sich, dass *Ae. albopictus* im Vergleich zu Stechmückenarten wie *Ae. aegypti, Ae. japonicus*, *Ae. triseriatus* oder *Cx. pipiens* geringere Fitnesseinbußen zu verzeichnen hat (Bevins, 2007; Armistead et al., 2008; Murrell & Juliano, 2008; Costanzo et al., 2011).

### 1.5.3 Parasiten und Pathogene

*Bacillus thuringiensis israelensis* sowie *B. thuringiensis sphaericus* sind grampositive, fakultativ anaerobe entomopathogene Bakterien, die ubiquitär in Böden und Gewässern vorkommen und mit sogenannten B.t.i.-Kristalloproteinen (auch Bt-Toxin genannt) assoziiert dormante Sporen produzieren, die für Stechmücken toxisch sind – nicht jedoch für Wirbeltiere (Kumar & Hwang, 2006). *Sacchoropolyspora spinosa*, ebenfalls ein Vertreter der grampositiven Bakterien, bildet das Toxin Spinosad, welches hochspezifisch auf Stechmückenlarven wirkt (Kumar & Hwang, 2006). Beide Toxine werden in der Stechmückenbekämpfung eingesetzt, wobei ersterem mengenmäßig eine weitaus größere Bedeutung zukommt (Kumar & Hwang, 2006). Oomycota (Scheinpilze), wie *Lagenidium giganteum*, leben parasitisch in Insektenlarven und infizieren Stechmückenlarven effektiv (Kumar & Hwang, 2006). Sie bilden zudem trockentolerante Stadien aus, die in den temporären Gewässern mehrere Jahre überdauern können (Kumar & Hwang, 2006). Die Sporozoide von Protozoen wie *Ascogregarina taiwanensis* oder *Ascogregarina culicis* sondern Toxine ab, die vor allem das Darmepithel von Stechmückenlarven angreifen und so massiv die Nahrungsverwertung einschränken (Kumar & Hwang, 2006). Schließlich sind Nematoden der Familie Mermithidae wie *Romanomermis iyengari* effektive Parasiten für viele Stechmückenlarven (Kumar & Hwang, 2006).

## 1.6 Charakterisierung der Puppe von *Aedes* (*Stegomyia*) *albopictus*

Die Puppe von *Ae. albopictus* ist durch die Verschmelzung von Kopf und Thorax von der Gestalt her kompakter als die Larve und nimmt keine Nahrung zu sich, wodurch eine Reihe von Insektiziden kaum nennenswerte Effekte auf dieses Entwicklungsstadium haben (Kettle, 1984). Abgesehen von Störungen, bei denen die Puppe sich mit schnellen Schlägen der am neunten und letzten Segment befindlichen Paddel in tiefere Gewässerschichten begibt, verharrt sie meist regungslos mit zwei vom Cephalothorax abgehenden Atemröhren an der Wasseroberfläche (Kettle, 1984). Unter bestimmten Bedingungen können die Puppen ihre komplette Entwicklungsdauer an Land verbringen, da entweder die Puppe selbst oder das sich darin befindliche adulte Tier eine gewachste und/oder im Vergleich zur Larve sklerotisiertere Cuticula besitzt (z.B. 2 Tage bei 26–30 °C und 87% relativer Luftfeuchte) (Kettle, 1984; Hawley, 1988). Die Emergenz erfolgt, indem sich Luft zwischen der Cuticula der Puppe und der Cuticula des mittlerweile entwickelten Adultus sowie im Mitteldarm bildet und das Tier sowohl auftreiben, als auch die Cuticula der umhüllenden Puppe aufreißen lässt (Kettle, 1984; Clements, 1993). Der Schlupf dauert wenige Minuten, und das adulte Tier ist nach rund einer Stunde und einer Diurese flugfähig (Kettle, 1984).

# 1.7 Charakterisierung des Imagos von *Aedes* (*Stegomyia*) *albopictus*

Die Flugreichweite ist bei *Ae. albopictus* sehr gering. Mark-Release-Capture-Studien zeigten eine maximale Flugreichweite der Weibchen von 525 m, Männchen kamen lediglich auf 225 m, und 90% der Tiere entfernten sich im Laufe ihres Lebens nicht mehr als 100 m vom Ort der Freisetzung (Estrada-Franco & Craig, 1995). Die Tiere halten sich meist nicht in Gebäuden auf, sondern suchen windgeschützte Vegetation als Ruheplatz auf (Estrada-Franco & Craig, 1995; Bonizzoni et al., 2013). Die Männchen ernähren sich als Imago ausschließlich von Pflanzensäften und Nektar (Clements, 1993; Estrada-Franco & Craig, 1995). Paarungsschwärme von Männchen bilden sich entweder an optischen Markierungen wie an Baumstümpfen, Metall-Containern oder in der Nähe von potentiellen Wirten (Estrada-Franco & Craig, 1995). Angezogen fühlen sich die Männchen durch das Fluggeräusch der Weibchen, eine Arterkennung erfolgt jedoch mit Chemorezeptoren (Hawley, 1988; Estrada-Franco & Craig, 1995). Wenn Männchen von *Ae. albopictus* artfremde Weibchen begatten, kann es zum sogenannten Satyrismus kommen (Estrada-Franco & Craig, 1995). Das bedeutet, dass das Sperma den weiblichen Genitaltrakt für weitere Besamungen durch das Kopulationshormon „Matrona" verschließt und das Weibchen somit sterilisiert (Estrada-Franco & Craig, 1995). Andererseits können Weibchen von *Ae. albopictus* nicht durch die Begattung von anderen Arten sterilisiert werden, wodurch dieses Phänomen eine Rolle bei der Verdrängung von Arten wie *Ae. aegypti* spielen kann (Estrada-Franco & Craig, 1995). Nach der regulären Paarung tritt das Weibchen von *Ae. albopictus* in den sogenannten gonotrophen Zyklus ein, sucht einen Wirt für eine Blutmahlzeit auf und führt je nach Temperatur zwischen 2 und 5 Tagen später eine Oviposition durch, wohingegen die Männchen nach rund 7 Paarungen nur noch eine geringe Lebenserwartung haben (siehe Abbildung 4) (Hawley, 1988; Estrada-Franco & Craig, 1995; Vinogradova, 2000; Delatte et al., 2009). Pro gonotrophem Zyklus kann ein Weibchen auch mehrere Blutmahlzeiten von unterschiedlichen Wirten zur Bildung der reifen Eier aufnehmen, was noch unbekannte epidemiologische Konsequenzen haben könnte, da die meisten Studien nur abgeschlossene gonotrophe Zyklen berücksichtigen und so Vektorkapazität und -kompetenz unterschätzt werden können (Estrada-Franco & Craig, 1995).

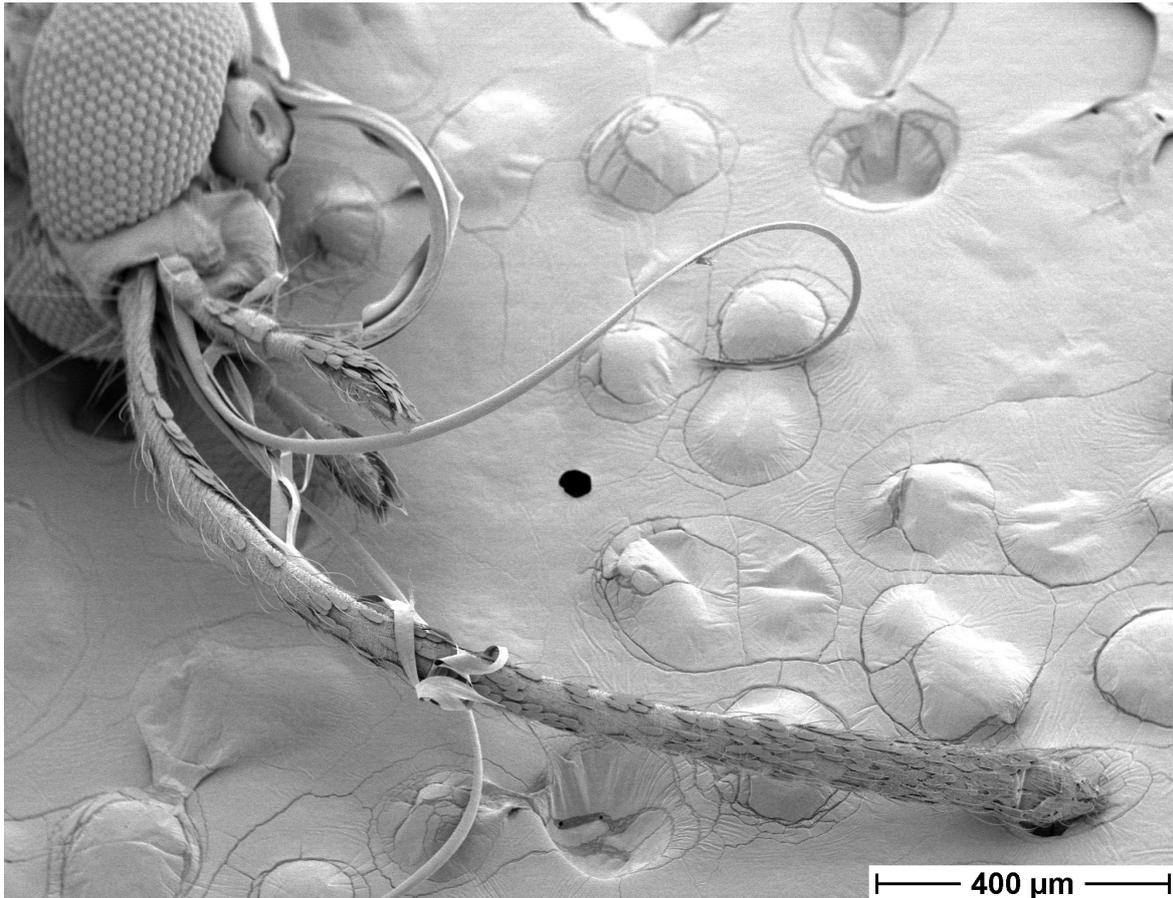

**Abbildung 4 (eigenes Werk, unveröffentlicht): Rasterelektronenmikroskopische Aufnahme der Mundwerkzeuge zur Aufnahme von Blutmahlzeiten eines adulten Weibchens von *Aedes albopictus*.**

## 1.7.1 Wirtssuche

Die adulten Weibchen von *Ae. albopictus* suchen sich recht opportunistisch ihre Blutmahlzeiten innerhalb einer großen Spannbreite vertebrater Wirte (Estrada-Franco & Craig, 1995). Abhängig von der Population und den verfügbaren Wirten besitzt *Ae. albopictus* jedoch auch klare Präferenzen: Die Zoophilie fokussiert sich vor allem auf eindeutige Warmblüter, wie Säugetieren gefolgt von Vögeln, jedoch sind auch vereinzelte Blutmahlzeiten bei endothermen Tieren, wie Reptilien, Amphibien und Schildkröten, dokumentiert worden; im Labor auch an Invertebraten, wie afrikanischen Landschnecken und Raupen des Seidenspinners, gezeigt worden (Hawley, 1988; Estrada-Franco & Craig, 1995; Paupy et al., 2009; Bonizzoni et al., 2013). Innerhalb der Klasse der Mammalia zeigt *Ae. albopictus* eine starke Anthropophilie, sticht nachweislich aber auch Kaninchen, Hirsche, Hunde, Katzen, Eichhörnchen, Opossums und Hornträger; in der Klasse Aves werden Sperlingsvögel, Taubenvögel und Schreitvögel bevorzugt (Estrada-Franco & Craig, 1995; Bonizzoni et al., 2013).

Die Weibchen stechen überwiegend außerhalb von Bauwerken und werden im Gegensatz zu *Ae. aegypti* als ektophil eingestuft (Estrada-Franco & Craig, 1995; Bonizzoni et al., 2013). Die Hauptaktivität ist fast ausschließlich tagsüber (Estrada-Franco & Craig, 1995). Dabei gibt es eine bimodale Präferenz der Morgen- und Abendstunden (Estrada-Franco & Craig, 1995). Das Aufsuchen eines Wirts erfolgt in zwei Phasen: Die erste Phase wird durch einen Anstieg des $CO_2$-Gehalts der Luft ausgelöst und veranlasst das ruhende Weibchen zu einem randomisierten Flug (Estrada-Franco & Craig, 1995). In der zweiten Phase nutzt das Weibchen ab einer Entfernung von 4 bis 5 m weitere Sinnesorgane, um durch Anstieg der Feuchte, visuelle Faktoren und volatile organische Chemikalien wie Fettsäuren (und weitere Metaboliten von mikrobiologischen Kulturen der Haut) den Wirt zu erkennen und direkt anzufliegen (siehe Abbildung 5) (Estrada-Franco & Craig, 1995; Knols, 1996; Fernández-Grandon et al., 2015).

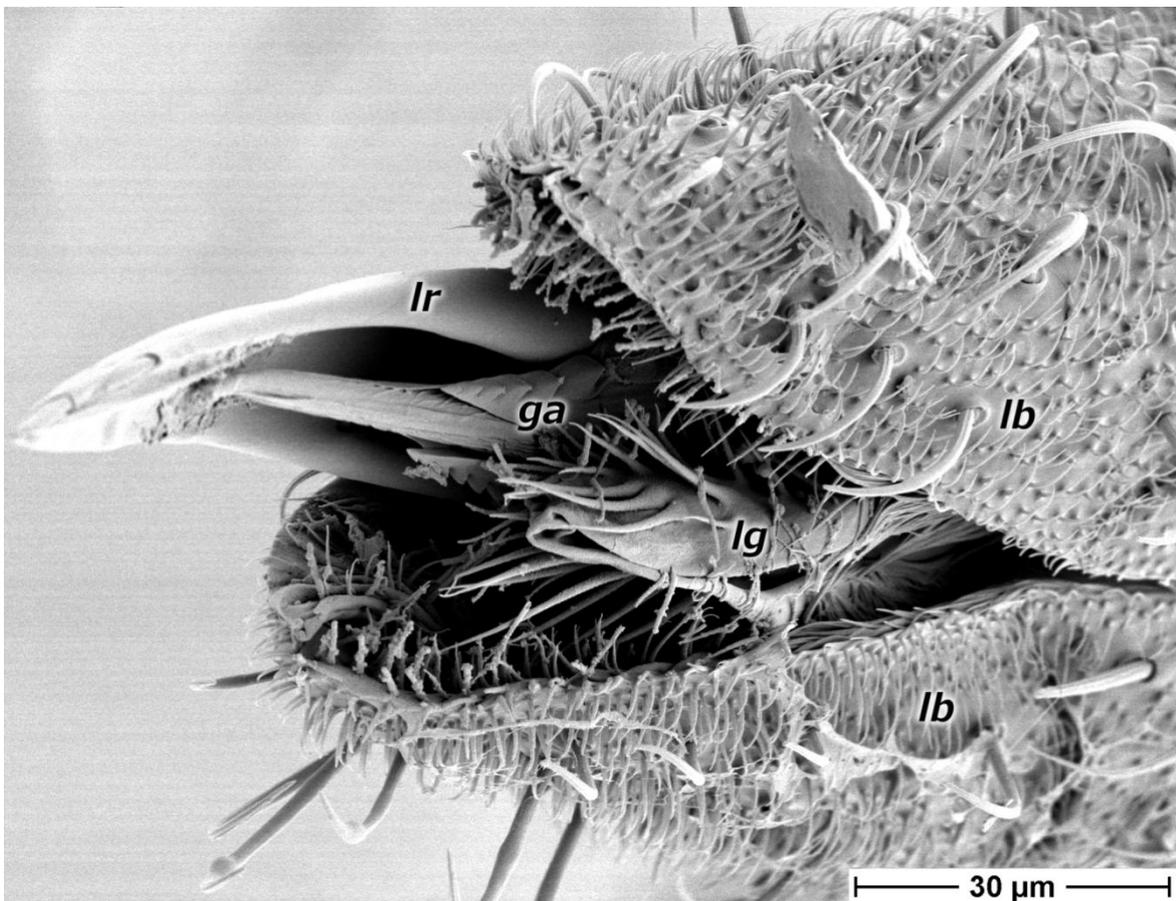

**Abbildung 5 (eigenes Werk, unveröffentlicht): Rasterelektronenmikroskopische Aufnahme des Stechapparats (Labium) von *Aedes albopictus* zur Penetrierung der Haut des Wirtes und Aufnahme der Blutmahlzeit. Im Inneren der beiden mit Sensillen ausgestatteten Labellae (*lb*) liegen die unterschiedlichen Kompartimente wie Ligula (*lg*), Labrum (*lr*) und die mit Widerhaken versehenen Galea (*ga*), die zum besseren Vortrieb in die Wirtshaut dienen (Beschriftung nach Christophers, 1960).**

## 1.7.2 Oviposition

Pro gonothrophischem Zyklus legt ein Weibchen abhängig von der Körpergröße und der Qualität und Größe der Blutmahlzeit (mind. 0,7 mg Blut, sonst keine Reifung, i.d.R. 0,8–2,5 mg) in seiner Lebensspanne in der Natur 51–74 Eier ab (bis zu 283–344 Eier unter Laborbedingungen) (besprochen in Hawley, 1988; Estrada-Franco & Craig, 1995; Delatte et al., 2009). Die Weibchen legen die Einzeleier meist tagsüber am ehesten auf dunklen und rauen Natursubstraten oberhalb der Wasserkante kleiner, temporärer und ebenerdig gelegener (bis max. 15 m Höhe) Gewässer mit mittlerer Saprobie ab (Hawley, 1988; Estrada-Franco & Craig, 1995). Dabei spielen, wie oben genannt, viele chemische Faktoren, wie Kairomone, Qualität der Nahrung für die Larven und die Sauerstoffsättigung, eine wichtige Rolle (Hawley, 1988; Estrada-Franco & Craig, 1995). Wenn es ihnen möglich ist, suchen die Weibchen pro Eiablage mehrere Gewässer auf, in denen sie Eier ablegen (Hawley, 1988; Estrada-Franco & Craig, 1995).

## 1.7.3 Vektorkompetenz und -kapazität

Als Vektor wird ein Organismus bezeichnet, der ein Pathogen von einem Wirt zum nächsten transferieren kann (Allaby, 2010). Die quantitative Isolation von Pathogenen aus Aufsammlungen (definiert als Vektorkapazität) ist ebenso wie auch der experimentelle Nachweis einer Übertragbarkeit der aufgenommenen Pathogene (definiert als Vektorkompetenz) aber noch kein lückenloser Beleg dafür, dass eine Stechmückenart auch eine signifikante Rolle bei der Übertragung der Pathogene in freier Wildbahn spielt (Mullen & Durden, 2009; Paupy et al., 2009; Vega-Rua et al., 2013).

Im Jahr 1882 veröffentlichte der spanisch-kubanische Augenarzt und Wissenschaftler Carlos Juan Finlay seine rein auf Observation und Deduktion beruhende Hypothese, dass das gefürchtete Gelbfieber in einer Klasse von Insekten gefunden werden kann, die durch eine Penetration in das Innere von Blutgefäßen das Blut zusammen mit den darin enthaltenen „infektiösen Partikeln" aufsaugt und sie dadurch vom Erkrankten zum Gesunden überträgt (Finlay, 1882; Bullock, 2011). Er identifizierte später *Ae. aegypti* als den einzigen potentiellen Überträger und wies durch Menschenversuche, u.a. an sich selbst, schließlich die Übertragung nach (Finlay, 1882; Bullock, 2011). Diesen Nachweis erbrachte er 11 Jahre vor Dmitri Iwanowskis Entdeckung, dass Viren Krankheitserreger sind, und 16 Jahre bevor Ronald Ross den experimentellen Nachweis lieferte, dass Malaria durch Stechmücken übertragen wird (Bullock, 2011). Letzteres leitete Ross von der in den 1890er Jahren von

Patrick Manson und Alphonse Laveran aufgestellten Hypothese ab, dass die Elephantiasis-Erkrankung eine von Stechmücken übertragene Fadenwurmerkrankung sei (Bullock, 2011). Ross gilt deshalb als Begründer der modernen Tropenmedizin, erhielt 1902 den Nobelpreis für Physiologie und Medizin, wurde in die Royal Society aufgenommen und zum Ritter geschlagen (Chernin, 1992; Bullock, 2011).

Lange Zeit wurde die Vektorkompetenz von *Ae. albopictus* jedoch von medizinischen Entomologen unterschätzt (Siler et al., 1926; Shroyer, 1986; Paupy et al., 2009). *Ae. albopictus* ist ein Vektor für Parasiten und sogenannte Arboviren (engl.: „arthropod-borne viruses", Viren, die von Arthropoden verbreitet werden) (Paupy et al., 2009). Es ist experimentell gezeigt worden, dass *Ae. albopictus* prinzipiell die Vektorkompetenz für 27 Viren besitzt, von denen 15 auch in Feldaufsammlungen in den Stechmücken nachgewiesen wurden (Tabelle 1) (Estrada-Franco & Craig, 1995; Moore & Mitchell, 1997; Paupy et al., 2009; Wong et al., 2013; Grard et al., 2014). Die belegte Vektorkapazität und -kompetenz sowie das Auftreten von Chikungunya- und Dengue-Epidemien zusammen mit infizierten *Ae. albopictus* in derselben Region deutet auf eine wichtige Rolle dieser Art bei der Übertragung mindestens dieser beiden bedeutenden humanpathogenen Viren hin (Paupy et al., 2009; Vega-Rua et al., 2013). Aufgrund des opportunistisch-zoophilen Stechverhaltens ist *Ae. albopictus* zusätzlich ein potentieller Brückenvektor zwischen unterschiedlichen Arten von Wirten, vor allem in neu besiedelten Gebieten, wie Zentralafrika, das ein bekannter „Hotspot" für Viren-Diversität und neu aufkommende, insbesondere zoonotische Viruserkrankungen ist (Paupy et al., 2009).

Dengue-Fieber ist die wohl bedeutsamste von *Ae. albopictus* übertragene Krankheit (Estrada-Franco & Craig, 1995). Hierzu gehören auch die besonders schweren Verlaufsformen des hämorrhagischen Dengue-Fiebers (DHF) und des Dengue-Schock-Syndroms (DSS), die beide am häufigsten bei Kindern auftreten und je nach Region eine Letalität von rund 10% haben können (Estrada-Franco & Craig, 1995): Wird man durch den Stich einer infizierten Mücke mit einem der vier Dengue-Virus-Serotypen infiziert (ein fünfter und wohl auf nichtmenschliche Primaten beschränkter Serotyp wird vermutet: Normile, 2013), so kann man gegen diesen bestimmten Serotyp immun werden. Bei Infektion mit einem anderen Dengue-Virus-Serotyp besteht dann aber das Risiko, dass anlässlich der ersten Infektion gebildete Antikörper als sogenannte infektionsverstärkende Antikörper (ADE) wirken und zu einem weitaus gravierenderen Krankheitsverlauf mit erhöhtem Risiko von DHF oder DSS beitragen (Fredericks & Fernandez-Sesma, 2014). Erstmals wurde die Übertragung der Erreger des Dengue-Fiebers durch *Ae. albopictus* im Jahr 1926 nachgewiesen (Siler et al., 1926); nach *Ae. aegypti* ist *Ae. albopictus* heute weltweit die wichtigste Überträgerin von Dengue-Viren. Aktuell geht die WHO weltweit von 22.000 Todesfällen durch Dengue-Fieber und knapp 100.000 Infektionen pro Jahr aus, wobei sich diese Zahl in den letzten 50 Jahren um das 30fache erhöhte (WHO, 2009). Neueren Modellrechnungen zufolge kommt es dagegen zu 390 Mio. (95% KI: 284–528 Mio.) Dengue-Infektionen pro Jahr (Brady et al., 2013). Die meisten unter diesen verlaufen asymptomatisch, aber 96 Mio. (95% KI: 67–136 Mio.) nehmen klinisch signifikante Verläufe (Bhatt et al., 2013).

Mit einer hohen Wahrscheinlichkeit wird auch Europa durch die erfolgte Invasion geeigneter Vektoren wieder eine Dengue-Region (Schaffner & Mathis, 2014). Einzelne Fälle lokal übertragener Dengue-Infektionen in Frankreich und Kroatien haben diese Möglichkeit bereits exemplarisch aufgezeigt (Schaffner & Mathis, 2014). Aktuell werden die weltweiten Kosten im Bruttosozialprodukt, die im Gesundheitswesen, durch krankheitsbedingten Ausfall von Arbeitskräften sowie durch Monitoring- und Bekämpfungsmaßnahmen entstehen, auf 39 Mrd. US-Dollar geschätzt (Selck et al., 2014). Ein Impfstoff gegen den Dengue-Virus (Dengvaxia®, Sanofi-Pasteur) durchlief bereits klinische Prüfungen an Kindern und später auch Erwachsenen im asiatisch-pazifischen Raum sowie Lateinamerika und wurde im Dezember 2015 in Mexiko trotz großer Kritik am geringen Impfschutz (z.B. nur 35% gegen den Serotyp DENV-2) sowie ethisch fragwürdiger Studiendesigns zugelassen (Sabchareon et al., 2012; Fredericks & Fernandez-Sesma, 2014; dpa, 2015). Andere Hersteller wie Takeda (DENVax®), National Institutes of Health (TetraVax-DV®), Fiocruz, GlaxoSmithKline, WRAIR (DPIV®), Merck & Co. und Vical befinden sich mit ihren Produkten in verschiedenen früheren Phasen der klinischen Prüfung (diskutiert in Fredericks & Fernandez-Sesma, 2014).

Die bisher zweitwichtigste von *Ae. albopictus* übertragene Virus-Erkrankung ist Chikungunya-Fieber (Fredericks & Fernandez-Sesma, 2014). Durch eine genetische Punkt-Mutation des Chikungunya-Virus und einer damit verbunden besseren Anpassung an den Vektororganismus scheint in manchen Regionen die Rolle des Hauptüberträgers von *Ae. aegypti* auf *Ae. albopictus* übergegangen zu sein (Bonizzoni et al., 2013). Unter anderem steigt deswegen in den letzten Jahren die Bedeutung von Chikungunya-Fieber, und es kam zu großen Epidemien mit millionenfachen Erkrankungsfällen (Higgs, 2014). Ein durch *Ae. albopictus* übertragener Ausbruch des Chikungunya-Fiebers in Italien im Jahr 2007 unterstreicht auf dramatische Weise die Bedeutung stabiler Populationen kompetenter Vektoren für die plötzliche Ausbreitung tropischer Pathogene in Europa (Angelini et al., 2007).

Im Herbst 2015 erschütterte Lateinamerika eine Zika-Virus-Epidemie zeitgleich mit einem erhöhten Auftreten u.a. von Fällen des Guillain-Barré-Syndroms und Mikrozephalie bei Neugeborenen, woraufhin im Februar 2016 der Generaldirektor der WHO den „öffentlichen Gesundheitsnotstand internationalen Ausmaßes" ausrief (WHO, 2016a, 2016b). Neben dem Hauptüberträger *Ae. aegypti* gilt *Ae. albopictus* erst seit kurzem als Vektor für das Zika-Virus, und seine Rolle in der Epidemiologie dieses Virus ist nicht abschließend geklärt (Grard et al., 2014; Diagne et al., 2015; Musso & Nhan, 2015; DVV e.V., 2016; WHO, 2016a). Anders als beim Guillain-Barré-Syndrom, in dessen Fall bei allen während der Zika-Epidemie Betroffenen Patienten auch Zika-Antikörper im Blut nachgewiesen werden konnten und eine sehr starke Korrelation eine Kausalität als sehr wahrscheinlich erscheinen lässt (Cao-Lormeau et al., 2016), sprechen bis jetzt lediglich einzelne Indizien für einen (mono)kausalen Zusammenhang zwischen Zika-Virus-Infektion und Mikrozephalie (DVV e.V., 2016; Oliveira Melo et al., 2016; Rasmussen et al., 2016). Das CDC schlägt in einer Metaanalyse auf Basis der „Shepard's Criteria for „Proof" of Teratogenicity in Humans" vor, dass ausreichend viele Evidenzen gefunden wurde, als dass man von einem kausalem Zusammenhang ausgehen kann (Rasmussen et al., 2016). Andere Erklärungsversuche für die gehäuften Fälle von Mikrozephalie vor allem in Brasilien kritisieren den Umstand, dass Zika-Epidemien in der Vergangenheit sowohl in Afrika als auch in Lateinamerika aufgetreten waren und damals keine nennenswerten Fehlentwicklungen dokumentiert wurden und auch bei der Zika-Epidemie 2015/16 in Lateinamerika das Auftreten von Zika-Infektionen und Mikrozephalie anders als beim Guillain-Barré-Syndrom nur partiell bis schlecht korreliert sind (PCST, 2016). Diskutiert wurden daher auch andere Erklärungsmöglichkeiten der Mikrozephalie, z.B. ein Zusammenhang mit Insektiziden wie Pyriproxyfen, welches seit 2014 vom brasilianischen Gesundheitsministerium zur Stechmückenbekämpfung dem Trinkwasser beigemischt wird (siehe Kapitel 1.8.4) (DVV e.V., 2016; PCST, 2016), oder Effekte von Co-Infektionen oder früherer Infektionen mit anderen Viren wie z.B. bestimmten Dengue-Virus-Serotypen (Branswell, 2016). Ein eindeutiger kausaler Zusammenhang ist jedoch auch hier noch nicht nachgewiesen.

Trotz der möglichen Vektorkompetenz für Gelbfieber sind die im Frühjahr 2016 stark steigenden Fallzahlen vor allem in Angola und die Ausbreitung auf die Demokratische Republik, Kongo und Kenia wahrscheinlich nicht auf *Ae. albopictus* zurück zu führen, da sie in den betroffenen Ländern über keine ausreichend großen Populationen zu verfügen scheint, sondern auf *Ae. aegypti* (WHO, 2016c).

Neben der horizontalen Übertragung von Wirt zu Wirt ist *Ae. albopictus* auch zu transovariellen, also vertikalen Übertragungswegen, in der Lage (Estrada-Franco & Craig, 1995).

So ist es Weibchen dieser Art möglich, Dengue-Virus, Japan-Enzephalitis-Virus, St.-Louis-Enzephalitis-Virus und St.-Angelo-Virus an ihre Nachkommen weiterzugeben, letzteres unter Laborbedingungen bis zu 38 Generationen in Folge (Estrada-Franco & Craig, 1995).

Die Erreger der von Zecken übertragenen Lyme-Borreliose (Bakterien wie *Borrelia afzelii*, *Borrelia bavariensis* und *Borrelia garinii*) wurden zwar in *Aedes*-Arten gefunden, bis jetzt gibt es jedoch noch keinen Nachweis, dass Stechmücken auch eine nennenswerte Vektorkompetenz besitzen (Melaun et al., 2016). Experimentell konnte gezeigt werden, dass *Ae. albopictus* in der Lage ist, Protozoen der Gattung *Plasmodium* zu übertragen: die Erreger der Geflügelmalaria *Plasmodium lophurae* und *Plasmodium gallinaceum* (Estrada-Franco & Craig, 1995). Veterinärmedizinisch bedeutsame und darüber hinaus fakultativ humanpathogene Parasiten, bei deren Übertragung *Ae. albopictus* zumindest im europäischen Kontext eine mittlerweile signifikante Rolle spielt, sind die Nematoden *Dirofilaria immitis* (Hunde-Herzwurm; Verursacher kardiovaskulärer Dirofilariose) und *Dirofilaria repens* (Verursacher kutaner Dirofilariose beim Menschen) (Cancrini et al., 1995; Genchi et al., 2009). Für *Setaria labiatopapillosa* (Setariose bei Ruminantia) wurde in *Ae. albopictus* die Entwicklung bis zur $L_3$ Larve dokumentiert (Cancrini et al., 1995) und für *Wuchereria bancrofti* (lymphatische Filariose und Elefantiasis) wurde lediglich eine äußerst geringe orale Aufnahme nachgewiesen (0,01%) womit eine Vektorkompetenz als sehr unwahrscheinlich gilt (Estrada-Franco & Craig, 1995; Pothikasikorn et al., 2008).

**Tabelle 1 (eigenes Werk, unveröffentlicht, auf Datenbasis von Shroyer, 1986; Paupy et al., 2009; Wong et al., 2013; Grard et al., 2014): Aufzählung und Systematik von Viren, die aus *Ae. albopictus* isoliert wurden oder für die im Labor eine Vektorkompetenz nachgewiesen wurde.**

| Familie | Genus | Serumsgruppe | Art | Feldisolation | Infektion | Übertragung |
|---|---|---|---|---|---|---|
| Flaviviridae | Flavivirus | | Dengue-Virus | + | + | + |
| | | | Gelbfieber-Virus | | + | + |
| | | | West-Nil-Virus | + | + | + |
| | | | Japan-Enzephalitis-Virus | + | + | + |
| | | | St.-Louis-Enzephalitis-Virus | | + | + |
| | | | Zika-Virus | + | + | n.d. |
| Togaviridae | Alphavirus | | Chikungunya-Virus | + | + | + |
| | | | Östliche Pferdeenzephalomyelitis-Virus | + | + | + |
| | | | Venezolanische-Pferdezeph.-Virus | | + | + |
| | | | Westliche-Pferdeenzeph.-Virus | | + | + |
| | | | Ross-River-Virus | | + | + |
| | | | Sindbis-Virus | | + | + |
| | | | Mayaro-Virus | | + | + |
| | | | Getah-Virus | | + | + |
| Bunyaviridae | Orthobunyavirus | Bunyamwera | Potosi-Virus | + | + | + |
| | | | Cache-Valley-Virus | + | n.d. | n.d. |
| | | | Tensaw-Virus | + | n.d. | n.d. |
| | | California | Keystone-Virus | + | + | - |
| | | | San-Angelo-Virus | | + | + |
| | | | La-Crosse-Virus | + | + | + |
| | | | Jamestown-Canyon-Virus | + | + | + |
| | | | Trivittatus-Virus | | + | - |
| | | Simbu | Oropouche-Virus | | + | - |
| | Phlebovirus | | Rifttalfieber-Virus | | + | + |
| Reoviridae | Orbivirus | | Orungo-Virus | | + | + |
| Nodaviridae | Picornavirus | | Nodamura-Virus | | + | n.d. |

# 1.8 Invasion durch *Aedes* (*Stegomyia*) *albopictus*

Eine biologische Invasion erfolgt in drei Stufen: Einschleppung, Etablierung und Expansion (Williamson, 1996). Sie werden im Folgenden erläutert.

## 1.8.1 Einschleppung

In den letzten vier Jahrzehnten zählte *Ae. albopictus* zu den sich am schnellsten ausbreitenden Tierarten der Welt (Benedict et al., 2007). Vom asiatischen Ursprungsgebiet aus besiedelte die Art in dieser Zeit außer der Antarktis alle Kontinente und insgesamt über 30 neue Länder (Benedict et al., 2007; Paupy et al., 2009). *Ae. albopictus* rangiert damit auf Platz vier der „World's Worst Invasive Alien Species" der „Global Invasive Species Database" und auf Platz 1 der invasiven Insekten (Scholte & Schaffner, 2007). Die Verbreitung über lange Strecken gelingt *Ae. albopictus* durch eine Kombination von Faktoren wie der Produktion von trockentoleranten, kältetoleranten und diapausierenden Eiern, einem hohen Adaptationspotential und der effektiven Nutzung von Anthrotelmata als Brutgewässer (Scholte & Schaffner, 2007). Lange Zeit wurde vermutet, dass Stechmücken sich durch den internationalen Luftverkehr verbreiten, da sie in Fahrwerksschächten sowie in Passagierkabinen gefunden wurden und geringe atmosphärische Drücke überleben können (Lounibos, 2002). Vermutlich fand auf solchem Wege die Einschleppung von *Anopheles gambiae* in den Nordosten von Brasilien statt, aber dies blieb eine Ausnahme (Lounibos, 2002). Empirische Untersuchungen des Luftfahrtverkehrs lassen diesen Einschleppungspfad als eher unwahrscheinlich erscheinen (Lounibos, 2002). Anders als bei der globalen Verbreitung von *Ae. aegypti* über Wasserreservoirs auf Sklavenschiffen im 15.–17. Jh. ist der wahrscheinlichste Ausbreitungsmechanismus für *Ae. albopictus* der internationale Warenhandel mit als Rohstoff begehrten Altreifen, die zwischen Kontinenten i.d.R. auf dem Seeweg transportiert werden (Hawley et al., 1987; Reiter & Sprenger, 1987; Reiter, 1998). Diese Anthrotelmata eignen sich aufgrund der Regenwasseraufnahme in jeglicher Lageposition sowie der leichten Aufheizung durch Sonneneinstrahlung und der damit verbundenen schnelleren Entwicklung besonders gut und sind weltweit bevorzugte Brutkontainer dieser Art sowie von *Ae. aegypti*. An der Wasserkante abgelegte Eier werden so samt geeignetem Brutgefäß für die Larven verbreitet.

Ein zweiter Einschleppungspfad vor allem in der Europäischen Union und Nordamerika ist die Einfuhr von sogenanntem „Lucky Bamboo" (Zierpflanzen-Stecklinge, u.a. *Dracaena sanderiana*) (Hofhuis et al., 2009), wodurch *Ae. albopictus* z.B. zeitweise in Gewächshäuser in den Niederlanden eingeführt wurde (Scholte & Schaffner, 2007; Scholte et al., 2007, 2008).

## 1.8.2 Etablierung

Im Jahr 1979 wurde in der Stadt Laç in Albanien die erste etablierte Population von *Ae. albopictus* in Europa nachgewiesen, die nach anekdotischen Überlieferungen ggf. schon vier Jahre früher dort aufgetreten sein könnte (Adhami & Reiter, 1998; Scholte & Schaffner, 2007). In Italien wurden im Jahr 1990 zum ersten Mal *Ae. albopictus* in einem Kindergarten in Genua entdeckt, gefolgt von einer dokumentierten Population im darauffolgenden Jahr (Sabatini et al., 1990; Scholte & Schaffner, 2007). Hierauf folgten Nachweise in Ländern wie Frankreich (1999), Belgien (2000), Montenegro (2001), der Schweiz und Griechenland (2003), Spanien (2004), Bosnien und Herzegowina, Kroatien, Slowenien, Serbien und den Niederlanden (2005) (Benedict et al., 2007; Scholte & Schaffner, 2007; Paupy et al., 2009). In Süddeutschland folgten einem ersten Fund von Eiern im Jahr 2007 (im Oberrheingraben an der Autobahn A5) weitere Funde von Adulten und die Entdeckungen von kleineren Populationen zwischen 2012 (ebenda) und 2014 (südlich von München) (Pluskota et al., 2008; Werner et al., 2012; Werner & Kampen, 2014). Schließlich wurde im August 2015 eine große und mehrjährige Population bei Freiburg entdeckt, die damit die Kriterien (Scholte & Schaffner, 2007) für die erste etablierte Population in Deutschland erfüllt (KABS e.V., 2015a, 2015b, 2015c, 2016; dpa, 2016; Pluskota et al., 2016).

In Südamerika wurde 1983 die Insel Trinidad erstmals nachweislich von *Ae. albopictus* besiedelt, im Jahre 1986 gefolgt von dem großen Flächenland Brasilien, anschließend Bolivien (1997), Argentinien, Kolumbien und Paraguay (1998) (Benedict et al., 2007; Paupy et al., 2009). 1946 und 1972 wurden schon Larven und Eier von *Ae. albopictus* in Schiffsladungen mit Altreifen aus Südostasien in den USA gefunden, dennoch zählt erst der Fund im Jahr 1985 in den USA als der erste dokumentierte Nachweis einer stabilen Population in Nordamerika, gefolgt von Mexiko (1988), Barbados und der Dominikanischen Republik (1993), Kuba, Guatemala, Honduras und El Salvador (1995), den Kaimaninseln (1997), Panama (2003) und Nicaragua (2003) (Benedict et al., 2007; Scholte & Schaffner, 2007; Paupy et al., 2009).

Auf dem afrikanischen Kontinent wurde *Ae. albopictus* bereits 1998 in Südafrika zum ersten Mal gefunden (Paupy et al., 2009). Dort wurde die Population jedoch vor ihrer Etablierung erfolgreich bekämpft, und die erste formal als etabliert geltende Population in Afrika somit erst 1991 in Nigeria dokumentiert, gefolgt von Kamerun, Äquatorial-Guinea und Gabun (Benedict et al., 2007; Paupy et al., 2009).

In Australien und Neuseeland wurden mehrere Male einzelne Funde dokumentiert, jedoch bildeten sie lediglich auf den Torres-Strait-Inseln stabile Populationen aus (Scholte & Schaffner, 2007).

## 1.8.3 Expansion

Als Infiltration wird die natürliche Bewegung zu (neuen) Habitaten durch eine natürliche Verbreitung wie Winddrift, Flug oder Zoochorie bezeichnet (Lounibos, 2002). Besonders detailliert ist die Expansion von *Ae. albopictus* nach der Einschleppung in den USA, Brasilien und Südeuropa dokumentiert (Moore & Mitchell, 1997; Lounibos et al., 2002; Scholte & Schaffner, 2007). Aufgrund des schwach ausgeprägten Flugverhaltens von *Ae. albopictus* (vgl. Kapitel 1.7) wird angenommen, dass die zurückgelegten Strecken in der Expansionsphase mittels der Verbreitung durch den Straßenverkehr (PKW, LKW) stattfinden (Scholte & Schaffner, 2007). Dieser anthropogene Verbreitungspfad entspricht jedoch nicht der Ursprungsdefinition der natürlichen Verbreitung während einer Infiltration und müsste strenggenommen daher erweitert werden.

Die klimatische Ausbreitungsgrenze für *Ae. albopictus* liegt in den USA grob bei der -5° C-Januar-Isotherme im Norden, was der minimalen Temperatur der diapausierenden und kälteadaptierten Eier entspricht (Lounibos, 2002). Die Januar-Isotherme ist jedoch aufgrund von Unterschieden in der Kontinentalität der Ausbreitungsgebiete von *Ae. albopictus* nicht aussagekräftig genug, weswegen andere Arbeitsgruppen die jährliche Durchschnittstemperatur als Maßstab nehmen, um so auch z.B. die wärmeren Vegetationsperioden und damit höheren Populationswachstumsraten in kontinentaleren Regionen mit zu berücksichtigen (Mogi et al., 2012). Für Japan wurde daher eine nördliche Expansion bis zu einer jährlichen Durchschnittstemperatur-Isotherme von 10 bis 11 °C und einer Januar-Isotherme von -2° C identifiziert und für die USA eine jährliche Durchschnittstemperatur-Isotherme von vergleichbaren 11 bis 12 °C festgestellt (Kobayashi et al., 2002; Mogi et al., 2012). Grundlage hierbei ist jedoch immer die Fähigkeit zur maximalen Diapause und Kälteadaptation der Eier (Lounibos, 2002; Mogi et al., 2012; Urbanski et al., 2012). Ob diese klimatischen Ausbreitungsgrenzen von Japan und den USA auf Europa zu extrapolieren sind, ist noch nicht abschließend geklärt, da Europa nördlich der Alpen zu der Köppen-Geiger-Klimaklassifikation der warm-temperaten Breitengrade (= C) mit voller Humidität (= f) und warmen, jedoch nicht heißen Sommermonaten (= b) gehört, die in den jetzigen Verbreitungsgebieten von *Ae. albopictus* nicht repräsentiert ist (hauptsächlich wurde *Ae. albopictus* in den Klimaklassifikationen Aw, Am, Af, Bsh, Bsk, Cfa, Csa, Cwb gefunden) (Kottek et al., 2006; Mogi et al., 2012).

Es wird vermutet, dass die jetzige nördliche Expansionsgrenze durch die globale Klimaerwärmung immer weiter verschoben werden wird (Benedict et al., 2007; Mogi et al., 2012). So scheint die in Japan dokumentierte Nordwärts-Expansion mit der Erhöhung der jährlichen

Durchschnittstemperatur von ca. 1 °C im Zeitraum von 30 Jahren in Zusammenhang zu stehen (Kobayashi et al., 2002). Ähnliches wird auch für Europa erwartet, und Verbreitungsmodelle auf Basis der IPCC-Klimaprognosen der Emissions-Szenarien A1B und des optimistischeren B1 sagen eine flächendeckende Etablierung in Deutschland bis zum Ende des laufenden Jahrhunderts voraus (Benedict et al., 2007; Fischer et al., 2011; Caminade et al., 2012; Suk & Semenz, 2013; Kraemer et al., 2015). Durch eine erhöhte Jahresmitteltemperatur im Zuge der globalen Erderwärmung für Nordeuropa (z.B. 2,38–5,38 °C bis zum Ende des laufenden Jahrhunderts) erhöht sich nicht nur die Populationswachstumsrate von Invertebraten wie *Ae. albopictus,* sondern verlängert sich auch die saisonale Aktivität der Art. Damit verbessern sich die Lebensbedingungen für die Tigermücke in Mitteleuropa insgesamt (Porter et al., 1991; Alto & Juliano, 2001a, 2001b; Benedict et al., 2007; Delatte et al., 2009; Caminade et al., 2012).

Bereits 2007 trat auf dem Festland von Europa eine Chikunguya-Epidemie in Ravenna (Italien) mit 166 nachgewiesenen Fällen auf, deren Grundlage das dortige Massenauftreten von *Ae. albopictus* war (Angelini et al., 2007). 2010 wurden die ersten autochthonen Fälle von Dengue-Fieber in Kroatien und Frankreich dokumentiert und auf *Ae. albopictus* als Vektor zurückgeführt, 2013 weiter erneut in Frankreich (La Ruche et al., 2010; Schmidt-Chanasit et al., 2010; Gjenero-Margan et al., 2011; Marchand et al., 2013; Schaffner & Mathis, 2014). Erkrankungen durch Dirofilarien werden aller Wahrscheinlichkeit nach durch den globalen Klimawandel zunehmen (Genchi et al., 2009). Die mit der Ausbreitung von Vektoren wie *Ae. albopictus* verbundenen medizinischen Risiken durch Arboviren werden von manchen Autoren zu den größten Gefahren durch die Klimaerwärmung gerechnet (Epstein et al., 1998; Stark et al., 2009). Um die Risiken für Bevölkerung und Umwelt frühzeitig zu erkennen und nötige Gegenmaßnahmen einzuleiten, müssen die komplexen Beziehungen zwischen Klima- und anderen Umweltveränderungen, einheimischen und invasiven Krankheitsvektoren und sich ausbreitenden Pathogenen besser erforscht werden (Lee II & Chapman, 2001; Stark et al., 2009).

### 1.8.4 Bekämpfung

Eine nachhaltige und integrierte Vektorbekämpfung besteht aus mehreren Komponenten, deren Anwendungen auf die lokalen Begebenheiten abgestimmt werden müssen und ein Zusammenspiel zwischen Behörden, kommunaler Partizipation und der Gesundheitsbildung erfordern (Estrada-Franco & Craig, 1995; Schaffner & Mathis, 2014). Hierzu gehört zuerst die Entfernung, Versiegelung oder regelmäßige Reinigung von potentiellen Brutgewässern in den betroffenen Gebieten (Estrada-Franco & Craig, 1995; Schaffner & Mathis, 2014). Seit dem Bekanntwerden der Rolle von Stechmücken als Krankheitsüberträger werden sie mit einer Reihe chemischer Insektizide bekämpft, die man in Larvizide und Adultizide einteilt, wobei manche Substanzen zur Bekämpfung beider Lebensstadien angewendet werden (Estrada-Franco & Craig, 1995; Bonizzoni et al., 2013). Eines der ältesten Insektizide ist das als Schweinfurter Grün oder auch Pariser Grün bezeichnete Kupfer(II)-arsenitacetat, das wirksam – aber auch mit großen Konsequenzen für Mensch und Natur – in den 1940er Jahren im Kampf gegen *An. gambiae* in Brasilien angewendet wurde (Lounibos, 2002). Häufig verwendete Juvenilhormon-Analoga, wie Methopren, Novaluron, Pyriproxifen, und die den Carbamaten zugehörigen Bendiocarb und Propoxur werden auf unterschiedliche Weise in die Brutgewässer gegeben und verhindern langanhaltend die Adoleszenz der Larven, welche im Puppenstadium verbleiben und schließlich verenden (Bonizzoni et al., 2013; Owusu et al., 2015). Außerdem werden Organophosphate, wie Temephos, Fenitrothion und Malathion, als Larvizid eingesetzt, welche genau wie die Carbamate die Acetylcholinesterase der Larven hydrolysieren, wodurch es zu einer Akkumulation von Acetylcholin im synaptischen Spalt kommt, die eine Dauererregung und damit Bewegungsunfähigkeit bis hin zum Tot auslöst (WHO, 1998; Basilua Kanza et al., 2013; Bonizzoni et al., 2013). Resistente Stämme weisen eine Punktmutation in einem der beiden Gene der Acetylcholinesterase (*ace-1*) auf, die zu einer Substitution der Aminosäure Glycin durch Serin führt (Weill et al., 2002; N'Guessan et al., 2003; Basilua Kanza et al., 2013). Sie besitzen dadurch eine insektizidinsensitive Acetylcholinesterase (iAChE) (Basilua Kanza et al., 2013).

Hinzu kommen die weltweit durch das Stockholmer Übereinkommen über persistente organische Schadstoffe geächtete, aber im Kampf gegen krankheitsübertragende Insekten weiterhin zugelassenen Chlorkohlenwasserstoffe DDT (Dichlordiphenyltrichlorethan) und Dieldrin, welche versprüht werden und in den Motoneuronen der Insekten Natriumkanäle öffnen und dadurch dosisabhängig zu Tremores und zum Tod führen (WHO, 1998; Mörner et al., 2002; Basilua Kanza et al., 2013; Bonizzoni et al., 2013). Mit demselben Wirkmechanismus, nur wesentlich weniger persistent, wirken das natürliche Pyrethrum und seine synthetischen Analoga der Pyrethroide, wovon vor allem Permethrin, Deltamethrin, λ-Cyhalothrin und in geringeren Mengen auch Cyfluthrin und Etofenprox (Pyrethroid-Ether) im Kampf gegen Stechmücken durch Vernebelung als Adultizid eingesetzt werden (WHO, 1998; Basilua Kanza et al., 2013). Pyrethroide werden auch stellenweise als Larvizid eingesetzt und sind im Privatanwenderbereich zur Bekämpfung von Stechmücken zugelassen (Sulaiman et al., 1991; Lawler et al., 2007; Whelan et al., 2009). Eine Punktmutation im Genom des Natriumkanals von Steckmückenarten wie *Ae. albopictus* führt zu einer Aminosäuresubstitution (Leucin durch Phenylalanin oder Serin) und damit zu einer erhöhten Kreuzresistenz gegenüber DDT und gewissen Pyrethroiden (Brogdon et al., 1999; Hemingway & Ranson, 2000; Ranson et al., 2000; Vontas et al., 2012; Basilua Kanza et al., 2013).

Durch den selektiven Druck auf die Populationen werden diese sogenannten „knock down resistance Allele" (*kdr*) zunehmend in Gebieten verbreitet, in denen Bekämpfungsmaßnahmen auf Basis der oben genannten Insektizide durchgeführt werden (Hemingway & Ranson, 2000; Ranson et al., 2000; Vontas et al., 2012; Basilua Kanza et al., 2013). Ferner stehen viele der Resistenzen in Kombination mit Veränderungen in den Genen für Detoxifikationsmechanismen, wie die Glutathion-S-Transferasen, Cytochrom-P450-Monooxygenasen und andere Esterasen (Vontas et al., 2012). Zurzeit sind die beiden Schnelltests der Centers for Disease Control and Prevention (CDC) und der Weltgesundheitsorganisation (WHO) zur Bestimmung der Insektizid-Sensitivität die einzigen beiden standardisierten biologischen Testsysteme, die Stechmücken im Fokus haben (Owusu et al., 2015).

Zudem werden die extrahierten Kristalloproteine von *Bacillus thuringiensis israelensis* sowie *B. thuringiensis sphaericus* zur Bekämpfung eingesetzt (als Bt-Toxin bekannt), die hochwirksam das Darmepithel der Larven schädigen und so zum Tod führen, ohne für andere Invertebraten toxisch zu sein (Estrada-Franco & Craig, 1995; Fillinger et al., 2003; Bonizzoni et al., 2013). Auswirkungen auf die Insekten-Biozönose von Überschwemmungsgebieten scheint es durch den großflächigen Einsatz von Bt-Toxinen nicht zu geben (Vinnersten et al., 2010; Duchet et al., 2015). Jedoch sind bereits schon länger Resistenzen bei Culiciden bekannt (Georghiou & Wirth, 1997). Außerdem zeigen Studien, dass die Vektorkontrolle mit Bt-Toxinen einen Einfluss auf die trophischen Ebenen des Ökosystems hat und in behandelten Gebieten Mehlschwalben signifikant weniger Bruterfolg aufwiesen (Gilbert, 2010; Poulin et al., 2010).

Hinzu kommen Feldversuche der Larvenbekämpfung mit Parasiten, Prädatoren und Pathogenen wie *Wolbachia*-Stämmen (welche durch die Hochregulierung des Immunsystems der Mücke auch die Übertragung der Chikungunya-Viren und Dengue-Viren reduzieren), Cyclopoiden Copepoden, *Toxorhynchites splendens* und die beiden Fischarten *Gambusia affinis* und *Gambusia holbrooki*, die jedoch nur vereinzelt zur flächendeckenden Bekämpfungsstrategie gehören oder kommerzielle Anwendung fanden (Estrada-Franco & Craig, 1995; Bonizzoni et al., 2013; Schaffner & Mathis, 2014).

Neu sind Versuche auf Basis genetischer Modifikationen, wobei man zwischen der Strategie der Populationsreduktion und der Populationssubstitution durch Stämme mit einer geringeren Vektorkapazität unterscheidet (Bonizzoni et al., 2013). Zur Populationsreduktion wird bei vielen Stechmückenarten bereits erfolgreich einerseits die „Sterile insect technique" (SIT) eingesetzt, bei der Männchen durch Chemikalien oder γ-Strahlung sterilisiert und anschließend freigesetzt werden und so bei einer Kopulation eine weitere Befruchtung durch ein fertiles Männchen verhindern (Benedict & Robinson, 2003; Bellini et al., 2007; Bellini & Albieri, 2010; Lacroix et al., 2012; Bonizzoni et al., 2013). Andererseits werden genetisch veränderte Männchen freigesetzt, die lebensunfähige Nachkommen produzieren (Release of Insects carrying a Dominant Lethal, RIDL) oder das Geschlechterverhältnis zugunsten der Männchen verschieben (Harris et al., 2011; Bonizzoni et al., 2013; Massonnet-Bruneel et al., 2013; Galizi et al., 2014). Gerade die Freilassung von genetisch veränderten Stechmücken wird aus ethischen Gesichtspunkten und einer noch nicht hinreichenden Risikobewertung in Bezug auf einen etwaigen horizontalen Gentransfer kritisch gesehen (Benedict & Robinson, 2003).

## 1.8.5 Verhinderung des Mücke-Mensch-Kontakts

Ein weiteres wichtiges Element der integrierten Vektorbekämpfung besteht aus der Verhinderung des Mücke-Mensch-Kontakts mittels Repellentien (Schaffner & Mathis, 2014). Von den Centers for Disease Control and Prevention (USA) werden lediglich vier Substanzen für den Einsatz als effektives Repellent empfohlen: DEET (N,N-Diethyl-Toluamide) als „Goldstandard", Picaridin (Icaridin), das bei *Aedes*-Arten ähnlich wirksam wie DEET ist, jedoch bei *Anopheles*-Arten nur einen geringen Schutz bietet, IR3535 (Ethylbutylacetylaminopropionat), welches als synthetisches Produkt weniger Nebenwirkungen, aber auch einen geringeren Schutz als DEET und Icaridin zeigt, und schließlich PMD (Para-Menthan-3,8-diol, Handelsname Citriodiol), das als schwer flüchtige Phase des Extraktes aus Zitroneneukalyptus (*Corymbia citriodora*) gewonnen wird und nach ersten Versuchen ähnlich effektiv wie DEET wirken soll (Rose & Kröckel, 2010). Zusätzlich werden mit Pyrethroiden (vor allem Permethrin) auch Kleidungsstücke, Vorhänge oder Mückennetze behandelt (sogenannte ITNs: Insecticide Treated Nets), welche langanhaltend einen Knock-down-Effekt aufweisen und aufgrund der hohen Toxizität für Invertebraten bei gleichzeitiger geringer Toxizität für Warmblüter bewirken, dass es beim Kontakt der Mücke mit dem Netz zu Krämpfen und zum Tod des Insekts kommt (Rose & Kröckel, 2010; Basilua Kanza et al., 2013; Schaffner & Mathis, 2014). Ätherische Öle aus Teebaum, Neembaum, Nelken, Knoblauch und Geranien haben keine wissenschaftlich nachgewiesene repellente Wirkung auf Stechmückenarten (Rose

& Kröckel, 2010; Davis, 2014). Einzelne Öle und ihre extrahierten Sesquiterpene vom Echten Sandelholz *Santalum album*, dem Westindischen Sandelholz *Amyris balsamifera,* dem Westindischen Zitronengras *Cymbopogon citratus* (Citronella-Öl) und anderen zeigen eine mäßige bis hohe repellente Wirkung auf Stechmücken (Paluch et al., 2009; Rose & Kröckel, 2010; Birkett et al., 2011; Carroll et al., 2011; Suwansirisilp et al., 2012; Chio et al., 2013; Kulkarni et al., 2013; Davis, 2014). Die neuere Forschung zeigt vielversprechende Ergebnisse bei der Identifizierung von repellenten Terpenen der Japanischen Schönbeere (*Callicarpa japonica*) und der Amerikanischen Schönbeere (*Callicarpa americana*), die nach anekdotischen Überlieferungen indigener Völker Nordamerikas zum Schutz von Nutztieren vor stechenden und beißenden Insekten eingesetzt wurde und deren aktive Inhaltsstoffe (Callicarpenal und Intermedeol) bei *Ae. aegypti* und *Ae. albopictus* ab einer Konzentration von 25 nmol/cm² abschreckend wirken (Cantrell et al., 2005; Carroll et al., 2007; Cantrell & Klun, 2011; Ling et al., 2011; Debboun et al., 2014).

# 1.9 Stand der Forschung und Implementierung der eigenen Arbeit in die Forschungsnetzwerke

Abbildung 6 zeigt eine Netzwerkanalyse der zitierten Quellen der drei im Annex A.1 folgenden Manuskripte und Veröffentlichungen sowie des Rahmentextes mit dem Programm Gephi (Version 0.8.2. beta, © Gephi contributors). In der Fruchtermann-Reingold-Darstellung wurde die Knotengröße aus der addierten Gewichtung der Anzahl der Publikationen jedes Autors und der Anzahl der Autoren pro Publikation berechnet. Knoten, welche unterhalb einer Gewichtung von fünf lagen, wurden zugunsten der Übersichtlichkeit herausgefiltert. Die Kantenstärke wurde in Abhängigkeit von der Anzahl der gemeinsamen Autorenschaften gewichtet. Die farblichen Cluster wurden nach der Methode von Latapy berechnet und eingefärbt (Latapy, 2008).

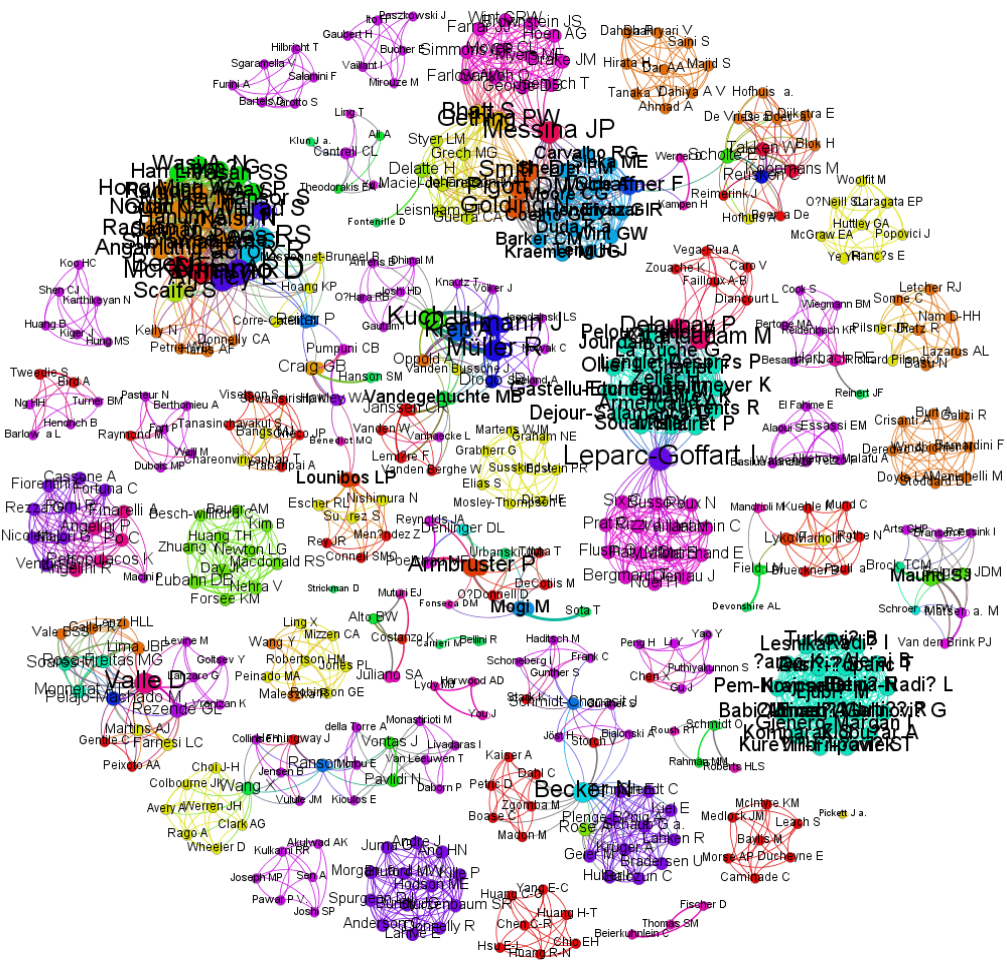

**Abbildung 6 (eigenes Werk, unveröffentlicht): Netzwerkanalyse der Autoren. Darstellung nach Fruchtermann-Reingold. Die Knotengröße berechnet sich aus der addierten Gewichtung der Anzahl der Publikationen jedes Autors und der Anzahl der Autoren pro Publikation. Knoten unterhalb einer Gewichtung von fünf wurden zwecks Übersichtlichkeit herausgefiltert. Die Kantenstärke wurde in Abhängigkeit von der Anzahl der gemeinsamen Autorenschaften gewichtet. Cluster wurden nach der Methode Latapy (2008) berechnet und eingefärbt. Als Datensatz lagen die kompletten Referenzen der drei Publikationen/Manuskripte aus dem Annex sowie des Rahmentextes vor (825 Knoten, von denen 57,0% dargestellt wurden; 3.180 Kanten, von denen 77,6% dargestellt wurden).**

Durch diese Netzwerkkarte der Coautorenschaft lassen sich die wichtigsten Quellen der Dissertationsschrift leicht und übersichtlich identifizieren, um daraus den aktuellen Stand der Wissenschaft zu erläutern. Im Anschluss ist es dann möglich sich hieraus entwickelnde aktuelle wissenschafts- und gesellschaftsrelevante Fragen zu formulieren und zu erörtern.

Eins der bedeutsamsten Cluster liegt im Norden der Netzwerkkarte und umfasst die Arbeitsgruppen um Bhatt und Gething von der Oxford University (GB), die ihre bedeutenden Ergebnisse zur Modellierung des Dengue-Infektionsrisikos in „Nature" veröffentlichen und zusammen mit dem Arbeitskreis um Delatte vom CIRAD Réunion (Frankreich) die Temperaturnische von *Ae. albopictus* modellierten, um das Ausbreitungspotential dieses Denguevektors vorhersagen zu können (Bhatt et al., 2013; Brady et al., 2013). Messina, ebenfalls aus dieser in Nature veröffentlichten Studie, arbeitete zusammen mit seinen Kollegen Smith, Pigoth und Golding von der Oxford University (GB) an einer Studie zur Modellierung der weltweiten Verbreitung der Vektoren für Arboviren, wie *Ae. albopictus* und *Ae. Aegypti*, mit (Kraemer et al., 2015). An dieser Studie war auch Schaffner von der Universität Zürich (Schweiz) beteiligt, der ein Spezialist zur Ausbreitung von *Aedes*-Arten in Europa ist und zusammen mit Kampen vom Friedrich-Löffler-Institut (FLI, Deutschland) und Werner vom Leibniz-Zentrum für Agrarlandschaftsforschung (ZALF, Deutschland) über die Ausbreitung auch von *Ae. albopictus* in Deutschland publiziert (Werner et al., 2012; Werner & Kampen, 2014). Schaffner verfasste zusammen mit Scholte von der Wageningen University (Niederlande) ein Kapitel im Buch mit dem Titel „Emerging Pest and Vector-borne Diseases in Europe" zur Ausbreitung von *Ae. albopictus* in Europa (Scholte & Schaffner, 2007). Scholte wiederum war an wichtigen Publikationen zur Einfuhr von *Aedes*-Eiern durch die den Import von Drachenbaumgewächsen nach Europa beteiligt (Scholte et al., 2007; Hofhuis et al., 2009).

Ein weiteres großes Cluster, das etwas südlich liegt, umfasst die Arbeiten von Leparc-Goffart vom Institut de recherche biomédicale des armées (Frankreich), wobei ihre bedeutenden Arbeiten zur Dokumentation der ersten autochtonen Dengue-Fälle auf dem europäischen Festland zu erwähnen wären (La Ruche et al., 2010; Marchand et al., 2013). Thematisch sehr nah ist das noch weiter südlich gelegene Cluster um Norbert Becker von der Kommunalen Aktionsgemeinschaft zur Bekämpfung der Schnakenplage (KABS e.V., Deutschland). Wichtige Publikationen aus den großflächigen Stechmücken-Monitorings zur Ersterfassung von invasiven Arten in Deutschland sowie deren Bekämpfung und zu den enthaltenen Viren verfasste er zusammen mit Kollegen wie Schmidt-Chanasit, Tannich und Krüger vom Bernhard-Nocht-Institut für Tropenmedizin (Deutschland) und Rose von der Biogents AG (Deutschland) (Fillinger et al., 2003; Pluskota et al., 2008; Becker et al., 2010, 2013; Jöst et al., 2010; Schmidt-Chanasit et al., 2010).

Diese Cluster vereinen die gemeinsame Fragestellungen zu der Ausbreitung von Krankheitsvektoren wie *Ae. albopictus* (siehe Kapitel 1.8.1 Einschleppung, 1.8.2 Etablierung und 1.8.3 Expansion) und den damit verbundenen medizinischen Risiken durch Arboviren sowie zu der Gefahren für die öffentliche Gesundheit durch die Klimaerwärmung (siehe Kapitel 1.7.3 Vektorkompetenz und -kapazität). Ziel ist es, die Risiken für die Bevölkerung frühzeitig zu erkennen und nötige Gegenmaßnahmen einzuleiten und es wird festgehalten, dass hierzu die komplexen Bedingungen zwischen globaler Klimaerwärmung und diesen Vektoren besser erforscht werden müssen (Epstein et al., 1998; Stark et al., 2009).

Etwas westlich davon liegt das Cluster um Ranson von der University of Wales Cardiff (GB) und dem Kollegen Wang von der University of Notre Dame sowie Vontas und Pavlidi von der University of Crete, die zur Entwicklung von Insektizid-Resistenzen von *Ae. albopictus* forschen. Zwischen diesem Cluster und dem von Armbruster liegt das kleinere Cluster von Alto, Juliano und Costanzo von der University of Florida (USA) und Muturi von der University of Illinois (USA), die zur temperaturabhängigen Entwicklung und intraspezifischen Konkurrenz u.a. in Kombination mit Insektizidstress von *Ae. albopictus* arbeiteten (Alto & Juliano, 2001a, 2001b; Costanzo et al., 2005b, 2011; Muturi & Alto, 2011; Muturi et al., 2011). Erwähnenswert sind auch die Cluster direkt nebenan um Harwood, You und Lydy von der Southern Illinois University (USA) und das um Maund, Belgers und Brock von der Jealott's Hill Research Station (GB), das im Osten der Netzwerkkarte liegt. In beiden Clustern wurde thematisch viel zur Ökotoxikologie von Pyrethroiden gearbeitet (Maund et al., 1998; Schroer et al., 2004; Roessink et al., 2005; Harwood et al., 2009; Weston et al., 2009).

Wie in Kapitel 1.8.4 Bekämpfung erwähnt, gibt es zurzeit lediglich zwei standardisierte biologische Testsysteme für Culicidae, wobei es sich bei beiden lediglich um akute Schnelltests zur Bestimmung der Insektizid-Sensitivität im Knock-down-Verfahren mit einer einzigen Wirkstoffkonzentration handelt (Owusu et al., 2015). Gerade um auch Bereiche der Sensitivität oberhalb und unterhalb der in den Schnelltests verwendeten Konzentration zu testen, wird die Entwicklung von standardisierten Dosis-Wirkungs-Testsystemen empfohlen (Owusu et al., 2015). Aber nicht nur im Bereich der Ökotoxikologie ist ein standardisiertes Bioassay notwendig, um die offenen Forschungsfragen an invasiven Stechmücken wie *Ae. albopictus* beantworten zu können.

In der ersten Publikation wurde ein solches Testsystem für quarantänebedürftige Arten beschrieben und sogleich mit einer Modellsubstanz, zwei Arten (*Ae. albopictus* und der einheimischen *Culex pipiens*), zwei Fütterungsarten und drei Temperaturregimen evaluiert (siehe Annex A.1). Diese Publikation baut auf einer Futterstudie zur standardisierten und optimalen Aufzucht der Larven von *Ae. albopictus* auf (Müller et al., 2013) und stellt ein wichtiges Fundament für die weiteren experimentellen Arbeiten an der Asiatischen Tigermücke dar.

Da mit diesem Dosis-Wirkungs-System nicht nur bloß ein singulärer Endpunkt wie der prozentuale Knock-down-Effekt eines Insektizids erfasst werden konnte, sondern komplette integrative Parameter, die in Konzentrations-Wirkungs-Regressionen der Mortalität eingingen oder weitere Life-History-Parameter wie mittlere Verpuppungsdauer und Eigelegegröße in der endlichen Wachstumsrate $\lambda'$ der Population vereinen, war es erstmals möglich, ein komplexes Bild der toxikologischen Effekte von Insektiziden auf Stechmücken zu zeichnen. Da Strechmückenbekämpfungsmaßnahmen mit Pestiziden immer einen intensiven Einfluss für die betroffenen Ökosysteme, aber auch das Grundwasser bedeuten können (Grube et al., 2011), sind diese Kenntnisse von entscheidender Bedeutung, um die Maßnahmen so nachhaltig wie möglich und zugleich so schonend wie möglich durchzuführen (Bakouri et al., 2007). Hieraus erwächst die Fragestellung, in wieweit sich die toxikologischen Effekte eines Insektizids zur Bekämpfung zwischen einheimischen und invasiven Arten unterscheiden können und ob dies zu Konsequenzen für eine optimale Bekämpfungsmaßnahme in Habitaten mit Mischpopulationen führen könnte.

Als Modellsubstanz wurde $\lambda$-Cyhalothrin verwendet, da es ein Vertreter der immer häufiger zur Stechmückenbekämpfung verwendeten Pyrethroide ist und selbst als Larvizid und Adultizid eingesetzt wird (Lawler et al., 2007; Whelan et al., 2009). Im Allgemeinen bilden Insektizide bei höheren Temperaturen – wie sie im Zuge des Kimawandels vor allem bei Extremwetterereignissen in Europa zunehmend zu erwarten sind – eine höhere toxische Eigenschaft aus (Cairns et al., 1975; Mayer & Ellersieck, 1986; Heugens et al., 2001; Müller et al., 2012). Auch Malation als Vertreter der Pyrethroide zeigt eine solche positive Temperatur-Toxizitäts-Korrelation (Muturi et al., 2011). Es stellt sich daher die Frage, ob dies ein generelles toxikologisches Verhalten bei den Vertretern der Pyrethroide ist und welche Auswirkungen die veränderte Toxizität auf Bekämpfungsstrategien haben könnte.

Recht zentral auf der Netzwerkkarte liegt das Cluster um Armbruster von der Georgetown University (USA), der zusammen mit den Instituts-Kollegen Poelchau, Urbansky und Raynolds von der Ohio State University (USA) das Transkriptom von diapausierenden Eiern von *Ae. albopictus* u.a. per *Deep Sequenzing* untersucht und zusammen mit Sota und Mogi von der Saga University (Japan) die nördlichen Ausbreitungsgrenzen der Art analysiert (Urbanski et al., 2010a, 2010b, 2012; Mogi et al., 2012; Reynolds et al., 2012; Poelchau et al., 2013; Huang et al., 2015). Die Arbeitsgruppe um Mogi ist bekannt für ihre umfangreichen Arbeiten zur Trocken- und Frosttoleranz der Art (Sota & Mogi, 1992a, 1992b; Mogi, 2011; Urbanski et al., 2012).

Nicht weit davon entfernt zieht sich ein großes Cluster gen Nordnordwest auf der Karte. Lounibos von der University of Florida ist besonders für seine experimentellen Arbeiten zur lichtinduzierten Diapause von *Ae. albopictus* sowie zur interspezifischen Konkurrenz bekannt (Lounibos et al., 2002, 2003, 2011; Armistead et al., 2008; Leisnham et al., 2009). Zusammen mit Hawley und Benedict vom Centers for Disease Control and Prevention (CDC) in den USA veröffentlichte er auch ein viel beachtetes Review mit dem Titel „Spread of the tiger: global risk of invasion by the mosquito *Aedes albopictus*" (Benedict et al., 2007). Hawley, der zusammen mit seinen Kollegen Pumpuni, Hanson und Craig von der University of Notre Dame (USA) ebenfalls wichtige experimentelle Arbeiten zur Aufklärung der Forsttoleranz der Art durchführte, ist auch bekannt für seine Mitwirkung an der Ersterfassung von *Ae. albopictus* in den USA und zusammen u.a. mit Reiter vom CDC für die Klärung des Eintragspfades durch die trocken- und frosttoleranten Eier in Autoreifen (Hawley et al., 1987, 1989; Reiter & Sprenger, 1987; Hanson, 1991; Hanson et al., 1993; Focks et al., 1994; Hanson & Craig, 1994, 1995a, 1995b; Reiter, 1998). Reiter wiederum ist beteiligt an den Freilassungsversuchen von genetisch veränderten Männchen zusammen mit Kollegen wie Nimmo oder Alphey vom Oxitec Limited (GB) (Harris et al., 2011; Lacroix et al., 2012; Massonnet-Bruneel et al., 2013).

Weitere wichtige Cluster sind z.B. das im Westen liegende Cluster um Valle und Rezende vom FIOCRUZ (Brasilien), die zusammen mit Kollegen zur embryonalen Entwicklung, Physiologie und Trockenresistenz von *Ae. albopictus* arbeiten (Monnerat et al., 1999; Valle et al., 1999; Rezende et al., 2008; Farnesi et al., 2009; Goltsev et al., 2009).

Wie in Kapitel 1.8 Invasion durch *Ae. albopictus* beschrieben, geht das enorme Invasionspotential dieser Art auf eine Kombination von Faktoren wie der Produktion von trockentoleranten (vgl. Kapitel 1.4.2), kältetoleranten und diapausierenden Eiern (vgl. Kapitel 1.4.3), einem hohen Adaptationspotential und der effektiven Nutzung von Anthrotelmata als Brutgewässer (vgl. Kapitel 1.2) zurück (Scholte & Schaffner, 2007).

In der zweiten Publikation wurde daher der Mechanismus der Kältetoleranz durch die beiden Teilmechanismen der Diapause und der Kältekklimatisierung sowie einer möglichen Interaktion ergründet. Nach der Falsifizierung der Hypothese durch Hanson & Craig (1995b), dass dieses Phänomen auf die erhöhte Einlagerung von Polyolen und eine damit verbundene Reduzierung des Punkts der unterkühlten Schmelze zurückzuführen sei, verfolgten die andere Arbeitsgruppen eine Hypothese, die einer Erhöhung von Fettsäuren im Embryo und einer damit verbundenen dickeren Wachsschicht im Chorion des Eis nachgeht (Urbanski et al., 2010a, 2010b; Reynolds et al., 2012; Poelchau et al., 2013). Obwohl Messungen von Lipidgehalten im Chorion zeigten, dass tropische Populationen sogar mehr Fettsäuren einlagern als Populationen aus gemäßigten Breiten (Urbanski et al., 2010b), lagen noch keine gesicherten Daten über die eigentlichen Schichtdicken des Chorions von Stechmücken vor (Farnesi et al., 2015). Daher drängte sich eine Überprüfung der Hypothese anhand einer Transmissionselektronenmikroskop-Studie auf. Gleichzeitig wurde die Verortung der Fettsäuren innerhalb des Chorions mit seinen Schichten vorgenommen und eine einheitliche Systematisierung der Chorion-Schichten abgehandelt (vgl. Annex A.1.2).

Das beachtliche Adaptationspotential von *Ae. albopictus*, das für den Invasionserfolg verantwortlich gemacht wird (Scholte & Schaffner, 2007), könnte auf einem epigenetischen Mechanismus beruhen. Neben der Histonmodifikation und der RNA-Inteferenz (RNAi) ist der bekannteste epigenetische Mechanismus die DNA-Methylierung (Reamon-Buettner et al., 2008; Feil & Fraga, 2011; Vandegehuchte & Janssen, 2011). Etwas zerstreut liegen nun die Cluster, die u.a. die Methylierung der DNA in Drosophila entdeckten, wie z.B. das am westlichen Rand liegende Cluster um Tweedie und Kollegen von der University of Edinburgh (GB) und Turner von der University of Birmingham (GB). Außerdem waren die Werke aus dem Cluster (westlich von Leparc-Goffart) mit Lyko und Marhold vom Deutschen Krebsforschungszentrum (DKFZ), Field vom Rothamsted Research (GB) und Mandrioli von der Università di Modena e Reggio Emilia (Italien) zur Epigenetik von Invertebraten eine wichtige Grundlage für diese Arbeiten. Auch das Netzwerk um Maleszka (zwischen Armbruster und Valle gelegen) von der Australian National University oder um Ye (nordöstlich verortet) von der Monash University (Australien) steuerten wichtige Erkenntnisse zur epigenetischen Veränderung des Phänotyps von Dipteren als Grundlage der Arbeiten bei. Und zu guter Letzt nicht zu vergessen das Netzwerk um Michiel Vandegehuchte vom Laboratory of Environmental Toxicology and Aquatic Ecology von der Ghent University sowie seine Kollegen vom Centre for Proteome Analysis and Mass Spectrometry der University of Antwerp, die viel zur epigenetischen Veränderung durch methylierende Agenzien bei *Daphnia magna* arbeiteten und sowohl theoretische Grundlagen lieferten als auch wissenschaftlich kooperierten. Deshalb liegt hier auch die stärkste Verbindung zum eigenen Netzwerk, das in der Karte mit Oehlmann, Müller, Kuch und Kreß u.a. enthalten ist.

Schon früh wurde in Zellkulturen von *Ae. albopictus* nachgewiesen, dass rund 0,17% des Cytosins der DNA methyliert vorlag (sowie im geringen Maße auch Adenin und Thymin) (Adams et al., 1979). Später wurden diese Befunde *in vivo* bestätigt (Nayak et al., 1991). Nachdem das wirbeltierähnliche methylierende Protein „DNA (5-cytosine) Methyltransferase 2" (kurz dDNMT 2) und das demethylierende Protein „Methyl-CpG-Binding-Domains" (dMBD2/3) in *Drosophila melanogaster* gefunden wurde (Hung et al., 1999; Tweedie et al., 1999), nahm die Zahl epigenetischer Untersuchungen bei Diptera stark zu. Bald schon wurde eine – wenn auch geringe – DNA-Methylierung in *D. melanogaster* gefunden (Gowher et al., 2000; Lyko et al., 2000). Zusätzlich wurde die Korrelation zwischen den DNMT2-ähnlichen Proteinen und der Quantität von 5'-Methylcytosin bei *Drosophila* erkannt und zudem in der Honigbiene die drei DNMT-Orthologe 2, 3a und 3b sowie ihre korrelierende Menge an 5'-Methylcytosin in der DNA (Kunert et al., 2003; Wang et al., 2006; Glastad et al., 2011). Diese Entdeckungen wurden für weitere Dipteren, wie *Ae. albopictus*, *Ae. aegypti* und *Anopheles gambiae*, bestätigt (Mandrioli & Volpi, 2003; Marhold et al., 2004b; Ye et al., 2013).

Auf Basis dieses Wissenstandes konnte daher davon ausgegangen werden, dass auch bei *Ae. albopictus* die Wahrscheinlichkeit für ein funktionierendes epigenetisches „Toolkit" vorhanden sein könnte, das der Grund für das hohe Adaptationspotential wäre. Daher wurde die Frage gestellt, ob eine mögliche epigenetische und transgenerationale Modulation dieses Phänotyps möglich sei, die die Grundlage für eine epigenetische Adaptation wäre. Aufbauend auf einer Transgenerationsstudie mit den beiden DNA-methylierenden Agenzien Genistein und Vinclozolin (Oppold et al., 2015) wurde die Veränderung der Kälte-Phänotypen von einer tropischen Population von *Ae. albopictus* in den beiden nichtexponierten nachfolgenden Generationen getestet (vgl. Annex A.1.3).

# 2 Diskussion

## 2.1 Erste Publikation

**Abstract:** The global spread of the Asian tiger mosquito *Aedes albopictus*, an urban pest as well as a vector for arboviruses, is a threat for public health. As control measures include the use of insecticides such as the pyrethroid λ-cyhalothrin, it is crucial to assess their efficiency and their potential impact on the biodiversity especially under climate change conditions. To evaluate the environmental risk, biotests are well established for non-target organisms but not yet for mosquitoes. We therefore developed a fulllifecycle biotest for mosquitoes kept under quarantine conditions based on the OECD guideline 219. Therewith we tested the effect of temperature and nutrition on the ecotoxicological response to λ-cyhalothrin on the mosquitoes *Ae. albopictus* and *Culex pipiens* by assessing sublethal and life history parameters. The efficiency of λ-cyhalothrin decreased in both mosquito species with increasing temperature and changed with feeding protocol. At effective concentrations for potential mosquito control in surface waters, λ-cyhalothrin poses a high risk for indigenous aquatic key role species inhabiting the same microhabitats. Those aspects should to be taken into account in vector control strategies.

**Keywords:** Climate change, Population growth, Pyrethroid, Reproduction, Vector control

**Key message:**

• We developed a not yet existing standardized chronic biotest for invasive vectors of human diseases like the Asian tiger mosquito Aedes albopictus.

• With this tool, we tested hypotheses about the speciesspecific response to an insecticide under multiple stressors like temperature and nutrition.

• We revealed a negative temperature toxicity relation, a food dependent efficacy and an environmental risk by applications of k-cyhalothrin.

• These findings and the biotest itself are valuable complements in investigating mosquito control measures and for an interlaboratory comparison.

Aufgrund der massiven Gefährdung der öffentlichen Gesundheit durch das Vektorpotential und die Vektorkapazität von *Ae. albopictus* werden in betroffenen Gebieten Bekämpfungsmaßnahmen für nötig gehalten. Das Erreichen von geeigneten Effektkonzentrationen zur nachhaltigen Bekämpfung der Art steht hier mit dem Schutz der Biodiversität in den betreffenden Ökosystemen im Zielkonflikt. Zur Bestimmung der Effektkonzentrationen bedarf es daher verlässlicher chronischer Testsysteme zur Untersuchung der spezifischen Dosis-Wirkungs-Beziehungen. Zusätzlich muss ein solches Testsystem einfach und leicht reproduzierbar sein, damit es eine hohe Vergleichbarkeit der gewonnenen Ergebnisse unterschiedlicher Labore weltweit ermöglicht. Und schließlich muss das Testsystem auch gewissen Grundsätzen der Quarantäne gerecht werden, wenn invasive Stechmückenarten in noch nicht besiedelten Gebieten untersucht werden sollen.

In einer ersten Evaluierung dieses Testsystems wurden Versuche zu artspezifischen Reaktionen der invasiven Asiatischen Tigermücke (*Ae. albopictus*) und der einheimischen gemeinen Stechmücke (*Cx. pipiens*) auf Vektor-Bekämpfungsmaßnahmen unter dem Einfluss multipler Stressoren durchgeführt. Zur Simulation einer solchen Bekämpfungsmaßnahme wurde die Modellsubstanz λ-Cyhalothrin verwendet, die nicht nur als Adultizid, sondern auch als Larvizid gegen Stechmücken eingesetzt wird und einen Vertreter der immer mehr verwendeten Pyretroide darstellt. Als multiple Stressoren wurde im Hinblick auf die globale Klimaerwärmung die Veränderung der ökotoxikologischen Sensitivität in Abhängigkeit von der Temperatur einerseits und andererseits unterschiedliche Nährstoffregime der Ökosysteme, simuliert durch unterschiedliche Fütterungen, gewählt.

Im Vergleich der beiden Arten zeigte sich, dass zwar die Effekt-Konzentrationen der $LC_{10}$ (letale Konzentration, bei der im Vergleich zur Kontrolle 10% der Individuen sterben), $LC_{50}$ und $LC_{90}$ sich nicht statistisch unterschieden, jedoch die Inklination der Konzentrations-Wirkungs-Beziehung bei *Ae. albopictus* steiler war als bei *Cx. pipiens*. Die Temperatur als Stressor (20 °C, 25 °C und 30 °C) zeigt eine klare negative Korrelation sowohl bei der getesteten einheimischen Stechmückenart bei der $LC_{10}$, $LC_{50}$ und $LC_{90}$ sowie bei der invasiven Asiatischen Tigermücke bei der $LC_{50}$, wobei eine Temperaturerhöhung die Insektizidwirkung für *Ae. albopictus* stärker reduziert als für *Cx. pipiens*. Aus weiteren erhobenen subletalen Endpunkten, wie dem Zeitpunkt der Eiablage, der Fertilität der Eier und den Gelegegrößen wurde die endliche Wachstumsrate λ' für beide Arten in der Kontrolle und unter Insektizideinfluss ($LC_{50}$) sowie bei den drei Temperaturregimen berechnet. Es zeigte sich, dass für *Ae. albopictus* bei einer Bekämpfungsmaßnahme, bei der nicht die vollen Zielkon-zentration zur Bekämpfung der Arten erreicht wurde (LC$_{50}$), eine geringere Reduktion der endlichen Wachstumsrate λ' erfährt als *Cx. pipiens* – und das bei allen drei getesteten Temperaturen. Der Einfluss der Fütterung machte sich durch eine Erhöhung der Sensitivität gerade bei geringen Insektizidkonzentrationen bemerkbar. Diese Ergebnisse zeigen, dass die Effektkonzentration weit über der von anderen aquatischen Schlüsselarten und insbesondere invertebraten Räubern von Stechmücken liegt und so das Ökosystem durch den Einsatz des Insektizids λ-Cyhalothrin vor allem die natürliche Top-down-Regulierung von Stechmückenpopulationen verliert.

Neben der Generierung von zuverlässigen Life-History-Parametern und Dosis-Wirkungs-Beziehungen des neu entwickelten Testsystems hat die Studie zusammenfassend ergeben, dass unzureichende Bekämpfungsmaßnahmen, wie sie durch steigende Temperaturen begünstigt werden, *Ae. albopictus* sogar bei der Einnischung in heimische Ökosysteme von Nutzen sein können. Der getestete Wirkstoff könnte sowohl die interspezifische Konkurrenz mit heimischen Stechmückenarten verringern als auch den Prädationsdruck von *Aedes*-Larven durch invertebrate Räuber deutlich einschränken und damit einen unterstützenden Effekt für Ausbreitung und Etablierung der invasiven Art haben.

## 2.2 Zweite Publikation

**Abstract:** The Asian tiger mosquito, *Aedes albopictus* (Diptera: Culicidae, SKUSE), is an important threat to public health due to its rapid spread and its potential as a vector. The eggs of *Ae. albopictu*s are the most cold resistant life stage and thus, the cold hardiness of eggs is used to predict the future occurrence of the species in distribution models. However, the mechanism of cold hardiness has yet to be revealed. To address this question, we analyzed the layers of diapausing and cold acclimatized eggs of a temperate population of *Ae. albopictu*s in a full factorial test design using transmission electron microscopy. We reviewed the hypotheses that a thickened wax layer or chorion is the cause of cold hardiness but found no evidence. As a result of the induced diapause, the thickness of the dark endochorion as a layer of high electron density and thus an assumed location for waxes was decreasing. We therefore hypothesized a qualitative alteration of the wax layer due to compaction. Cold acclimation was causing an increase in the thickness of the middle serosa cuticle indicating a detachment of serosa membrane from the endochorion as a potential adaptation strategy to isolate inoculating ice formations in the inter-membranous space.

**Keywords**: Chorion, cold tolerance, compaction, freeze avoidance, wax layer, winter survival.

Populationen von *Aedes albopictus* aus gemäßigten Breiten verfügen über die Fähigkeit, durch eine Diapause und durch eine Kälteakklimatisierung der Eier eine Kältetoleranz (*cold hardiness*) aufzubauen und so Temperaturen von unter 0 °C im Ei-Stadium zu überleben. Die Ausbreitungsgrenzen von *Ae. albopictus* sind stark mit dieser Frosttoleranz der Eier verbunden. Eine genaue Kenntnis der beiden Mechanismen der Kältetoleranz ist also zentral, um zu verstehen, wie sich die ursprünglich tropisch verbreitete Art so schnell an Regionen in gemäßigten Breiten anpassen konnte und dort stabile Populationen aufbaute.

Unklar ist bisher, wie die Kältetoleranz der Eier von *Ae. albopictus* entsteht. Nachdem Hanson and Craig (1995b) zeigen konnten, dass eine erhöhte Einlagerung von Polyolen und die damit verbundenen Reduzierung des Supercooling-Points nicht der zugrunde liegende Mechanismus ist, wird aktuell die Hypothese verfolgt, dass durch einen verstärkten Fettsäuremetabolismus des Embryos die Wachsschicht des Chorions des Eis verstärkt wird (Urbanski et al., 2010a, 2010b; Reynolds et al., 2012; Poelchau et al., 2013).

Um einen Beitrag zur Untersuchung dieser Hypothese zu leisten, wurden in einer explorativen gekreuzten Zweifaktoren-Studie mit (nicht-) diapausierenden und (nicht-) kälteadaptierten Eiern (= erklärende Variablen) mittels transmissionselektronenmikroskopischer Aufnahmen die unterschiedlichen Chorion-Schichten vermessen (= abhängige Variablen) und die Daten in einer Hauptkomponentenanalyse korreliert. Da in der Literatur keine einheitliche Nomenklatur der Chorion-Schichten verwendet wird, wurde auf Basis der eigenen Ergebnisse, beruhend auf den unterschiedlichen Elektronendichten der Chorionschichten, eine Erweiterung der Chorion-Nomenklatur vorgeschlagen. Demnach umfasst das Chorion von außen nach innen folgende Schichten: Exochorion (EX), Endochorion, gegliedert in ein dunkles Endochorion (DE) und ein helles Endochorion (LE), gefolgt von der Serosa-Cuticula, die in die drei Schichten äußere Serosa-Cuticula (oSC), mittlere Serosa-Cuticula (mSC) und innere Serosa-Cuticula (iSC) gegliedert wurde, und die Serosa-Membran (SM) als innerste Schicht.

Die Hauptkorrelationen wurden anschließend in Einzeltests genauer untersucht, wobei festgestellt wurde: (i) Der einzige signifikante Unterschied zwischen diapausierenden und nicht diapausierenden Eiern bestand in der Reduktion des dunklen Endochorions (DE), das aufgrund der hohen Elektronendichte auch als der Ort für die intensivste Fettsäureeinlagerung und damit die Wachsschicht identifiziert wurde; (ii) Kälteadaptation führte zu einer Vergrößerung einer als Intermembranraum interpretierten mittleren Serosa-Cuticula (mSC); (iii) das helle Endochorion (LE) unterschied sich in seiner Dicke zwischen der Oberseite und der Un-terseite des Eis und (iv) die Variabilität des hellen Endochorions (LE) und der Serosa-Membran (SM) zwischen den Eiern war hoch.

Zuerst konnte gezeigt werden, dass die Kälteadaptation des Eis und die Diapause zwei unterschiedliche physiologische Prozesse sind. Außerdem resultierte aus den Ergebnissen eine Falsifizierung der Hypothese, dass sich im Zuge der Diapause oder Kälteadaptation die Wachsschicht des Chorions verdicken würde. Statt dessen werden die Ergebnisse im Sinne einer Vernetzung von Proteinen und Chitin mit den Fettsäuren sowie einer Veränderung der Sättigungen der Fettsäurereste interpretiert, die zu einer Kompaktierung des dunklen Endochorions (DE) führen könnte, was zu demselben gesuchten Effekt der gesteigerten Frosttoleranz und sogar Trockentoleranz führen könnte. Des Weiteren wurde die Vermutung aufgestellt, dass die durch die Kälteadaptation ausgelöste Weitung der als Intermembranraum interpretierten Schicht der mittleren Serosa-Cuticula (mSC) ein Mechanismus zur Vermeidung von inokulierenden Eisformationen sein könnte.

## 2.3 Dritte Publikation

Aljoscha Kreß, Ann-Marie Oppold, Ulrich Kuch, Jörg Oehlmann, Ruth Müller. **Cold tolerance of the Asian tiger mosquito *Aedes albopictus* and its response to epigenetic alterations.**

**Status:** under rev. in Journal of Insect Physiology

**Abstract:** Phenotypic plasticity is considered as one of the key traits responsible for the establishment of the invasive mosquito *Aedes albopictus* that is an important vector of viral and parasitic pathogens. The successful spread to higher altitudes and latitudes may be explained especially by the ability to rapidly induce a heritable low-temperature phenotype (cold hardiness in eggs). In view of the low genetic diversity of founder populations, an epigenetic short-term mechanism has been suggested to drive its diversification. We investigated if random epigenetic alterations promote the cold hardiness of eggs of *Ae. albopictus* individuals out of a transgenerational study with two epigenetic agents (genistein and vinclozolin). Therefore, we evaluated changes in lethal time for 50% of pharate larvae ($Lt_{50}$) of eggs exposed to -2°C in two subsequent generations using a new dose-response test design. Associated to the epigenetic change in the two subsequent offspring generations, we were able to detect a significant diversification of cold hardiness of the eggs (up to 64.5%). This effect size of the epigenetically modulated cold hardy effects is likely to have an impact on the spatial distribution of the species. Our results provide a framework for further research on epigenetic temperature adaptation of invasive species to better explain and predict their rapid range expansion.

**Keywords.** cold hardiness, distribution limits, winter survival, phenotypic plasticity, *Stegomyia albopicta*

Populationen von *Ae. albopictus* sind in der Lage, in wenigen Generationen einen Kälte-Phänotyp auszuprägen und auch wieder zurückzubilden, der befähigt ist, diapausierende Eier zu produzieren, die außerdem zur Kälteakklimatisierung in der Lage sind. Dieses Anpassungspotential von *Ae. albopictus* an die klimatischen Bedingungen der neubesiedelten Gebiete wird neben weiteren Faktoren für den Invasionserfolg verantwortlich gemacht (Scholte & Schaffner, 2007).

Eine solche Veränderung wäre mit einer Selektion in der Allelfrequenz der Gründerpopulation begründbar (Bergland et al., 2014), jedoch zeigen Release-Capture-Studien mit tropischen Männchen in gemäßigten Klimaregionen, dass bereits nach einem Winter die tropischen Allele eliminiert waren (Hanson et al., 1993). Jedoch gilt ein konstanter Eintrag von neuen Allelen in die Gründerpopulationen als unwahrscheinlich, und invasive Populationen haben eine auffallen geringe genetische Diversität (Black et al., 1988; Focks et al., 1994; Birungi & Munstermann, 2002; Lounibos et al., 2003). Die schnelle Ausprägung des Kälte-Phänotyps wäre jedoch auch mit einer sogenannte „epigenetischen Temperatur-Adaptation" zu begründen (Tzschentke & Basta, 2002). Dieser Prozess, der sich sowohl von der rein phänotypischen Adaptation an Temperaturen innerhalb einer Lebensspanne des Organismus (Akklimatisierung) sowie von der rein genetischen Adaptation an Temperaturen abgrenzt, beschreibt die vererbbare und durch Umwelteinflüsse veränderbare Mediation zwischen vorhandenen Allelen des Genotyps und dem resultierenden Phänotyp (Tzschentke & Basta, 2002; Gilbert & Epel, 2009; Glastad et al., 2011; Roberts & Gavery, 2012). Der Prozess ist reversibel und basiert nicht auf einer Veränderung der DNA-Sequenz (Russo et al., 1996; Tzschentke & Basta, 2002).

Der Zusammenhang zwischen der Plastizität des Phänotyps und der Veränderung des epigenetischen Systems wurde jedoch seitdem mehrfach für Invertebraten und Insekten demonstriert (Hick et al., 1996; Ono et al., 1999; Mandrioli & Volpi, 2003; Marhold et al., 2004a, 2004b; Schaefer & Lyko, 2007; Maleszka, 2008; Lyko et al., 2010; Roberts & Gavery, 2012; Ye et al., 2013; Wang et al., 2013), und auch die eigenen Ergebnisse zeigen, dass sich z.B. nach einer epigenetischen Veränderung die Insektizidsensitivität von *Ae. albopictus* in den nachfolgenden Generationen verändert (Oppold et al., 2015).

Es stellt sich daher die Frage, ob es möglich ist, eine vererbliche Diversifikation des Kälte-Phänotyps nach einer randomisierten epigenetischen Änderung der DNA-Methylierung in *Ae. albopictus* festzustellen. Hierzu wurden die als epigenetische Agenzien bekannten Substanzen Genistein und Vinclozolin zur Vorexposition der tropischen parentalen $F_0$-Generation ver-wendet (vgl.: Oppold et. al 2015), und im Anschluss die nichtexponierten $F_1$- und $F_2$-Generationen auf eine mögliche Veränderung in der Kältetoleranz hin überprüft. Dazu wurde ein neues und hochsensibles Versuchsdesign zur Bestimmung der mittleren Überlebensdauer bei Frostexposition der Eier ($Lt_{50}$) entwickelt.

Es zeigte sich, dass sowohl die epigenetischen Agenzien als auch das verwendete Lösemittel Aceton einen signifikanten Einfluss auf die Frosttoleranz der Eier haben und sich die Halbwertszeiten der Frostexposition signifikant mit über 50% unterschieden. Es ließ sich keine einfache Korrelation zwischen dem reinen Methylierungsgehalt des Cytosins der DNA und des Kälte-Phänotyps nachweisen, sondern ein substanzspezifischer Effekt wurde aufgedeckt. Scheinbar abhängig vom Wirkmechanismus unterscheiden sich die hypermethylierenden und hypomethylierenden Agenzien in ihrer transgenerationalen Wirkung auf die Frosttoleranz von *Ae. albopictus*. Diese Ergebnisse bilden damit die Grundlage für weitere Forschungen an der epigenetischen Temperatur-Adaptation von *Ae. albopictus* und können so dazu beitragen, den Invasionserfolg dieses Vektors zu erklären und vorher zusagen.

## 2.4 Methoden und Schlüsselergebnisse und ihre Bedeutung für die Wissenschaft

Neben der Entwicklung von neuen **Methoden** wie dem Testsystem für quarantänebedürftige invasive Arten zur Generierung von Life-History-Parametern und Dosis-Wirkungs-Beziehungen sowie dem Versuchsdesign zur Bestimmung der mittleren Überlebensdauer bei Frostexposition der Eier ($Lt_{50}$) wurde eine Erweiterung der Nomenklatur der Chorion-Schichten von *Ae. albopictus* vorgenommen. Die **wissenschaftlichen Schlüsselergebnisse** liegen in der Risikobewertung der Anwendung von λ-Cyhalothrin bei Bekämpfungsmaßnahmen, der weiteren Analyse der physiologischen Grundlage der Kältetoleranz der Eier von *Ae. albopictus* und in der Grundlage für einen möglichen epigenetischen Adaptations-Mechanismus der Kältetoleranz von *Ae. albopictus*.

### 2.4.1 Testsystem für Life-History-Parameter und Dosis-Wirkungs-Beziehungen

Wie sich anhand der sehr zuverlässigen Vertrauensbereiche der generierten Daten zeigen lässt, eignet sich das Testsystem zur Generierung von Life-History-Parametern und Dosis-Wirkungs-Beziehungen, um umfangreiche Studien an quarantänebedürftigen invasiven Arten wie *Ae. albopictus* durchzuführen. Dieser Versuchsaufbau leitet sich von der OECD-Guideline 219 zur chronischen Testung von Chemikalien mit der Zuckmücke *Chironomus riparius* ab und wurde explizit für Stechmücken mit ihren Besonderheiten, wie etwa dem gonotrophischen Zyklus, angepasst (Kreß et al., 2014). In der rund ein Jahr später erschienenen Studie von Owusu et al. (2015) von der Swiss TPH und der Uni Basel wurden das „CDC bottle assay" und das „WHO susceptibility assay" als die beiden einzigen standardisierten Testverfahren für Insektizidsensitivität von Culiciden anhand mehrerer Arten (*Aedes aegypti*, *Anopheles stephensi*, *An. gambiae* und *An. arabiensis*) und mehrerer Insektizide (Permethrin, λ-Cyhalothrin, DDT, Bendiocarb und Malathion) verglichen. Die Testsysteme wiesen gravierende Schwächen beim Detektieren von niedrigen sowie starken Veränderungen in den Sensitivitäten gegenüber den getesteten Insektiziden auf. Daraus schlossen Owusu et al. (2015), dass neue und standardisierte biologische Testsysteme gebraucht werden, die in der Lage sind, komplette Dosis-Wirkungs-Beziehungen zu erfassen. Das aus der OECD-Guideline 219 und für Culiciden abgeleitete Testsystem würde hierfür eine sehr geeignete Grundlage bieten. Der einfache und kostengünstige Aufbau, der dennoch die Generierung einer Vielzahl von Endpunkten ermöglicht, ist erstmals ein Angebot an andere wissenschaftliche Arbeitsgruppen und Institutionen, sich daran zu orientieren, um vergleichbarere Ergebnisse für physiologische oder ökotoxikologische Studien unter Quarantänebedingungen zu generieren. Hierzu trägt die Ermittlung von optimalen Fütterungsbedingungen für die Larven entscheidend bei (Müller et al., 2013). Auch andere Studien auf Basis des „WHO susceptibility assay" wie die von Tikar et al. (2008) mit *Aedes aegypti* (Temephos, Fenthion, Malathion und DDT) oder von Kanza et al (2013) *Anopheles gambiae* (Pyrethroid, DDT und Malathion) würden von dem neuentwickelten Testdesign profitieren. Aber auch komplexere Studien wie die von Muturi et al. 2011, die die dosisabhängige Malathion-Sensitivität von *Culex restuans* und *Ae. albopictus* bei unterschiedlichen Temperaturen untersuchte, haben gewisse Defizite. So leidet diese Studie an den fehlenden Quarantänevorkehrungen, die in Gebieten ohne bisher vorkommende Populationen von invasiven Stechmückenarten eine exakte Reproduktion zu einem Risiko werden lässt. Die KABS e.V. sowie das angeschlossene Institut für Dipterologie (IfD) sehen ihre Aufgaben u.a. in der experimentellen Optimierung der Bekämpfungspraxis und untersucht z.B. die Wirksamkeit von Bt-Toxinen auf Anopheles (Fillinger et al., 2003), wofür sich das Testsystem hervorragend eignen würde. Auch für das Department System-Ökotoxikologie des Helmholtz-Zentrums für Umweltforschung wäre das Testsystem ein Ansatzpunkt für weitere Forschung. Hier wurden schon mehrere Arbeiten zu den Effekten von Insektiziden auf Culiciden sowie eine Risikobewertung zu den Effekten auf die Biozönose der Brutgewässer durchgeführt (Meyabeme Elono et al., 2010; Meyabeme Elono, 2011).

## 2.4.2 Nomenklatur der Chorion-Schichten

Die Erweiterung der **Nomenklatur der Chorion-Schichten** von *Ae.-albopictus*-Eiern bietet die Grundlage für einen weiteren wissenschaftlichen Erkenntnisgewinn, was die Physiologie des Chorions angeht. Die metrische Aufklärung der Chorionschichten durch eine transmissionselektronenmikroskopische Studie trug dazu bei, weitere strukturelle Differenzierungen der Serosa-Cuticula sowie des Endochorions aufzudecken, die bei der Beantwortung der Frage zur Entstehung der Kältetoleranz und Trockentoleranz schon jetzt entscheidende Erkenntnisse lieferten. Wie Valle et al. 1999 ausführten, bestanden lange Zeit konzeptionelle Fehler in der Literatur, da der Aufbau der Eihülle von Stechmücken 1:1 von höheren Dipterengattungen wie *Drosophila* abgeleitet wurde, die als Standardorganismus der Genetik und Entwicklungsbiologie intensiv studiert wurden. Aber selbst nach der begrifflichen Korrektur von Valle et al 1999 und Monnerat et al. 1999 wurde das Chorion nicht bis ins Detail untersucht. So fokussierten sich auch die Autoren der beiden Studien lediglich auf das Exochorion mit seinen Tuberkelstrukturen. Die Serosa-Membran sowie die Serosa-Cuticula blieben dagegen gänzlich unberücksichtigt. Auch der in Monnerat et al. 1999 sich schon andeutenden Differenzierung zwischen einem dunklen Endochorion (DE) und einem hellen Endochorion (LE) in *Anopheles albitarsis* wurde nicht weiter nachgegangen. Auf Grundlage der nun gewonnenen wesentlich differenzierteren Aufschlüsselung der Chorion-Schichten sind daher vergleichende Studien zwischen den Arten von hohem Interesse, um weitere Unterschiede im strukturellen Aufbau zu erkennen und diese mit den Eigenschaften der Eier – wie Kälte- und Trockentoleranz – zu korrelieren.

## 2.4.3 Versuchsdesign zur Bestimmung der Halbwertszeit bei Frostexposition der Eier

Auch das neuentwickelte Versuchsdesign zur Bestimmung der Halbwertszeit bei Frostexposition der Eier bietet ein großes Potential, vergleichbarere Ergebnisse zwischen den internationalen Laboren zu generieren. Vorangegangene Studien weisen teilweise erhebliche methodische Schwierigkeiten auf, da sie z.B. ohne Replikate (Pumpuni et al., 1992; Hanson et al., 1993), mit teilweiser Replikation (Hanson & Craig, 1994) oder mit geringer Replikation (Mogi, 2011) arbeiteten und andere wiederum Probleme mit der Überrepräsentation von singulären Schlupfereignissen von Kohorten hatten (Thomas et al., 2012). Pumpuni et al. 1992 nutzten 900 bis 3,000 Eier ohne Replikation für jeweils eine von 14 getesteten Populationen von *Ae. albopictus* zur Bestimmung der kritischen Hell-dunkel-Zeit zur Auslösung der Diapause. Ohne Angaben der Standardabweichung oder des Standardfehlers ist die Interpretation der Datenreihen daher stark erschwert. Dahingegen geben Hanson et al. (1993) für ihre Untersuchung der Schlupfrate von hybridisierenden tropischen und gemäßigten Mischpopulationen in einer Release-Capture-Studie 33 bis 351 Eier pro Datenpunkt an; leider auch hier ohne Replikation und Vertrauensbereiche der Datenreihen. In der Studie von Hanson & Craig (1994) gab es nun eine dreifache Replikation mit 101 bis 344 Eiern pro Datenpunkt, um die Veränderung der Kältetoleranz durch die Diapause und/oder Kälteakklimatisierung festzustellen, jedoch wurde die tropische Vergleichspopulation nicht repliziert. In einer späteren Studie von Mogi (2011) wurde eine vierfache Replikation verwendet, um die Kältetoleranz von *Ae. albopictus* entlang eines Nord-Süd-Gradienten in Japan zu bestimmen. Das bewusste Weglassen der Standardabweichung in den Ergebnisabbildungen („SD is not shown") suggeriert jedoch auch hier eine Schwierigkeit im Umgang mit den Daten. Schließlich wendeten Thomas et al. (2012) ein neues Verfahren zur Bestimmung der temperaturabhängigen maximalen Überlebensdauer an. Hierbei wird jedoch ein singuläres Schlupfereignis in einer entsprechenden Temperatur-Dauer-Exposition schon als erfolgreiche Überdauerung des Temperaturstres-ses gewertet, was eine statistische Unterrepräsentierung der restlichen ungeschlüpften Eier in dem Datensatz zur Folge hat. Das nun entwickelte Verfahren bietet dagegen erstmals die Möglichkeit, einen leicht verständlichen und robusten Wert mit Validitätskriterien (95%-Vertrauensbereich) unter Berücksichtigung des kompletten Datensatzes zu generieren.

### 2.4.4 Risiko bei Bekämpfungsmaßnahmen

Es konnte gezeigt werden, dass unzureichende Bekämpfungsmaßnahmen, wie sie durch steigende Temperaturen wahrscheinlicher werden, bei der Einnischung von *Ae. albopictus* in heimischen Ökosystemen sogar von Nutzen sein können. Der Einsatz von Insektiziden, wie dem verwendeten Pyrethroid λ-Cyhalothrin, führt mit Gewissheit zu einer starken Schädigung von aquatischen Schlüsselarten wie den invertebraten Topprädatoren. Die massive Störung der Biozönose könnte auch Auswirkungen auf nichtaquatische Lebensstadien und Arten wie Libellen und Vögel haben, die als Prädatoren in Frage kämen. Ferner zeigt sich, dass durch den Einsatz von Pyrethroiden auch die Gefahr bestehen könnte, die interspezifische Konkurrenz mit den einheimischen Stechmückenarten – die in der Regel sogar den größeren Faktor für die Beeinflussung der Populationswachstumsrate der Larven darstellt – zugunsten der invasiven Art zu verändern. Daher kann die prinzipielle Etablierung der Art in heimischen aquatischen Ökosystemen durch eine falsche Stechmückenbekämpfung sogar gefördert werden. Diese Erkenntnisse weisen darauf hin, wie wichtig der Blick über die reine Testung von Wirkstoffschwellenwerten hinaus auf das gesamte Ökosystem ist.

Erste Studien zum Einfluss von z.B. Bt-Toxinen auf die Biozönose wurden wie bereits oben erwähnt im Department System-Ökotoxikologie des Helmholtz-Zentrums für Umweltforschung durchgeführt (Meyabeme Elono et al., 2010; Meyabeme Elono, 2011). Zudem wurden wie oben erwähnt erste Untersuchungen zur temperaturabhängigen Toxizität von Malathion an zwei Stechmücken-Arten getestet (Muturi et al., 2011). Andere Arten wie die Büschelmücke *Chironomus dilitus* oder *Hyalella azteca* (Amphipoda) zeigten ähnliche negative Temperatur-Toxizitäts-Korrelationen wie die getesteten Stechmücken (Harwood et al., 2009; Weston et al., 2009). Insgesamt ist der Wissensstand jedoch viel zu gering, um eine ausreichend gesicherte Effektkonzentration bei Bekämpfungsmaßnahmen benennen zu können, die einerseits ausreichend Wirkung zeigt und so gut wie möglich das betroffene Ökosystem schützt. Es empfiehlt sich daher, den Temperatur-Wirkungs-Zusammenhang von Insektiziden, die zur Bekämpfung von invasiven Stechmückenarten eingesetzt werden, komplett anhand repräsentativer Arten zu evaluieren, um die Maßnahmen genauer auf die jeweilige Witterungsbedin-gung abstimmen zu können. Insgesamt sollte die Einbeziehung von biotischen und abiotischen Faktoren in der Stechmückenbekämpfung weiter intensiviert werden, um die Konsequenzen unseres menschlichen Handelns auf die empfindlichen Ökosysteme genauer vorhersagen zu können.

Ferner zeigte sich, dass bei Vergleichen von Insektizidsensitivitäten die Temperatur ein überaus empfindlicher Einflussfaktor ist (Kreß, 2011). So wurde in einer Studie von Schroer et al. (2004) die Toxizität von λ-Cyhalothrin auf exemplarische Arten eines Kleinstgewässers im Labor mit denen in einem Mesokosmos verglichen, um im Anschluss das zur Risikobewertung vorgeschlagene Species-Sensitivity-Distribution-Modell (SSD) von Campbell et al. (1998) zu evaluieren. Es zeigte sich in dieser Studie, dass der im Labor gewonnene $HC_5$-Wert (Hazardous Concentration = schädliche Konzentration für 5% der getesteten Arten) von 2,7 ng/l sowohl unter dem im Freiland gewonnen $HC_5$-Wert von 4,1 ng/l als auch dem $HC_{10}$-Wert von 5,1 ng/l liegt und das Verfahren daher zuverlässige Werte zur Risikobewertung generiert. Jedoch wurden die Labordaten bei konstanten Temperaturen von 20 °C gewonnen, wohingegen im Freiland zum Zeitpunkt der Initalbehandlung eine Temperatur von 20 bis 24 °C vorlag (Leistra et al., 2004). Durch die hohe Akkumulation im Sediment sowie den hohen Abbauraten von λ-Cyhalothrin können die ersten 24 Stunden als das kritische Zeitfenster der Toxizitätsbelastung für die getesteten pelagischen Organismen angenommen werden (Leistra et al., 2004; Lawler et al., 2007). Benutzt man nun den Mittelwert aus den gewonnen Faktoren der temperaturabhängigen λ-Cyhalothrin-Toxizität und den einzigen sonst noch vorliegenden Daten von Harwood et al. (2009) und Weston et al. (2009), so würde sich in einem 4-°C-Worst-Case-Szenario die Toxizität um den Faktor 1,7 verändern. Wendet man nun diesen Faktor auf den Labor-$HC_5$-Wert von 2,7 ng/l an, so erhöht er sich theoretisch auf 5,6 ng/l. Damit liegt er nicht nur weit über dem im Freiland gewonnenen $HC_5$-Wert von 4,1 ng/l und würde so zu einer ökotoxikologischen Risiko-Unterschätzung im SSD-Verfahren führen, sondern auch über dem als noch sicherer eingestuften $HC_{10}$-Wert von 5,1 ng/l. Durch diese Rechnung zeigt sich erneut, wie wichtig das Wissen um die temperaturabhängigen Toxizität von Insektiziden zur Risikobewertung ist, es stellt jedoch auch komplette Studienergebnisse, die Grundlage für neue Risikobewertungsverfahren sind, in Frage.

## 2.4.5 Auswirkung von Kältetoleranz auf das Chorion

Die Ergebnisse der transmissionselektronenmikroskopischen Studie zeigen, dass die Wachsschicht womöglich gar nicht, wie in der Literatur angenommen, in der Serosa-Cuticula selbst liegt, sondern eher ein Teilbereich des Endochorions ist. Dieser elektronendichte Bereich wächst nicht mit zunehmender Kältetoleranz durch die induzierte Diapause sowie die Kälteadaptation, sondern die Diapause scheint zu einer Reduzierung der Eihüllendicke in diesem Bereich zu führen. Dies wurde mit einer Kompaktierung der Schicht erklärt. Auf Grundlage dieser Erkenntnis können weitere Studien, z.B. über die Sättigung oder die Quervernetzung der Fettsäuren, die Frage zur physiologischen Etablierung der Kältetoleranz abschließend beantworten. In der Studie von Farnesi et al. 2015 wird diskutiert, dass die Trockentoleranz von Stechmückeneiern vermutlich mit der Dicke des Chorions zusammenhängt und es keine vollständigen Daten zu den Schichtdicken gibt, was die Interpretation der dort gewonnenen Erkenntnisse zur Chitin-Einlagerung erschwert. Die nun erhobenen Daten liefern daher das Grundgerüst zur Neuinterpretation dieser und vorangegangener Studien (Rezende et al., 2008; Goltsev et al., 2009). Hier wurde jeweils das Exo- und Endochorion von *Ae. albopictus* mittels Natriumhypochlorit verdaut, um an die Serosa-Cuticula zu gelangen und diese z.B. auf ihren Chitingehalt zu untersuchen. Problematisch erscheint nun, dass nach den jetzt gewonnenen Erkenntnissen die Serosa-Cuticula nicht der Einlagerungsort der Lipide und damit der Wachse zu sein scheint und daher die Chitinisierung dieser Schicht als eher unbedeutend gedeutet werden kann. Dies würde erklären, warum in der Studie von Farnesi et al. (2015) die Chitinisierung nur teilweise mit der Trockentoleranz der getesteten Arten korrelierte und den Schluss nahelegte, dass noch weitere Faktoren entscheidend sein müssen. Eine weitere Suche nach einer Verdickung oder einer Anreicherung von Fettsäuren wie von Farnesi et al. (2015) in Aussicht gestellt, scheint jedoch in Anbetracht des aktuellen Erkenntnisstandes nicht mehr zielführend.

Gleichzeitig wurde gezeigt, dass bei Vermessungsstudien die Orientierung der Eihülle von entscheidender Bedeutung ist, da je nach räumlicher Lage vor allem das Endochorion Unterschiede aufweist, die nicht zu unterschätzen sind und in weiteren Studien zur Vermessung des Chorions stets berücksichtigt werden müssten. Ferner zeigte sich, dass der Mechanismus der Kälteadaptation sich physiologisch von dem der Diapause unterscheidet, auch wenn beide u.a. die Erhöhung der Kältetoleranz als Folge haben. Bei der Kälteadaptation scheint eher ein Mechanismus zur Ablösung der Serosa-Cuticula vom Endochorion stattzufinden. In diesem Intermembranraum könnten inokulierende Eisformationen, die bereits die kompaktierte

Wachsschicht des Endochorions durchdrungen haben, wachsen, ohne Schäden an Geweben anzurichten. Die Akkumulation von Wassermolekülen am wachsenden Eiskristall führt sogar noch zu einer Aufsalzung des verbleibenden Lumens und damit zu einer Erhöhung des Frostpunktes für das übrige Gewebe. Solche Mechanismen wurden bereits für andere Organismen beschrieben, bleiben bis jetzt jedoch bei der Klärung der Frage, wie *Ae. albopictus* seine besondere Kältetoleranz physiologisch erreicht, unberücksichtigt. Daher helfen die Erkenntnisse dieser Arbeit auch, z.B. Kandidatengene bei qRT-PCR-Untersuchungen festzulegen oder bei Auswertungen von Deep-Sequenzing-Daten eine Vorauswahl zu treffen. Dies bietet somit Anknüpfungspunkte z.B. für die Arbeitsgruppen von der Georgetown sowie Ohio State University oder die Arbeitsgruppe vom FIOCRUZ. Erstere z.B. haben eine Deep-Sequenzing-Studie von diapausierenden und nichtdiapausierenden Eiern von *Ae. albopictus* durchgeführt und ein genetisches Tool-Kit der Glukoneogenese und Zellproliferation identifiziert (Poelchau et al., 2013), das zwar die höhere Lipideinlagerung diapausierender Eier erklären kann (Reynolds et al., 2012), jedoch wurden hier komplette Eier sequenziert und nicht zwischen Larve und der extralarvalen Serosa-Membran differenziert. Somit ist zu vermuten, dass wichtige Signale des unabhängigen Serosa-Membran-Transkriptoms im Rauschen des embryonalen Transkriptoms untergegangen sind. Hier empfiehlt sich daher eine Deep-Sequenzing-Studie der alleinigen Serosa-Membran, um wichtige physiologische Erkenntnisse zur Diapause und auch zur Kälte-Akklimatisierung zu gewinnen.

## 2.4.6 Epigenetische Temperatur-Adaptation und Verbreitungsmodelle

Die Grundlagenarbeiten zur epigenetischen Temperatur-Adaptation in *Ae. albopictus* bieten das größte Anknüpfungspotential. Es wurde gezeigt, dass die schnelle epigenetische Temperatur-Adaptation an gemäßigte Breiten in Form der Etablierung eines Kälte-Phänotyps durch ein epigenetisches System moduliert werden könnte. Es trifft wohl auf generelles Verständnis, dass Individuen von *Ae. albopictus* eine reduzierte Fitness aufweisen würden, wenn sie in einem tropischen Habitat diapausierende und kältetolerante Eier produzieren würden oder in gemäßigten Breiten mit Minustemperaturen im Winter das phänotypische Potential zur Kältetoleranz sowie die Hibernation durch die Diapause ungenutzt bliebe (Focks et al., 1994). Dies ist in den Tropen und Subtropen durch die unangepasste Energieallokation in phänotypische Fähigkeiten wie die Kältetoleranz zu erklären sowie durch die verminderte Populations-Wachstumsrate durch die Diapause in einer potentiellen Vegetationsperiode und in den gemäßigten Breiten durch die erhöhte Mortalität im Winter.

Es zeigt sich also, das *Ae. albopictus* die biologische Reaktionsnorm für sowohl tropische, subtropische als auch gemäßigte Breiten besitzt. Um jedoch eine Maladaptation zu verhindern, ist der mehrfach dokumentierte rasche Wechsel zwischen den beiden Phänotypen in neu besiedelten Habitaten von größtem Vorteil für die Fitness der jeweiligen Population (Focks et al., 1994). Dieses in der Literatur sehr unterschiedlich bezeichnete Anpassungsvermögen (Focks et al., 1994; Kobayashi et al., 2002; Lounibos et al., 2003; Urbanski et al., 2012; Bonizzoni et al., 2013) scheint nach den nun vorliegenden Erkenntnissen eine epigenetische Temperatur-Adaptation zu sein, bei der ein Umwelteinfluss eine vererbbare, aber reversible Kurzzeit-Anpassung auslöst, ohne dass sich die Allelfrequenz der Population oder das Genom an sich verändern (Russo et al., 1996; Martienssen, 2008; Roberts & Gavery, 2012).

Wie in Kraemer et al. (2015) zusammengefasst, gingen die ersten Verbreitungsmodelle von *Ae. albopictus* (und *Ae. aegypti*) lediglich von Temperaturrestriktionen aus gewonnen Laborversuchen aus und fokussierten sich auf einzelne Regionen. Nawrocki & Hawley (1987) legten eine minimale Durchschnittstemperatur des kältesten Monats von -5 °C an, wohingegen Kobayashi et al. (2002) dies später bei -2 °C anlegten. Benedict et al. (2007) modellierten erstmals anhand von einem großen Set von Funddaten und interpolierten diese Ergebnisse auf die Klimavorhersagen des IPCC. Weitere Studien erweiterten die Verbreitungsmodelle um Faktoren wie den Niederschlag (Caminade et al., 2012). Brady et al. (2013) erweiterten die Studien um Überlebensdaten aus dem Feld zunächst auch nur lokal und schließlich für eine weltweite Untersuchung (Brady et al., 2014). Neuere Studien beziehen nun anthropogene Faktoren wie die Urbanisierung sowie interspezifische Konkurrenz mit ein (Kraemer et al., 2015).

Jedoch gehen die meisten Modellrechnungen über die zu erwartende Verbreitung von *Ae. albopictus* zurzeit von nur einem statischen Phänotyp und daher von einem Nischen-Konservatismus aus. Daher sind diese neuen Erkenntnisse und Hypothesen ggf. von Relevanz für zukünftige Berechnungen. Die phänotypische Plastizität ist jedoch ein sehr entscheidender Faktor für die Ausbreitungsmodelle von invasiven Arten, wie *Ae. albopictus,* deren Beachtung für die Qualität der Studien unerlässlich ist (Medley, 2010; Mogi et al., 2012; Jia et al., 2016). So zeigte eine Studie von Medley (2010), dass Modelle für *Ae. albopictus* auf Basis der Nischen im asiatischen Ursprungsgebiet nur schlecht die aktuellen Verbreitungen in den neuen Verbreitungsgebieten widergeben und vice versa. Medley schließt daraus, dass der geringe Nischenkonservatismus zu besonderen Schwierigkeiten bei der Modellentwicklung für *Ae. albopictus* führt. Jedoch wird das Adaptationspotential seitdem in keinem Modell für die Artverbreitung mehr berücksichtigt (Kraemer et al., 2015; Jia et al., 2016).

## 2.4.7 Eingehende Darstellung der epigenetischen Temperatur-Adaptation und Erweiterung des Konzepts

Der Prozess der epigenetischen Temperatur-Adaptation steht, wie bereits erwähnt, zwischen der rein phänotypischen Adaptation an Temperaturen innerhalb der Lebensspanne eines Organismus (Akklimatisierung) und der rein genetischen Adaptation an Temperaturen (Tzschentke & Basta, 2002). Er beschreibt eine vererbbare und durch Umwelteinflüsse veränderbare und reversible Mediation zwischen vorhandenen Allelen des Genotyps und dem resultierenden Phänotyp (Tzschentke & Basta, 2002; Gilbert & Epel, 2009; Glastad et al., 2011; Roberts & Gavery, 2012). Um den Mechanismus der epigenetischen Temperatur-Adaptation genauer zu beschreiben, werden im Folgenden bestimmte Annahmen generalisierend und daher auch simplifiziert dargestellt. Daher bleiben z.B. weitere, ggf. gleichzeitig oder synergistisch auftretende genetische Adaptationsmechanismen wie die Mutation, Alleldrift und -shift teils ebenso unberücksichtigt wie die Funktion der Epigenetik bei Ontogenese und DNA-Reparatur.

In einem stabilen Stadium ist die biologische Reaktionsnorm einer Population aufgrund der reduzierten Alleldiversität kleiner als die einer gesamten Art, da sie nur einen Ausschnitt bildet (siehe Abbildung 7a). Die Reaktionsnorm einer Population ist daher auf die Reaktionsnormen der Individuen der Population (in Abbildung 7 der Übersicht halber als kleine Kreise dargestellt; in der Realität jedoch auch größer und vielförmiger) und auf das vorhandene Set der Allele begrenzt. Zusätzlich greifen Mechanismen der phänotypischen Kanalisierung, hierbei werden z.B. „leckende" Transkripte durch microRNAs unterdrückt, oder die epigenetische Gen-Stilllegung puffert die Plastizität des Phänotyps zusätzlich ab, indem die Merkmale eines Allels, die aufgrund limitierter Ressourcen keinen Fitnessvorteil bringen, nicht ausgeprägt werden (Hornstein & Shomron, 2006; Reamon-Buettner et al., 2008; Posadas & Carthew, 2014). Reduzierte genetische Diversität und phänotypische Kanalisierung tragen daher zu einer reduzierten Reaktionsnorm der Population im Vergleich zur gesamten Art bei.

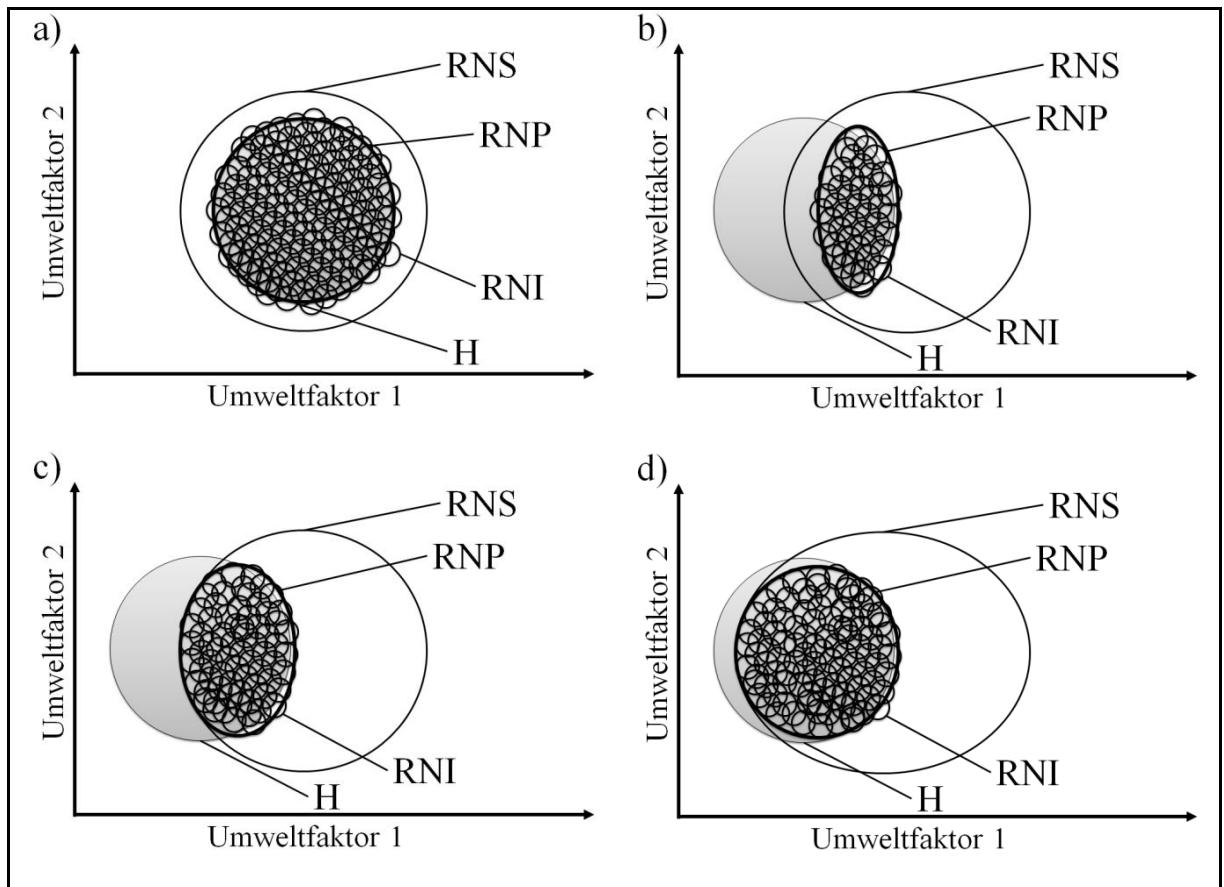

**Abbildung 7 (eigenes Werk, unveröffentlicht): Universelles und vereinfachtes Schema der epigenetischen Adaptation in Form einer schnellen Antwort des vererbbaren Phänotyps auf veränderte Umwelteinflüsse. RNS = Reaktionsnorm der Spezies, RNP = Reaktionsnorm der Population, RNI = Reaktionsnorm des Individuums, H = Habitat. a) Stabiler Ursprungszustand, b) Veränderung des Habitats entlang des Umweltfaktors 1, gefolgt von natürlicher Selektion, c) schnelle epigenetische Adaptation durch Aktivierung stillgelegter Gene, gefolgt von Erhöhung der phänotypischen Plastizität, d) langfristige Adaptation durch Mutation und natürliche Selektion.**

Im Falle einer schlagartigen Veränderung des Habitats (oder auch eines Habitatwechsels z.B. durch Ausbreitung) kommt es durch natürliche Selektion zum Aussterben aller Individuen, deren Reaktionsnormen außerhalb der neuen Habitatseigenschaften liegen. In Abbildung 7 ist das in der Realität durch Vielzahl von Faktoren definierte Habitat als ein zweidimensionales Habitatsmodell dargestellt. Hier könnte der Umweltfaktor 1 z.B. die Temperatur beschreiben, die in Abbildung 7b geringer wird. Durch die dadurch verursachte Reduktion der genetischen Diversität aufgrund des Verlustes von Allelen ergibt sich eine Reduktion der Reaktionsnorm der kompletten Population (siehe Abbildung 7b).

Nach einer Phase der Akklimatisierung der Individuen könnte es in einem nächsten Schritt entweder zu einer ungezielten und transgenerationalen epigenetischen Regulierung durch diffuse Stressoren kommen, oder einer durch Umwelteinflüsse ausgelösten gezielten epigenetischen Aktivierung stillgelegter Allele vorhandener Gene und einer damit verbundenen transgenerationalen und gerichteten epigenetischen Adaptation an das neue Habitat (Abbildung 7c). Es zeigte sich bei mehreren Arthropoden, dass Haushaltsgene in der Regel weniger von epigenetischer Regulation betroffen sind (Wang et al., 2013). Zu vermuten ist, dass zufällige genomweite Änderungen von hohen Risiken einer negativen Selektion begleitet wären, so wie z.B. randomisierte Strahlenschäden in der Regel auch keine positiv selektierten Mutationen hervorbringen. Eine gerichtete Adaptation, ausgelöst durch spezifische Umwelteinflüsse, könnte mit der vermittelnden Rolle von microRNA der siRNAs erklärt werden, da small-RNAs (i) durch Umwelteinflüsse wie Temperaturstress beeinflusst werden, (ii) die Fähigkeit der Zielorterkennung auf der DNA besitzen, (iii) mit DNMT interferieren und dadurch die Möglichkeit haben, auf den Methylierungsgrad der DNA Einfluss zu nehmen, und (iv) trans-generational wirken (Lippman & Martienssen, 2004; Hilbricht et al., 2008; Feil & Fraga, 2011; Ito et al., 2011; Vandegehuchte & Janssen, 2011; Puthiyakunnon et al., 2013; Elgart et al., 2015). Es kann also davon ausgegangen werden, dass sich der Mechanismus im Falle einer durch spezifische Umweltfaktoren ausgelösten, gerichteten epigenetischen Adaptation maßgeblich auf sogenannte genetische Tool-Kits beschränkt, wie das im Falle des kältetoleranten Phänotyps bei *Ae. albopictus* hypothetisierte Pepck/PCNA-Tool-Kit, welches durch die Diapause reguliert wird (Poelchau et al., 2013). Zudem zeigte sich jüngst, das in der Wespe *Nasonia vitripennis* die photoperiodisch induzierte Diapause unter epigenetischer Regulation steht (Pegoraro et al., 2016) und die ersten microRNAs wurden bereits für *Ae. albopictus* charakterisiert (Puthiyakunnon et al., 2013). In beiden möglichen Fällen, der gerichteten oder der ungerichteten epigenetischen Adaptation, könnte sich die phänotypische Plastizität der Population durch epigenetische Regulation erhöhen und so die Reaktionsnorm der Population im Rahmen der Habitateigenschaften verbreitern.

Im nächsten Schritt könnte eine erhöhte Mutationsrate der Gene eine Rolle spielen, da methylierte Nukleotide zu Fehlern in der DNA-Reparatur führen und so eine ungerichtete Erhöhung der phänotypischen Plastizität stattfindet (Chahwan et al., 2010). Diese neuen Genotypen und ihre korrespondierenden Phänotypen mit ihren neuen Reaktionsnormen bieten dann das Potential – wenn sie innerhalb der Habitateigenschaft liegen und nicht negativ selektiert werden –, die Reaktionsnorm der Population oder gar der gesamten Art zu erweitern (Abbildung 7d).

Dieses Konzept der epigenetischen Adaptation könnte zusätzlich herangezogen werden, um das Phänomen der hohen Diversität der Biotypen der Mitglieder des *Culex-pipiens*-Komplexes zu erklären, die z.T. auch als (ökophysiologische) Rassen, Varietäten oder Ökotypen bezeichnet werden (Vinogradova, 2000) und genetisch eng miteinander verwandt sind (Farajollahi et al., 2011; Boffelli & Martin, 2012). Der *Cx.-pipiens*-Komplex besteht aus der nördlichen Hausstechmücke *Cx. pipiens pipiens* sensu stricto (Linnaeus 1758) zusammen mit ihrer hypogäischen Form *Cx. p. pipiens* forma *molestus* (Forskal 1775) und der asiatischen Subspezies *Cx. pipiens pallens* (Coquillett 1898) (Vinogradova, 2000; Farajollahi et al., 2011; Becker et al., 2012). Außerdem gehören diesem Komplex wahrscheinlich die südliche Hausstechmücke *Cx. quinquefasciatus* (Say 1823) und die in Australien verbreitete *Cx. australicus* (Dobrotworsky 1965) und (strittig) *Cx. globocoxitus* (Dobrotworsky 1953) an (Vinogradova, 2000; Farajollahi et al., 2011; Harbach, 2012). Neben der geographischen Verbreitung der einzelnen Taxa gibt es eine Reihe von Unterschieden zwischen den Biotypen, z.B. im Verhalten und der Physiologie, kaum jedoch in der Morphologie und Genetik (Vinogradova, 2000; Farajollahi et al., 2011; Boffelli & Martin, 2012): (i) bevorzugtes Larvenhabitat (hypogäisch versus epigäisch; rural versus urban) oder ähnlich wie bei *Ae. albopictus*, (ii) die Hibernationsstrategie (heterodynamisch versus homodynamisch; Winterruhe der Larven versus Diapause der Eier), (iii) gonotrophischer Zyklus (fakultativ autogen versus nichtautogen), (iv) Paarungsschwarm-Struktur (eurygam versus stenogam) und, letztlich von besonderer Bedeutung für Vektorkompetenz und -kapazität, (v) Wirtssuche und Wirtspräferenz (ornithophil versus mammalophil und im Speziellen anthropophil; aktophil versus endophil) (Vinogradova, 2000; Farajollahi et al., 2011). Diese teilweise weder genetisch noch enzymatisch zu unterscheidenden Biotypen passen sich womöglich durch eine schnelle epigenetische Adaptation an vorhandene Habitate an, indem spezifische Tool-Kits von Genen stillgelegt oder reaktiviert werden, gefolgt von genetischen Mutationen und Bifurkationen der Klade bei räumlicher Isolation oder assoziativer Paarung (Boffelli & Martin, 2012). In anderen Arten ist vor allem die Anthropophilie untersucht worden. *Aedes aegypti* kann, abgesehen von einer nicht zwangsläufig eindeutigen Färbung, auch anhand von olfaktometrischen Studien in die weltweit verbreitete anthropophile Unterart *Ae. aegypti aegypti* und die weniger anthropophile und ausschließlich im ruralen Kenia vorkommenden Unterart *Ae. aegypti formosus* unterteilt werden (McBride et al., 2014; McBride, 2016). Dabei sind 14 chemosensorische Gene der Antennen bei *Ae. aegypti aegypti* in ihrer Funktion signifikant hoch reguliert, wovon einige für Ausprägungen von Rezeptorproteinen verantwortlich sind, die auf Bestandteile des menschlichen Schweißes reagieren (McBride et al., 2014). In der Gattung *Anopheles* ragen die beiden einzigen anthropophilen Arten *An. gambiae* und *An. coluzzii* (die früher als S- und M-Form von *An. gambiae* geführt wurden) aus einem morphologisch identischen Artkomplex heraus, dessen übrige Mitglieder als weniger anthropophil gelten (McBride, 2016). Es zeigt sich, dass in den beiden anthropophilen Arten im Vergleich zur nichtanthropophilen Art *An. quadriannulatus* die Gene für die Ausprägung humanspezifischer Rezeptoren der Antennen und Maxillar-Palpen stärker exprimiert werden und die nicht humanspezifischen Rezeptoren geringer; und das bei gleicher Anzahl und einem evolutionär gesehen stabilen Repertoire an chemosensorischen Genen zwischen den Arten (Rinker et al., 2013). Außerdem wurde gezeigt, dass die chemosensorischen Gene einer erhöhten Evolutionsrate unterliegen (Rinker et al., 2013), was für methylierte Gene typisch ist (Chahwan et al., 2010). Wie eingangs erwähnt, kann die epigenetische Adaptation parallel und synergistisch mit anderen Adaptationsmechanismen, wie z.B. der Veränderung der Allelfrequenz, eine weitere Erklärung für sehr schnelle Adaptationen sein. Diese Beispiele von Biotypen bei Stechmücken könnten also im eigentlichen Sinne auch sogenannte Epitypen darstellen, also Populationen, die im Genotyp nicht signifikant abweichen, deren daraus resultierende Phänotypen sich jedoch aufgrund unterschiedlicher epigenetischer Information z.B. im Verhalten oder in der Physiologie unterscheiden (Meagher, 2010).

Außerdem zeigten Studien eine Korrelation zwischen der DNA-Methylierung in Esterase-Genen und einer Überproduktion an Esterase-Proteinen in organophosphatresistenten Stämmen von den beiden Blattlausarten *Schizaphis graminum* und *Myzus persicae* (Field et al., 1989; Hick et al., 1996; Ono et al., 1999). Auch wenn in den organophosphatresistenten Stämmen von *Cx. quinquefasciatus* keine auffälligen Methylierungsmuster gefunden wurden und man von einer Gen-Amplifizierung in diesen Stämmen ausgeht (Qiao & Raymond, 1995; Field et al., 2004), so zeigen die eigenen Befunde doch, dass eine Änderung des epigenetischen Musters in *Ae. albopictus* zu einer geringeren Sensitivität gegenüber einem Insektizid führt (Oppold et al., 2015). Daher ist es auch empfehlenswert, die epigenetische Adaptation als einen nicht unwesentlichen Prozess der Anpassung an Insektizide und der schnellen Resistenzbildung bei weiteren Forschungsfragen zu berücksichtigen (Bass & Field, 2011; Pavlidi et al., 2012; Vontas et al., 2012).

Die Interpolation von der epigenetischen Temperatur-Adaptation auf weitere mögliche epigenetisch regulierte Tool-Kits in Culiciden ist lediglich ein erster Schritt. Weitere epigenetisch vermittelte phänotypische Plastizitäten anderer Arten könnten durch epigenetische Adaptation erklärbar sein – eine monokausale Erklärung für schnelle Adaptationsprozesse wird jedoch als unrealistisch eingeschätzt. Gleichwohl wird durch die gewonnenen Erkenntnisse womöglich ein weiterer Mechanismus der Adaptation hinzugefügt. Die epigenetisch vermittelte Plastizität des Phänotyps einer Art und seiner Nischen-verschiebung sollte daher in Risikobewertungen der Klimafolgenforschung Berücksichtigung finden.

## 2.5 Nutzen für die Gesellschaft

Die Bedeutung der Stechmücken für die Biodiversität steht selten im Blick der Gesellschaft. Duman und Kollegen (1991) beschreiben in ihrem Werk „Adaptations of insects to subzero temperatures" recht nachvollziehbar, dass diese seit 79 Mio. Jahren existierende Tierfamilie (Poinar & Zavortinik, 2000) einen nicht unerheblichen Anteil der Biozönose einiger Ökosysteme ausmacht: „The burst of insect activity in the Arctic in summer, notorious in many areas because of the abundance of biting dipterans, illustrates the tremendous overwintering success of these populations." Daher ist der Einfluss der knapp 3.350 Arten der Familie Culicidae (Harbach, 2016) auf die Stoffwechselkreisläufe und damit die Ökosystemdienstleistungen nicht zu vernachlässigen, gerade auch in Anbetracht dessen, dass durch den Parasitismus der Weibchen im gonotrophen Zyklus essentielle Spurenstoffe wie Stickstoff und Phosphor aus dem Blut der Wirte z.B. in die meist oligotrophen Ökosysteme der arktischen Zonen eingebracht werden. Auch die Auswirkungen von Stechmücken auf Tiere, welche als bevorzugte Wirte für Blutmahlzeiten dienen, können erheblich sein und ganze Populationen regulieren. So kann z.B. das Auftreten von großen Stechmückenschwärmen in nordamerikanischen Karibu-Herden zu starkem Blutverlust bis hin zu einer erhöhten Mortalität der Kälber führen (Mörschel & Klein, 1997; Culler et al., 2015). Als Überträger von zoonotischen Pathogenen hatten Stechmücken daher im Laufe der Evolution auch indirekt erheblichen Einfluss auf die Fitness, Dynamik und räumlich-zeitliche Ausbreitung von Tierpopulationen und damit die Zusammensetzung von Biozönosen. Darüber hinaus dienen viele Wild- und Nutztiere nicht nur als Wirte, sondern auch als Reservoire für die zoonotischen Pathogene und können von großer veterinärmedizinischer Bedeutung sein. Ihre zentralste Bedeutung für den Menschen und damit auch für die Beachtung der Stechmücken in der Gesellschaft haben diese Insekten jedoch durch ihre Rolle als Überträger zahlreicher Humanpathogene erhalten. Die Malaria als Erreger des Sumpf-, Faul-, Wechsel-, Marschen- oder auch Römischen Fiebers wirkte sich nicht nur populationsdezimierend auf die Zivilbevölkerung aus, sondern konnte kriegsentscheidend sein (Schäfer, 2014). Wichtige Persönlichkeiten der europäischen Geschichte litten oder starben an der Malaria, bis die Krankheit durch die Trockenlegung von Sümpfen und Feuchtgebieten wie etwa bei der Altarmbegradigung durch Johann Gottfried Tulla in Europa nahezu ausgelöscht wurde (Schäfer, 2014). Neben Parasiten wie den Erregern von Malaria und Lymphatischer Filariose haben in den vergangenen Jahrzehnten besonders von Stechmücken übertragene Virus-Erkrankungen wie Dengue-Fieber, Chikungunya-Fieber, Gelbfieber und Japan-Enzephalitis an Bedeutung gewonnen. Im globalen gesellschaftspolitischen Diskurs sind neben diesen zurzeit vor allem die weltweite Ausbreitung des Zika-Virus sowie ein Wiederaufkommen des Gelbfiebers im südlichen Afrika relevant (WHO, 2016c).

Schon zu den Anfangszeiten der Stechmückenbekämpfung mittels DDT und Dieldrin wurde die Hoffnung geweckt, damit das Ende von insektenübertragbaren Krankheiten eingeläutet zu haben (Dunlap, 2014). Der intensive Einsatz von Insektiziden bei der Bekämpfung von Krankheits-Vektoren ist jedoch ein massiver Eingriff in das ausdifferenzierte Gefüge vor allem der aquatischen Ökosysteme. Das Buch „Silent Spring" von Rachel Carson über Auswirkungen des Gebrauchs von Pestiziden führte ab 1962 zu einem kritischeren Bewusstsein in weiten Teilen der Bevölkerung über die Risiken und Konsequenzen etwa des Einsatzes von DDT auf unterschiedliche Ökosysteme (Carson et al., 2002; Murphy, 2007; Dunlap, 2014). Die im Rahmen der eigenen Untersuchungen gewonnenen Erkenntnisse nähren zusätzliche Befürchtungen, dass selbst Bekämpfungsmaßnahmen mit als wenig bedenklich geltenden Insektiziden wie Pyrethroiden – sogar solchen späterer Entwicklungsgenerationen – noch unbekannte Risiken für die öffentliche Gesundheit bergen können (wie z.B. die Förderung der Etablierung einer invasiven Art von Krankheitsvektoren).

Dennoch gibt es immer wieder aus der Politik oder der Wissenschaft Befürworter einer Intensivierung des insektizidbasierten Kampfes gegen Stechmücken, um die in die Hunderttausende gehenden Todesfälle und in die Millionen gehenden Krankheitsfälle sowie die damit in Zusammenhang stehenden hohen gesamtgesellschaftlichen monetären Aufwendungen zu reduzieren (Fang, 2010). Der von einigen Wissenschaftlern geforderte Schutz des Menschen vor Arboviren durch eine komplette Ausrottung von Stechmücken weltweit (Fang, 2010) kann im weitesten Sinne als Vorschlag eines Geo-Engineering der Biosphäre bezeichnet werden, dem jegliche Umweltrisikobewertung fehlt. Trotz des z.T. spezifischen Parasitismus gewisser Arten bei ihrem gonotrophen Zyklus wird eine einfache Substitution der Ökosystemfunktionen einer ganzen Tierfamilie durch andere Tiergruppen angenommen, ohne dass es hierfür belastbare Hinweise gäbe. Auch stellt sich bei Experimenten, in denen GVOs zur Bekämpfung von Stechmücken in Ökosysteme entlassen werden (z.B. der RIDL), die Frage, ob die durchgeführte Risikobewertung für einen so intensiven und irreversiblen Eingriff in die Biosphäre ausreichend war. Gerade auch im Hinblick darauf, dass neben ohnehin schon starken anthropogenen Belastungen wie durch Landnutzung nun auch noch der globale Klimawandel mit steigenden Temperaturen, veränderten Niederschlagsstrukturen und einer Häufung von Extremwetterereignissen zu einer weiteren Intensivierung abiotischer Stressoren auf die Ökosysteme führt, sind Geo-Engineering-Maßnahmen, welche die Biosphäre betreffen, mit besonderer Skepsis zu betrachten (Cairns et al., 1975; Deutsche Bundesregierung, 2008).

Die Anpassungsfähigkeit von Vektoren wie *Ae. albopictus* an Stressoren wie Kälte oder auch Insektizide scheint angesichts der in dieser Arbeit gewonnenen Erkenntnisse höher zu sein als bisher erwartet. Die schnelle Entwicklung von Resistenzen gegenüber etablierten, aber auch neuen Insektiziden bis hin zu Kreuzresistenzen deutet dabei auf ein „Wettrennen" zwischen Insektizidherstellern und Stechmücken hin. Dieses wird zu Lasten der Biodiversität, Funktion und Resilienz der Ökosysteme gehen, wenn Umweltrisikobewertungen und ihre Empfehlungen zugunsten des Akutschutzes der öffentlichen Gesundheit vor Infektionskrankheiten vernachlässigt werden (Stolz, 2008). Allerdings werden in diesem Zusammenhang häufig auch die nachfolgenden gesundheitlichen Auswirkungen für die Menschen nicht immer oder erst spät mitbedacht. So brachte z.B. der Diskurs über das Zika-Virus und die Ursache der gehäuft beobachteten Schädelfehlbildungen (Mikrozephalie) bei Neugeborenen eine neue Diskussion über Wirkungen und Nebenwirkungen insektizidbasierter Stechmücken-Bekämpfung in Gang (PCST, 2016). Diese Entwicklung weist auf ein weiteres relevantes, aber bisher vergleichsweise wenig beachtetes Problem für die öffentliche Gesundheit in verschiedenen Regionen hin.

Eine Fokussierung der Forschung auf die Entwicklung von Impfstoffen erscheint daher als nachhaltigerer Ansatz. Wenn z.B. *Ae. albopictus* aufgrund einer impfstoffbasierten Auslöschung der Pathogene schlichtweg ihre Gefährlichkeit für die menschliche Gesundheit als Vektor verliert, stellt diese Art in neubesiedelten – aber auch endemischen – Regionen lediglich eine Belästigung dar (mit einer geringen Gefahr von Sekundärinfektionen aufgekratzter Mückenstiche). Die hohe Effektivität der Gelbfieberimpfung als Beispiel, die nur einmal verabreicht werden muss und eine sehr hohe Immunisierungsquote von über 99% aufweist, hatte in vielen Ländern einen drastischen Rückgang der Krankheitsfälle zu Folge (WHO, 2013). Die Anwendung von Repellentien zur Verhinderung des Mensch-Mücke-Kontakts oder die gelegentliche Ausbringung von hochspezifischen Insektiziden, wie solchen auf Basis von B.t.-Toxinen, dürften für den Umgang mit den dann nur noch als Lästlinge zu bezeichnenden Arten ausreichend sein, selbst wenn auch dort von einer Resistenzentwicklung ausgegangen werden kann (Bates et al., 2005). In Anbetracht der enormen finanziellen Aufwendungen, die bis jetzt schon erbracht werden, um Stechmücken zu bekämpfen, ist auch im Hinblick auf die Schädigungen der Ökosysteme und eine Beeinträchtigung ihrer Dienstleistungen durch die Bekämpfungsmaßnahmen eine Abkehr von der bisherigen Strategie zum Schutz der Bevölkerung zu empfehlen. Bis zur Etablierung eines flächendeckenden Impfschutzes für die gefährlichsten Krankheitserreger nicht nur im finanzstarken sogenannten globalen Westen, sondern viel dringender auch mittels Generika im ärmeren sogenannten globalen Süden, bedarf es jedoch übergangsweise der bereits bestehenden Maßnahmen, die mit größter Sorgfalt und unter Risikoabwägung für Mensch und Natur angewendet werden sollten.

# 3 Quellenverzeichnis

Adams, R.L., McKay, E.L., Craig, L.M., Burdon, R.H., 1979. Methylation of mosquito DNA. Biochim. Biophys. Acta 563, 72–81. doi:10.1016/0005-2787(79)90008-X

Adhami, J., Reiter, P., 1998. Introduction and establishment of *Aedes* (*Stegomyia*) *albopictus* skuse (Diptera: Culicidae) in Albania. J. Am. Mosq. Control Assoc. 14, 340–3.

Allaby, M., 2010. A dictionary of ecology - oxford reference, 4th ed. Oxford University Press, Oxford. doi:10.1093/acref/9780199567669.001.0001

Alto, B.W., Juliano, S.A., 2001a. Temperature effects on the dynamics of *Aedes albopictus* (Diptera: Culicidae) populations in the laboratory. J. Med. Entomol. 38, 548–556. doi:10.1603/0022-2585-38.4.548

Alto, B.W., Juliano, S.A., 2001b. Precipitation and temperature effects on populations of *Aedes albopictus* (Diptera: Culicidae): implications for range expansion. J. Med. Entomol. 38, 646–656. doi:10.1603/0022-2585-38.5.646

Angelini, R., Finarelli, A., Angelini, P., Po, C., Petropulacos, K., Macini, P., Fiorentini, C., Fortuna, C., Venturi, G., Romi, R., Majori, G., Nicoletti, L., Rezza, G., Cassone, A., 2007. An outbreak of chikungunya fever in the province of Ravenna, Italy. Eur. Commun. Dis. Bull. 12, 3260.

Armistead, J.S., Arias, J.R., Nishimura, N., Lounibos, L.P., 2008. Interspecific larval competition between *Aedes albopictus* and *Aedes japonicus* (Diptera: Culicidae) in Northern Virginia. J. Med. Entomol. 45, 629–637. doi:10.1093/jmedent/45.4.629

Bakouri, H. El, Ouassini, A., Aguado, J.M., García, J.U., 2007. Endosulfan sulfate mobility in soil columns and pesticide pollution of groundwater in northwest Morocco. Water Environ. Res. 79, 2578–2584. doi:10.2175/106143007X184528

Basabose, K., 1996. Larvivorous potential of different stages of *Culex tigripes* (Diptera, Culicidae) in the prospective of its use in biological control of malaria vectors. Trop. 14, 13–15.

Basilua Kanza, J.P., El Fahime, E., Alaoui, S., Essassi, E.M., Brooke, B., Nkebolo Malafu, A., Watsenga Tezzo, F., 2013. Pyrethroid, DDT and malathion resistance in the malaria vector *Anopheles gambiae* from the Democratic Republic of Congo. Trans. R. Soc. Trop. Med. Hyg. 107, 8–14. doi:10.1093/trstmh/trs002

Bass, C., Field, L.M., 2011. Gene amplification and insecticide resistance. Pest Manag. Sci. 67, 886–90. doi:10.1002/ps.2189

Bates, S.L., Zhao, J.-Z., Roush, R.T., Shelton, A.M., 2005. Insect resistance management in GM crops: past, present and future. Nat. Biotechnol. 23, 57–62. doi:10.1038/nbt1056

Beckel, W.E., 1958. Investigations of permeability, diapause, and hatching in the eggs of the mosquito *Aedes hexodontus* DYAR. Can. J. Zool. 36, 541–554. doi:10.1139/z58-050

Becker, N., Geier, M., Balczun, C., Bradersen, U., Huber, K., Kiel, E., Krüger, A., Lühken, R., Orendt, C., Plenge-Bönig, A., Rose, A., Schaub, G. a., Tannich, E., 2013. Repeated introduction of *Aedes albopictus* into Germany, July to October 2012. Parasitol. Res. 112, 1787–1790. doi:10.1007/s00436-012-3230-1

Becker, N., Jöst, A., Weitzel, T., 2012. The *Culex pipiens* complex in Europe. J. Am. Mosq. Control Assoc. 28, 53–67. doi:10.2987/8756-971X-28.4s.53

Becker, N., Petric, D., Zgomba, M., Boase, C., Madon, M., Dahl, C., Kaiser, A., 2010. Mosquitoes and their control, 2nd ed. Springer Science & Business Media, Heidelberg, Dordrecht, London, New York.

Bellini, R., Albieri, A., 2010. Dispersal and survival of *Aedes albopictus* (Diptera: Culicidae) males in Italian urban areas and significance for sterile insect technique application. J. Med. Entomol. 47, 1082–1091. doi:10.1603/ME09154

Bellini, R., Calvitti, M., Medici, A., Carrieri, M., 2007. Use of the sterile insect technique against *Aedes albopictus* in Italy: First results of a pilot trial, in: Vreysen, M.J.B., Robinson, A.S., Hendrichs, J. (Eds.), Area-Wide Control of Insect Pests. pp. 505–515.

Benedict, M.Q., Levine, R.S., Hawley, W.A., Lounibos, L.P., 2007. Spread of the tiger: Global risk of invasion by the mosquito *Aedes albopictus*. Vector Borne Zoonotic Dis. 7, 76–85. doi:10.1089/vbz.2006.0562

Benedict, M.Q., Robinson, A., 2003. The first releases of transgenic mosquitoes: An argument for the sterile insect technique. Trends Parasitol. 19, 349–355. doi:10.1016/S1471-4922(03)00144-2

Bergland, A.O., Behrman, E.L., O'Brien, K.R., Schmidt, P.S., Petrov, D.A., 2014. Genomic evidence of rapid and stable adaptive oscillations over seasonal time scales in *Drosophila*. PLoS Genet. 10, e1004775. doi:10.1371/journal.pgen.1004775

Bevins, S.N., 2007. Timing of resource input and larval competition between invasive and native container-inhabiting mosquitoes (Diptera: Culicidae). J. Vector Ecol. 32, 252. doi:10.3376/1081-1710(2007)32[252:TORIAL]2.0.CO;2

Bhatt, S., Gething, P.W., Brady, O.J., Messina, J.P., Farlow, A.W., Moyes, C.L., Drake, J.M., Brownstein, J.S., Hoen, A.G., Sankoh, O., Myers, M.F., George, D.B., Jaenisch, T., Wint, G.R.W., Simmons, C.P., Scott, T.W., Farrar, J.J., Hay, S.I., 2013. The global distribution and burden of dengue. Nature 496, 504–507. doi:10.1038/nature12060

Birkett, M.A., Hassanali, A., Hoglund, S., Pettersson, J., Pickett, J.A., 2011. Repellent activity of catmint, *Nepeta cataria*, and iridoid nepetalactone isomers against Afro-tropical mosquitoes, ixodid ticks and red poultry mites. Phytochemistry 72, 109–114. doi:10.1016/j.phytochem.2010.09.016

Birungi, J., Munstermann, L.E., 2002. Genetic structure of *Aedes albopictus* (Diptera: Culicidae) populations based on mitochondrial ND5 sequences: Evidence for an independent invasion into Brazil and United States. Ann. Entomol. Soc. Am. 95, 125–132. doi:10.1603/0013-8746(2002)095[0125:GSOAAD]2.0.CO;2

Black, W.C., 2004. Learning to use *Ochlerotatus* is just the beginning. J. Am. Mosq. Control Assoc. 20, 215–6.

Black, W.C., Ferrari, J.A., Rai, K.S., Sprenger, D., 1988. Breeding structure of a colonising species: *Aedes albopictus* (Skuse) in the United States. Heredity (Edinb). 60, 173–181. doi:10.1038/hdy.1988.29

Boffelli, D., Martin, D.I.K., 2012. Epigenetic inheritance: a contributor to species differentiation? DNA Cell Biol. 31 Suppl 1, S11–6. doi:10.1089/dna.2012.1643

Bonizzoni, M., Gasperi, G., Chen, X., James, A.A., 2013. The invasive mosquito species *Aedes albopictus*: current knowledge and future perspectives. Trends Parasitol. 29, 460–8. doi:10.1016/j.pt.2013.07.003

Brady, O.J., Golding, N., Pigott, D.M., Kraemer, M.U.G., Messina, J.P., Reiner, R.C., Scott,

T.W., Smith, D.L., Gething, P.W., Hay, S.I., 2014. Global temperature constraints on *Aedes aegypti* and *Ae. albopictus* persistence and competence for dengue virus transmission. Parasit. Vectors 7, 338. doi:10.1186/1756-3305-7-338

Brady, O.J., Johansson, M.A., Guerra, C.A., Bhatt, S., Golding, N., Pigott, D.M., Delatte, H., Grech, M.G., Leisnham, P.T., Maciel-de-Freitas, R., Styer, L.M., Smith, D.L., Scott, T.W., Gething, P.W., Hay, S.I., 2013. Modelling adult *Aedes aegypti* and *Aedes albopictus* survival at different temperatures in laboratory and field settings. Parasit. Vectors 6, 351. doi:10.1186/1756-3305-6-351

Branswell, H., 2016. Dengue could be the surprise culprit making Zika worse [WWW Document]. STATnew.com. URL https://www.statnews.com/2016/02/17/zika-dengue-infections/ (accessed 5.28.16).

Brogdon, W.G., McAllister, J.C., Corwin, A.M., Cordon-Rosales, C., 1999. Oxidase-based DDT–pyrethroid cross-resistance in Guatemalan *Anopheles albimanus*. Pestic. Biochem. Physiol. 64, 101–111. doi:10.1006/pest.1999.2415

Bullock, J.D., 2011. Introduction to: The mosquito hypothetically considered as an agent in the transmission of yellow fever poison, by Carlos Juan Finlay, 1881, Delta Omega Classics.

Cairns, J., Heath, A.G., Parker, B.C., 1975. The effects of temperature upon the toxicity of chemicals to aquatic organisms. Hydrobiologia 47, 135–171. doi:10.1007/BF00036747

Caminade, C., Medlock, J.M., Ducheyne, E., McIntyre, K.M., Leach, S., Baylis, M., Morse, A.P., 2012. Suitability of European climate for the Asian tiger mosquito *Aedes albopictus*: recent trends and future scenarios. J. R. Soc. Interface 9, 2708–17. doi:10.1098/rsif.2012.0138

Campbell, P., Arnold, D., Brock, T., 1998. Higher-tier aquatic risk assessment for pesticides, in: SETAC-Europe/OECD/EC Conference Paper, Lacanau Océan, France, 19-22 April 1998.

Cancrini, G., Pietrobelli, M., Frangipane di Regalbono, A.F., Tampieri, M.P., della Torre, A., 1995. Development of Dirofilaria and Setaria nematodes in Aedes albopictus. Parassitologia 37, 141–5.

Cantrell, C.L., Klun, J.A., 2011. Callicarpenal and intermedeol: Two natural arthropod feedingdeterrent and repellent compounds identified from the southern folk remedy plant, *Callicarpa americana*, in: Paluch, G.E., Coats, J.R. (Eds.), Recent Developments in Invertebrate Repellents, ACS Symposium Series. American Chemical Society, Washington, DC, pp. 47–58. doi:10.1021/bk-2011-1090

Cantrell, C.L., Klun, J.A., Bryson, C.T., Kobaisy, M., Duke, S.O., 2005. Isolation and identification of mosquito bite deterrent terpenoids from leaves of American (*Callicarpa americana*) and Japanese (*Callicarpa japonica*) beautyberry. J. Agric. Food Chem. 53, 5948–5953. doi:10.1021/jf0509308

Cao-Lormeau, V.-M., Blake, A., Mons, S., Lastère, S., Roche, C., Vanhomwegen, J., Dub, T., Baudouin, L., Teissier, A., Larre, P., Vial, A.-L., Decam, C., Choumet, V., Halstead, S.K., Willison, H.J., Musset, L., Manuguerra, J.-C., Despres, P., Fournier, E., Mallet, H.-P., Musso, D., Fontanet, A., Neil, J., Ghawché, F., 2016. Guillain-Barré Syndrome outbreak associated with Zika virus infection in French Polynesia: a case-control study. Lancet 387, 1531–1539. doi:10.1016/S0140-6736(16)00562-6

Carrieri, M., Bacchi, M., Bellini, R., Maini, S., 2003. On the competition occurring between *Aedes albopictus* and *Culex pipiens* (Diptera: Culicidae) in Italy. Environ. Entomol. 32, 1313–1321. doi:10.1603/0046-225X-32.6.1313

Carroll, J.F., Cantrell, C.L., Klun, J.A., Kramer, M., 2007. Repellency of two terpenoid compounds isolated from *Callicarpa americana* (Lamiaceae) against *Ixodes scapularis* and *Amblyomma americanum* ticks. Exp. Appl. Acarol. 41, 215–224. doi:10.1007/s10493-007-9057-2

Carroll, J.F., Tabanca, N., Kramer, M., Elejalde, N.M., Wedge, D.E., Bernier, U.R., Coy, M., Becnel, J.J., Demirci, B., Başer, K.H.C., Zhang, J., Zhang, S., 2011. Essential oils of *Cupressus funebris*, *Juniperus communis*, and *J. chinensis* (Cupressaceae) as repellents against ticks (Acari: Ixodidae) and mosquitoes (Diptera: Culicidae) and as toxicants against mosquitoes. J. Vector Ecol. 36, 258–68. doi:10.1111/j.1948-7134.2011.00166.x

Carson, R.L., Lear, L.J., Wilson, E.O., 2002. Silent Spring. Houghton Mifflin Harcourt, Boston, New York.

Chahwan, R., Wontakal, S.N., Roa, S., 2010. Crosstalk between genetic and epigenetic information through cytosine deamination. Trends Genet. 26, 443–8. doi:10.1016/j.tig.2010.07.005

Chernin, E., 1992. Sir Patrick Manson: Physician to the colonial office, 1897-1912. Med. Hist. 36, 320–31.

Chio, E.H., Yang, E.-C., Huang, H.-T., Hsu, E.-L., Chen, C.-R., Huang, C.-G., Huang, R.-N., 2013. Toxicity and repellence of Taiwanese indigenous djulis, *Chenopodium formosaneum*, against *Aedes albopictus* (Diptera: Culicidae) and *Forcipomyia taiwana* (Diptera: Ceratopogonidae). J. Pest Sci. (2004). 86, 705–712. doi:10.1007/s10340-013-0500-3

Christophers, S.R., 1960. *Aedes Aegpti* (L.) the yellow ferver mosquito: Its life history, bionomics and structure, Cambridge At The Universit Press. Cambridge University Press, Cambridge.

Clements, A.N., 1993. The biology of mosquitoes, Vol. I; Development, nutrition and reproduction, first. ed, Parasitology Today. Chapman & Hall, Michigan. doi:10.1016/0169-4758(93)90183-G

Costanzo, K.S., Kesavaraju, B., Juliano, S.A., 2005a. Condition-specific competition in container mosquitoes: The role of noncompeting life-history stages. Ecology 86, 3289–3295. doi:10.1890/05-0583

Costanzo, K.S., Mormann, K., Juliano, S.A., 2005b. Asymmetrical competition and patterns of abundance of *Aedes albopictus* and *Culex pipiens* (Diptera: Culicidae). J. Med. Entomol. 29, 997–1003. doi:10.1016/j.biotechadv.2011.08.021.Secreted

Costanzo, K.S., Muturi, E.J., Lampman, R.L., Alto, B.W., 2011. The effects of resource type and ratio on competition With *Aedes albopictus* and *Culex pipiens* (Diptera: Culicidae). J. Med. Entomol. 48, 29–38. doi:10.1603/ME10085

Culler, L.E., Ayres, M.P., Virginia, R.A., 2015. In a warmer Arctic, mosquitoes avoid increased mortality from predators by growing faster. Proc. Biol. Sci. 282, 20151549–. doi:10.1098/rspb.2015.1549

Davis, S., 2014. Don' t let the bugs bite! A review of insect repellents. Prof Nurs Today 18, 21–23.

Debboun, M., Frances, S.P., Strickman, D., 2014. Insect repellents handbook, 2nd ed. CRC

Press, Boca Raton.

Delatte, H., Gimonneau, G., Triboire, A., Fontenille, D., 2009. Influence of temperature on immature development, survival, longevity, fecundity, and gonotrophic cycles of *Aedes albopictus*, vector of chikungunya and dengue in the Indian Ocean. J. Med. Entomol. 46, 33–41. doi:10.1603/033.046.0105

Deutsche Bundesregierung, 2008. Deutsche Anpassungsstrategie an den Klimawandel. Beschluss des Bundeskabinettes vom 17. Dezember 2008.

Diagne, C.T., Diallo, D., Faye, O., Ba, Y., Faye, O., Gaye, A., Dia, I., Weaver, S.C., Sall, A.A., Diallo, M., 2015. Potential of selected Senegalese *Aedes* spp. mosquitoes (Diptera: Culicidae) to transmit Zika virus. BMC Infect. Dis. 15, 492. doi:10.1186/s12879-015-1231-2

dpa, 2015. Mexiko lässt Impfstoff gegen Dengue-Fieber zu - Gesundheit-News Süddeutsche.de [WWW Document]. URL http://www.sueddeutsche.de/news/gesundheit/gesundheit-mexiko-laesst-impfstoff-gegen-dengue-fieber-zu-dpa.urn-newsml-dpa-com-20090101-151210-99-223865 (accessed 1.10.16).

dpa, 2016. Gesundheit: Mückenfachmann: Tigermücke wegen Zika-Verdachts bekämpfen - FOCUS Online [WWW Document]. Website. URL http://www.focus.de/regional/baden-wuerttemberg/gesundheit-mueckenfachmann-tigermuecke-wegen-zika-verdachts-bekaempfen_id_5258988.html (accessed 3.4.16).

Duchet, C., Franquet, E., Lagadic, L., Lagneau, C., 2015. Effects of *Bacillus thuringiensis israelensis* and spinosad on adult emergence of the non-biting midges *Polypedilum nubifer* (Skuse) and *Tanytarsus curticornis* Kieffer (Diptera: Chironomidae) in coastal wetlands. Ecotoxicol. Environ. Saf. 115, 272–8. doi:10.1016/j.ecoenv.2015.02.029

Duman, J.G., Wu, D.W., Xu, L., Tursman, D., Olsen, T.M., 1991. Adaptations of insects to subzero temperatures. Q. Rev. Biol. doi:10.1086/417337

Dunlap, T., 2014. DDT: Scientists, citizens, and public policy. Princeton University Press.

DVV e.V., 2016. Das Zika-Virus – Facts and Fiction. Deutsche Vereinigung zur Bekämpfung der Viruskrankheiten e.V. [WWW Document]. URL http://www.dvv-ev.de/news/Zika-Virus.pdf (accessed 2.3.16).

Elgart, M., Snir, O., Soen, Y., 2015. Stress-mediated tuning of developmental robustness and plasticity in flies. Biochim. Biophys. Acta 1849, 462–6. doi:10.1016/j.bbagrm.2014.08.004

Epstein, P.R., Diaz, H.F., Elias, S., Grabherr, G., Graham, N.E., Martens, W.J.M., Mosley-Thompson, E., Susskind, J., 1998. Biological and physical signs of climate change: Focus on mosquito-borne diseases. Bull. Am. Meteorol. Soc. 79, 409–417. doi:10.1175/1520-0477(1998)079<0409:BAPSOC>2.0.CO;2

Estrada-Franco, J.G., Craig, G.B., 1995. Biology, disease relationship, and control of *Aedes albopictus*, Technical Paper 42. Pan American Health Organisation, Washington, D.C.

Fang, J., 2010. Ecology: A world without mosquitoes. Nature 466, 432–4. doi:10.1038/466432a

Farajollahi, A., Fonseca, D.M., Kramer, L.D., Marm Kilpatrick, A., 2011. "Bird biting" mosquitoes and human disease: a review of the role of *Culex pipiens* complex mosquitoes in epidemiology. Infect. Genet. Evol. 11, 1577–85.

doi:10.1016/j.meegid.2011.08.013

Farnesi, L.C., Martins, A.J., Valle, D., Rezende, G.L., 2009. Embryonic development of *Aedes aegypti* (Diptera: Culicidae): influence of different constant temperatures. Mem. Inst. Oswaldo Cruz 104, 124–126. doi:10.1590/S0074-02762009000100020

Farnesi, L.C., Menna-Barreto, R.F.S., Martins, A.J., Valle, D., Rezende, G.L., 2015. Physical features and chitin content of eggs from the mosquito vectors *Aedes aegypti*, *Anopheles aquasalis* and *Culex quinquefasciatus*: Connection with distinct levels of resistance to desiccation. J. Insect Physiol. doi:10.1016/j.jinsphys.2015.10.006

Feil, R., Fraga, M.F., 2011. Epigenetics and the environment: emerging patterns and implications. Nat. Rev. Genet. 13, 97–109. doi:10.1038/nrg3142

Fernández-Grandon, G.M., Gezan, S.A., Armour, J.A.L., Pickett, J.A., Logan, J.G., 2015. Heritability of attractiveness to mosquitoes. PLoS One 10, e0122716. doi:10.1371/journal.pone.0122716

Field, L.M., Devonshire, A.L., Ffrench-Constant, R.H., Forde, B.G., 1989. Changes in DNA methylation are associated with loss of insecticide resistance in the peach-potato aphid *Myzus persicae* (Sulz.). FEBS Lett. 243, 323–327. doi:10.1016/0014-5793(89)80154-1

Field, L.M., Lyko, F., Mandrioli, M., Prantera, G., 2004. DNA methylation in insects. Insect Mol. Biol. 13, 109–115. doi:10.1111/j.0962-1075.2004.00470.x

Fillinger, U., Knols, B.G.J., Becker, N., 2003. Efficacy and efficiency of new *Bacillus thuringiensis* var *israelensis* and *Bacillus sphaericus* formulations against *Afrotropical anophelines* in Western Kenya. Trop. Med. Int. Health 8, 37–47. doi:10.1046/j.1365-3156.2003.00979.x

Finlay, C.J., 1882. El mosquito hipoteticamente considerado como agente de trasmision de la fiebre amarilla. An. la Real Acad. Ciencias Médicas, Físicas y Nat. la Habana 18, 147–169.

Fischer, D., Thomas, S.M., Niemitz, F., Reineking, B., Beierkuhnlein, C., 2011. Projection of climatic suitability for *Aedes albopictus* Skuse (Culicidae) in Europe under climate change conditions. Glob. Planet. Change 78, 54–64. doi:10.1016/j.gloplacha.2011.05.008

Focks, D.A., Linda, S.B., Craig, G.B., Hawley, W.A., Pumpuni, C.B., 1994. *Aedes albopictus* (Diptera: Culicidae): a statistical model of the role of temperature, photoperiod, and geography in the induction of egg diapause. J. Med. Entomol. 31, 278–286. doi:10.1093/jmedent/31.2.278

Fredericks, A.C., Fernandez-Sesma, A., 2014. The burden of dengue and chikungunya worldwide: Implications for the southern United States and california. Ann. Glob. Heal. 80, 466–475. doi:10.1016/j.aogh.2015.02.006

Furneaux, P.J.S., McFarlane, J.E., 1965a. A possible relationship between the occurrence of catecholamines and water absorption in insect eggs. J. Insect Physiol. 11, 631–635. doi:10.1016/0022-1910(65)90145-9

Furneaux, P.J.S., McFarlane, J.E., 1965b. Identification, estimation, and localization of catecholamines in eggs of the house cricket, *Acheta domesticus* (L.). J. Insect Physiol. 11, 591–600. doi:10.1016/0022-1910(65)90141-1

Galizi, R., Doyle, L.A., Menichelli, M., Bernardini, F., Deredec, A., Burt, A., Stoddard, B.L., Windbichler, N., Crisanti, A., 2014. A synthetic sex ratio distortion system for the control of the human malaria mosquito. Nat. Commun. 5, 3977. doi:10.1038/ncomms4977

Genchi, C., Rinaldi, L., Mortarino, M., Genchi, M., Cringoli, G., 2009. Climate and *Dirofilaria* infection in Europe. Vet. Parasitol. 163, 286–292. doi:10.1016/j.vetpar.2009.03.026

Georghiou, G., Wirth, M., 1997. Influence of exposure to single versus multiple toxins of *Bacillus thuringiensis* subsp. *israelensis* on development of resistance in the mosquito *Culex quinquefasciatus* (Diptera: Culicidae). Appl. Envir. Microbiol. 63, 1095–1101.

Gilbert, N., 2010. Mosquito spray affects bird reproduction. Nat. News 4–7. doi:10.1038/news.2010.296

Gilbert, S.F., Epel, D., 2009. Ecological developmental biology: Integrating epigenetics, medicine, and evolution. Sinauer Associates Inc., Sunderland, USA.

Gjenero-Margan, I., Aleraj, B., Krajcar, D., Lesnikar, V., Klobučar, A., Pem-Novosel, I., Kurečić-Filipović, S., Komparak, S., Martić, R., Duričić, S., Betica-Radić, L., Okmadžić, J., Vilibić-Čavlek, T., Babić-Erceg, A., Turković, B., Avsić-Županc, T., Radić, I., Ljubić, M., Sarac, K., Benić, N., Mlinarić-Galinović, G., 2011. Autochthonous dengue fever in Croatia, August-September 2010. Euro Surveill. Bull. Eur. sur les Mal. Transm. = Eur. Commun. Dis. Bull. 16, 19805.

Glastad, K.M., Hunt, B.G., Yi, S. V, Goodisman, M.A.D., 2011. DNA methylation in insects: on the brink of the epigenomic era. Insect Mol. Biol. 20, 553–65. doi:10.1111/j.1365-2583.2011.01092.x

Goltsev, Y., Rezende, G.L., Vranizan, K., Lanzaro, G., Valle, D., Levine, M., 2009. Developmental and evolutionary basis for drought tolerance of the *Anopheles gambiae* embryo. Dev. Biol. 330, 462–70. doi:10.1016/j.ydbio.2009.02.038

Gowher, H., Leismann, O., Jeltsch, A., 2000. DNA of *Drosophila melanogaster* contains 5-methylcytosine. EMBO J. 19, 6918–23. doi:10.1093/emboj/19.24.6918

Grard, G., Caron, M., Mombo, I.M., Nkoghe, D., Mboui Ondo, S., Jiolle, D., Fontenille, D., Paupy, C., Leroy, E.M., 2014. Zika virus in Gabon (Central Africa) – 2007: A new threat from *Aedes albopictus*? PLoS Negl. Trop. Dis. 8, e2681. doi:10.1371/journal.pntd.0002681

Grube, A., Donaldson, D., Kiely, T., Wu, L., 2011. Pesticides industry sales and usage. US EPA, Washington, DC.

Hanson, S.M., 1991. Thesis: Cold hardiness of *Aedes albopictus* eggs. University of Notre Dame du Lac, Notre Dame.

Hanson, S.M., Craig, G.B., 1994. Cold acclimation, diapause, and geographic origin affect cold hardiness in eggs of *Aedes albopictus* (Diptera: Culicidae). J. Med. Entomol. 31, 192–201.

Hanson, S.M., Craig, G.B., 1995a. *Aedes albopictus* (Diptera: Culicidae) eggs: field survivorship during northern Indiana winters. J. Med. Entomol. 32, 599–604. doi:10.1093/jmedent/32.5.599

Hanson, S.M., Craig, G.B., 1995b. Relationship between cold hardiness and supercooling point in *Aedes albopictus* eggs. J. Am. Mosq. Control Assoc. 11, 35–8.

Hanson, S.M., Mutebi, J.P., Craig, G.B., Novak, R.J., 1993. Reducing the overwintering

ability of *Aedes albopictus* by male release. J. Am. Mosq. Control Assoc. 9, 78–83.

Harbach, R.E., 2012. *Culex pipiens*: species versus species complex taxonomic history and perspective. J. Am. Mosq. Control Assoc. 28, 10–23. doi:10.2987/8756-971X-28.4.10

Harbach, R.E., 2016. Family Culicidae Meigen, 1818 [WWW Document]. Mosq. Taxon. Invent. URL http://mosquito-taxonomic-inventory.info/family-culicidae-meigen-1818 (accessed 5.23.16).

Harris, A.F., Nimmo, D., McKemey, A.R., Kelly, N., Scaife, S., Donnelly, C.A., Beech, C., Petrie, W.D., Alphey, L., 2011. Field performance of engineered male mosquitoes. Nat. Biotechnol. 29, 1034–7. doi:10.1038/nbt.2019

Harwood, A.D., You, J., Lydy, M.J., 2009. Temperature as a toxicity identification evaluation tool for pyrethroid insecticides: Toxicokinetic confirmation. Environ. Toxicol. Chem. 28, 1051–8. doi:10.1897/08-291.1

Harwood, R.F., Horsfall, W.R., 1959. Development, structure, and function of coverings of eggs of floodwater mosquitoes. III. functions of coverings. Ann. Entomol. Soc. Am. 52, 113–116. doi:10.1093/aesa/52.2.113

Hawley, W.A., 1988. The biology of *Aedes albopictus*. J. Am. Mosq. Control Assoc. Suppl. 1, 1–39.

Hawley, W.A., Pumpuni, C.B., Brady, R.H., Craig, G.B., 1989. Overwintering survival of *Aedes albopictus* (Diptera: Culicidae) eggs in Indiana. J. Med. Entomol. 26, 8.

Hawley, W.A., Reiter, P., Copeland, R.S., Pumpuni, C.B., Craig, G.B., 1987. *Aedes albopictus* in North America: probable introduction in used tires from northern Asia. Science 236, 1114–1116. doi:10.1126/science.3576225

Hemingway, J., Ranson, H., 2000. Insecticide resistance in insect vectors of human disease. Annu. Rev. Entomol. 45, 371–391. doi:10.1146/annurev.ento.45.1.371

Hennig, W., Davis, D.D., Zangerl, R., 1999. Phylogenetic systematics. University of Illinois Press, Urbana, Chicago.

Heugens, E.H.W., Jan Hendriks, A., Dekker, T., van Straalen, N.M., Admiraal, W., 2001. A review of the effects of multiple stressors on aquatic organisms and analysis of uncertainty factors for use in risk assessment. Crit. Rev. Toxicol. 31, 247–284. doi:10.1080/20014091111695

Hick, C.A., Field, L.M., Devonshire, A.L., 1996. Changes in the methylation of amplified esterase DNA during loss and reselection of insecticide resistance in peach-potato aphids, *Myzus persicae*. Insect Biochem. Mol. Biol. 26, 41–47. doi:10.1016/0965-1748(95)00059-3

Higa, Y., 2011. Dengue vectors and their spatial distribution. Trop. Med. Health 39, 17–27. doi:10.2149/tmh.2011-S04

Higgs, S., 2005. Journal policy on names of aedine mosquito genera and subgenera. Vector Borne Zoonotic Dis. 5, 93–4. doi:10.1089/vbz.2005.5.93

Higgs, S., 2014. Chikungunya virus: a major emerging threat. Vector Borne Zoonotic Dis. 14, 535–6. doi:10.1089/vbz.2014.14.8.edit

Hilbricht, T., Varotto, S., Sgaramella, V., Bartels, D., Salamini, F., Furini, A., 2008. Retrotransposons and siRNA have a role in the evolution of desiccation tolerance leading to resurrection of the plant *Craterostigma plantagineum*. New Phytol. 179, 877–887.

doi:10.1111/j.1469-8137.2008.02480.x

Hinton, H., 1981. Biology of insect eggs. Volume I, Volume II, Volume III. Elsevier Science & Technology Books, New York.

Hinton, H.E., Service, M.W., 1969. The surface structure of aedine eggs as seen with the scanning electron microscope. Ann. Trop. Med. Parasitol. 63, 409–11.

Hofhuis, A., Reimerink, J., Reusken, C., Scholte, E.-J., de Boer, A., Takken, W., Koopmans, M., 2009. The hidden passenger of lucky bamboo: do imported *Aedes albopictus* mosquitoes cause dengue virus transmission in the Netherlands? Vector Borne Zoonotic Dis. 9, 217–20. doi:10.1089/vbz.2008.0071

Hornstein, E., Shomron, N., 2006. Canalization of development by microRNAs. Nat. Genet. 38 Suppl, S20–S24. doi:10.1038/ng1803

Huang, X., Poelchau, M.F., Armbruster, P., 2015. Global transcriptional dynamics of diapause induction in non-blood-fed and blood-fed *Aedes albopictus*. PLoS Negl. Trop. Dis. 9, e0003724. doi:10.1371/journal.pntd.0003724

Hung, M.S., Karthikeyan, N., Huang, B., Koo, H.C., Kiger, J., Shen, C.J., 1999. *Drosophila* proteins related to vertebrate DNA (5-cytosine) methyltransferases. Proc. Natl. Acad. Sci. U. S. A. doi:10.1073/pnas.96.21.11940

ICZN, 1999. International code of zoological nomenclature, 4. ed. International Trust for Zoological Nomenclature, c/o The Natural History Museum, London.

Imai, C., Maeda, O., 1976. Several factors effecting on hatching of *Aedes albopictus* eggs. Med. Entomol. Zool. 27, 367–372.

Ito, H., Gaubert, H., Bucher, E., Mirouze, M., Vaillant, I., Paszkowski, J., 2011. An siRNA pathway prevents transgenerational retrotransposition in plants subjected to stress. Nature 472, 115–119. doi:10.1038/nature09861

Jia, P., Lu, L., Chen, X., Chen, J., Guo, L., Yu, X., Liu, Q., 2016. A climate-driven mechanistic population model of *Aedes albopictus* with diapause. Parasit. Vectors 9, 175. doi:10.1186/s13071-016-1448-y

Jöst, H., Bialonski, A., Storch, V., Günther, S., Becker, N., Schmidt-Chanasit, J., 2010. Isolation and phylogenetic analysis of sindbis viruses from mosquitoes in Germany. J. Clin. Microbiol. 48, 1900–1903. doi:10.1128/JCM.00037-10

Juliano, S.A., 1998. Species introduction and replacement among mosquitoes: Interspecific resource competition or apparent competition? Ecology 79, 255–268. doi:10.1890/0012-9658(1998)079[0255:SIARAM]2.0.CO;2

KABS e.V., 2015a. Update 1 über die Population der Asiatischen Tigermücke (Aedes albopictus) in Freiburg vom 9. August 2015 [WWW Document]. Website. URL http://www.kabsev.de/1/1_6/1_6_3/update09082015.php (accessed 2.9.15).

KABS e.V., 2015b. Update 3 über die Population der Asiatischen Tigermücke (Aedes albopictus) in Freiburg vom 16. November 2015 [WWW Document]. Website. URL http://www.kabsev.de/1/1_6/1_6_3/update16112015.php (accessed 3.4.16).

KABS e.V., 2015c. Update 2 über die Population der Asiatischen Tigermücke (Aedes albopictus) in Freiburg vom 10. September 2015 [WWW Document]. Website. URL http://www.kabsev.de/1/1_6/1_6_3/update10092015.php (accessed 3.4.16).

KABS e.V., 2016. Aktuelles von der KABS e.V. [WWW Document]. Website. URL

http://www.kabsev.de/7/7_1/7_1_0/6.php (accessed 3.4.16).

Kettle, D.S., 1984. Medical and veterinary entomology, 2nd ed. CABI, Wallington.

Knols, B.G., 1996. On human odour, malaria mosquitoes, and limburger cheese. Lancet 348, 1322. doi:10.1016/S0140-6736(05)65812-6

Kobayashi, M., Nihei, N., Kurihara, T., 2002. Analysis of northern distribution of *Aedes albopictus* (Diptera: Culicidae) in Japan by geographical information system. J. Med. Entomol. 39, 4–11. doi:10.1603/0022-2585-39.1.4

Kottek, M., Grieser, J., Beck, C., Rudolf, B., Rubel, F., 2006. World Map of the Köppen-Geiger climate classification updated. Meteorol. Zeitschrift 15, 259–263. doi:10.1127/0941-2948/2006/0130

Kraemer, M.U.G., Sinka, M.E., Duda, K.A., Mylne, A., Shearer, F.M., Barker, C.M., Moore, C.G., Carvalho, R.G., Coelho, G.E., Van Bortel, W., Hendrickx, G., Schaffner, F., Elyazar, I.R., Teng, H.-J., Brady, O.J., Messina, J.P., Pigott, D.M., Scott, T.W., Smith, D.L., Wint, G.W., Golding, N., Hay, S.I., 2015. The global distribution of the arbovirus vectors *Aedes aegypti* and *Ae. albopictus*. Elife 4, e08347. doi:10.7554/eLife.08347

Kreß, A., 2011. Interaktive Effekte von Temperatur und Insektiziden auf die invasive Tigermücke *Aedes albopictus* und die einheimische Stechmücke *Culex pipiens*. Johann Wolfgang Goethe-Universität Frankfurt am Main.

Kreß, A., Kuch, U., Oehlmann, J., Müller, R., 2014. Impact of temperature and nutrition on the toxicity of the insecticide λ-cyhalothrin in full-lifecycle tests with the target mosquito species *Aedes albopictus* and *Culex pipiens*. J. Pest Sci. (2004). 87, 739–750. doi:10.1007/s10340-014-0620-4

Kreß, A., Kuch, U., Oehlmann, J., Müller, R., 2016a. Effects of diapause and cold acclimation on egg ultrastructure reveal new insights to the cold hardiness mechanisms of the Asian tiger mosquito *Aedes* (*Stegomyia*) *albopictus*. J. Vector Ecol. 41, 1–9.

Kreß, A., Oppold, A.-M., Kuch, U., Oehlmann, J., Müller, R., 2016b. Cold tolerance of the Asian tiger mosquito *Aedes albopictus* and its response to epigenetic alterations.

Kulkarni, R.R., Pawar, P. V., Joseph, M.P., Akulwad, A.K., Sen, A., Joshi, S.P., 2013. *Lavandula gibsoni* and *Plectranthus mollis* essential oils: chemical analysis and insect control activities against *Aedes aegypti*, *Anopheles stephensi* and *Culex quinquefasciatus*. J. Pest Sci. (2004). 86, 713–718. doi:10.1007/s10340-013-0502-1

Kumar, R., Hwang, J.S., 2006. Larvicidal efficiency of aquatic predators: A perspective for mosquito biocontrol. Zool. Stud. 45, 447–466. doi:10.1007/s12630-011-9635-y

Kunert, N., Marhold, J., Stanke, J., Stach, D., Lyko, F., 2003. A Dnmt2-like protein mediates DNA methylation in *Drosophila*. Development 130, 5083–90. doi:10.1242/dev.00716

La Ruche, G., Souarès, Y., Armengaud, A., Peloux-Petiot, F., Delaunay, P., Desprès, P., Lenglet, A., Jourdain, F., Leparc-Goffart, I., Charlet, F., Ollier, L., Mantey, K., Mollet, T., Fournier, J.P., Torrents, R., Leitmeyer, K., Hilairet, P., Zeller, H., Van Bortel, W., Dejour-Salamanca, D., Grandadam, M., Gastellu-Etchegorry, M., 2010. First two autochthonous dengue virus infections in metropolitan France, September 2010. Eurosurveillance 15, 19676.

Lacroix, R., McKemey, A.R., Raduan, N., Kwee Wee, L., Hong Ming, W., Guat Ney, T., Rahidah A.A., S., Salman, S., Subramaniam, S., Nordin, O., Hanum A.T., N., Angamuthu, C., Marlina Mansor, S., Lees, R.S., Naish, N., Scaife, S., Gray, P., Labbé,

G., Beech, C., Nimmo, D., Alphey, L., Vasan, S.S., Han Lim, L., Wasi A., N., Murad, S., 2012. Open field release of genetically engineered sterile male *Aedes aegypti* in Malaysia. PLoS One 7, e42771. doi:10.1371/journal.pone.0042771

Latapy, M., 2008. Main-memory triangle computations for very large (sparse (power-law)) graphs. Theor. Comput. Sci. 407, 458–473. doi:10.1016/j.tcs.2008.07.017

Lawler, S.P., Dritz, D.A., Christiansen, J.A., Cornel, A.J., 2007. Effects of lambda-cyhalothrin on mosquito larvae and predatory aquatic insects. Pest Manag. Sci. 63, 234–40. doi:10.1002/ps.1279

Lee II, H., Chapman, J.W., 2001. Nonindigenous species - An emerging issue for the EPA: A landscape in transition, effects of invasive species on ecosystems, human health, and EPA goals. Landscape 2, 54.

Leisnham, P.T., Lounibos, L.P., O'Meara, G.F., Juliano, S.A., 2009. Interpopulation divergence in competitive interactions of the mosquito *Aedes albopictus*. Ecology 90, 2405–2413. doi:10.1890/08-1569.1

Leistra, M., Zweers, A.J., Warinton, J.S., Crum, S.J.H., Hand, L.H., Beltman, W.H.J., Maund, S.J., 2004. Fate of the insecticide lambda-cyhalothrin in ditch enclosures differing in vegetation density. Pest Manag. Sci. 60, 75–84. doi:10.1002/ps.780

Ling, T., Xu, J., Smith, R., Ali, A., Cantrell, C.L., Theodorakis, E.A., 2011. Synthesis of (-)-callicarpenal, a potent arthropod-repellent. Tetrahedron 67, 3023–3029. doi:10.1016/j.tet.2011.02.078

Lippman, Z., Martienssen, R., 2004. The role of RNA interference in heterochromatic silencing. Nature 431, 364–70. doi:10.1038/nature02875

Livdahl, T., Willey, M., 1991. Prospects tor an Invasion: Competition Between *Aedes albopictus* and Native *Aedes triseriatus*. Science (80-. ). 253, 189–191. doi:10.1126/science.1853204

Lounibos, L.P., 2002. Invasions by insect vectors of human disease. Annu. Rev. Entomol. 47, 233–266. doi:10.1146/annurev.ento.47.091201.145206

Lounibos, L.P., Escher, R.L., Lourenço-De-Oliveira, R., 2003. Asymmetric Evolution of Photoperiodic Diapause in Temperate and Tropical Invasive Populations of *Aedes albopictus* (Diptera: Culicidae). Ann. Entomol. Soc. Am. 96, 512–518. doi:10.1603/0013-8746(2003)096[0512:AEOPDI]2.0.CO;2

Lounibos, L.P., Escher, R.L., Nishimura, N., 2011. Retention and adaptiveness of photoperiodic egg diapause in florida populations of invasive *Aedes albopictus*. J. Am. Mosq. Control Assoc. 27, 433–436. doi:10.2987/11-6164.1

Lounibos, L.P., Suárez, S., Menéndez, Z., Nishimura, N., Escher, R.L., O'Connell, S.M., Rey, J.R., 2002. Does temperature affect the outcome of larval competition between Aedes aegypti and Aedes albopictus? J. Vector Ecol. 27, 86–95.

Lyko, F., Foret, S., Kucharski, R., Wolf, S., Falckenhayn, C., Maleszka, R., 2010. The honey bee epigenomes: Differential methylation of brain DNA in queens and workers. PLoS Biol. 8. doi:10.1371/journal.pbio.1000506

Lyko, F., Ramsahoye, B.H., Jaenisch, R., 2000. DNA methylation in *Drosophila melanogaster*. Nature 408, 538–840. doi:10.1002/(SICI)1097-0177(199909)216:1<1::AID-DVDY1>3.0.CO;2-T

Maleszka, R., 2008. Epigenetic integration of environmental and genomic signals in honey bees. Epigenetics 3, 188–192. doi:10.4161/epi.3.4.6697

Mandrioli, M., Volpi, N., 2003. The genome of the lepidopteran *Mamestra brassicae* has a vertebrate-like content of methyl-cytosine. Genetica 119, 187–91.

Marchand, E., Prat, C., Jeannin, C., Lafont, E., Bergmann, T., Flusin, O., Rizzi, J., Roux, N., Busso, V., Deniau, J., Noel, H., Vaillant, V., Leparc-Goffart, I., Six, C., Paty, M.C., 2013. Autochthonous case of dengue in France, October 2013. Eurosurveillance 18, 1–6. doi:10.2807/1560-7917.ES2013.18.50.20661

Marhold, J., Kramer, K., Kremmer, E., Lyko, F., 2004a. The *Drosophila* MBD2/3 protein mediates interactions between the MI-2 chromatin complex and CpT/A-methylated DNA. Development 131, 6033–6039. doi:10.1242/dev.01531

Marhold, J., Rothe, N., Pauli, A., Mund, C., Kuehle, K., Brueckner, B., Lyko, F., 2004b. Conservation of DNA methylation in dipteran insects. Insect Mol. Biol. 13, 117–23. doi:10.1111/j.0962-1075.2004.00466.x

Martienssen, R., 2008. Great leap forward? Transposable elements, small interfering RNA and adaptive Lamarckian evolution. New Phytol. 179, 570–2. doi:10.1111/j.1469-8137.2008.02567.x

Massonnet-Bruneel, B., Corre-Catelin, N., Lacroix, R., Lees, R.S., Hoang, K.P., Nimmo, D., Alphey, L., Reiter, P., 2013. Fitness of transgenic mosquito *Aedes aegypti* males carrying a dominant lethal genetic system. PLoS One 8, e62711. doi:10.1371/journal.pone.0062711

Maund, S.J., Hamer, M.J., Warinton, J.S., Kedwards, T.J., 1998. Aquatic ecotoxicology of the pyrethroid insecticide lambda-cyhalothrin: Considerations for higher-tier aquatic risk assessment. Pestic. Sci. 54, 408–417. doi:10.1002/(SICI)1096-9063(199812)54:4<408::AID-PS843>3.0.CO;2-T

Mayer, F., Ellersieck, M., 1986. Manual of acute toxicity: Interpretation and data base for 410 chemicals and 66 species of freshwater animals. United States Dep. Inter. U.S. Fish Wildl. Serv.

McBride, C.S., 2016. Genes and odors underlying the recent evolution of mosquito preference for humans. Curr. Biol. 26, R41–6. doi:10.1016/j.cub.2015.11.032

McBride, C.S., Baier, F., Omondi, A.B., Spitzer, S.A., Lutomiah, J., Sang, R., Ignell, R., Vosshall, L.B., 2014. Evolution of mosquito preference for humans linked to an odorant receptor. Nature 515, 222–7. doi:10.1038/nature13964

Meagher, R.B., 2010. The evolution of epitype. Plant Cell 22, 1658–66. doi:10.1105/tpc.110.075481

Medley, K.A., 2010. Niche shifts during the global invasion of the Asian tiger mosquito, *Aedes albopictus* Skuse (Culicidae), revealed by reciprocal distribution models. Glob. Ecol. Biogeogr. 19, 122–133. doi:10.1111/j.1466-8238.2009.00497.x

Melaun, C., Zotzmann, S., Santaella, V.G., Werblow, A., Zumkowski-Xylander, H., Kraiczy, P., Klimpel, S., 2016. Occurrence of *Borrelia burgdorferi* s.l. in different genera of mosquitoes (Culicidae) in Central Europe. Ticks Tick. Borne. Dis. 7, 256–63. doi:10.1016/j.ttbdis.2015.10.018

Meyabeme Elono, A.L., 2011. Mosquito control: Improving aquatic risk assessment and efficiency of mosquito control practices.

Meyabeme Elono, A.L., Liess, M., Duquesne, S., 2010. Influence of competing and predatory invertebrate taxa on larval populations of mosquitoes in temporary ponds of wetland areas in Germany. J. Vector Ecol. 35, 419–27. doi:10.1111/j.1948-7134.2010.00101.x

Mogi, M., 2011. Variation in cold hardiness of nondiapausing eggs of nine *Aedes* (*Stegomyia*) species (diptera: culicidae) from eastern Asia and Pacific islands ranging from the tropics to the cool-temperate zone. J. Med. Entomol. 48, 212–222. doi:10.1603/ME10196

Mogi, M., Armbruster, P., Fonseca, D.M., 2012. Analyses of the northern distributional limit of *Aedes albopictus* (Diptera: Culicidae) with a simple thermal index. J. Med. Entomol. 49, 1233–43. doi:10.1603/ME12104

Monnerat, A.T., Soares, M.J., Lima, J.B.P., Rosa-Freitas, M.G., Valle, D., 1999. *Anopheles albitarsis* eggs: ultrastructural analysis of chorion layers after permeabilization. J. Insect Physiol. 45, 915–922.

Moore, C., Mitchell, C., 1997. *Aedes albopictus* in the United States: ten-year presence and public health implications. Emerg. Infect. Dis. 3, 329–334.

Mori, A., Oda, T., Wada, Y., 1981. Studies on the egg diapause and overwintering of *Aedes albopictus* in Nagasaki. Trop. Med. 23, 79–90.

Mörner, J., Bos, R., Fredrix, M., 2002. Reducing and eliminating the use of persistent organic perticides. UNEP Chem. is a unit UNEP's Technol. Ind. Econ. Div.

Mörschel, F.M., Klein, D.R., 1997. Effects of weather and parasitic insects on behavior and group dynamics of caribou of the Delta Herd, Alaska. Can. J. Zool. 75, 1659–1670. doi:10.1139/z97-793

Mullen, G.R., Durden, L.A. (Eds.), 2009. Medical and veterinary entomology, 2. ed. Academic Press, London, San Diego.

Müller, R., Knautz, T., Völker, J., Kreß, A., Oehlmann, J., Kuch, U., Oehlmann, J., 2013. Appropriate larval food quality and quantity for *Aedes albopictus* (diptera : culicidae). J. Med. Entomol. 50, 668–673. doi:10.1603/ME12094

Müller, R., Seeland, A., Jagodzinski, L.S., Diogo, J.B., Nowak, C., Oehlmann, J., 2012. Simulated climate change conditions unveil the toxic potential of the fungicide pyrimethanil on the midge *Chironomus riparius*: a multigeneration experiment. Ecol. Evol. 2, 196–210. doi:10.1002/ece3.71

Murphy, P.C., 2007. What a book can do: The publication and reception of silent spring. Univ of Massachusetts Press, Massachusetts.

Murrell, E.G., Juliano, S.A., 2008. Detritus type alters the outcome of interspecific competition between *Aedes aegypti* and *Aedes albopictus* (diptera: culicidae). J. Med. Entomol. 45, 375–83.

Musso, D., Nhan, T.-X., 2015. Emergence of Zika virus. Clin. Microbiol. Open Access 04. doi:10.4172/2327-5073.1000222

Muturi, E.J., Alto, B.W., 2011. Larval environmental temperature and insecticide exposure alter *Aedes aegypti* competence for rboviruses. Vector Borne Zoonotic Dis. 11, 1157–1163. doi:10.1089/vbz.2010.0209

Muturi, E.J., Lampman, R., Costanzo, K., Alto, B.W., 2011. Effect of temperature and insecticide stress on life-history traits of *Culex restuans* and *Aedes albopictus* (Diptera: Culicidae). J. Med. Entomol. 48, 243–250. doi:10.1603/ME10017

N'Guessan, R., Darriet, F., Guillet, P., Carnevale, P., Traore-Lamizana, M., Corbel, V., Koffi, A.A., Chandre, F., 2003. Resistance to carbosulfan in *Anopheles gambiae* from Ivory Coast, based on reduced sensitivity of acetylcholinesterase. Med. Vet. Entomol. 17, 19–25. doi:10.1046/j.1365-2915.2003.00406.x

Nawrocki, S.J., Hawley, W.A., 1987. Estimation of the northern limits of distribution of *Aedes albopictus* in North America. J. Am. Mosq. Control Assoc. 3, 314–317.

Nayak, a. R., Nair, J.S., Hegde, M.V., Ranjekar, P.K., Pant, U., 1991. Genome analysis of two mosquito species. Insect Biochem. 21, 803–808. doi:10.1016/0020-1790(91)90122-U

Neal, L.E., Gon III, S.M. 'Ohukaniʻōhiʻa, 2007. 22. family culicidae, Catalog of the Diptera of the Australasian and Oceanian Regions. Bishop Museum.

Normile, D., 2013. Tropical medicine. Surprising new dengue virus throws a spanner in disease control efforts. Science 342, 415. doi:10.1126/science.342.6157.415

Oliveira Melo, A.S., Malinger, G., Ximenes, R., Szejnfeld, P.O., Alves Sampaio, S., Bispo De Filippis, A.M., 2016. Zika virus intrauterine infection causes fetal brain abnormality and microcephaly: Tip of the iceberg? Ultrasound Obstet. Gynecol. doi:10.1002/uog.15831

Ono, M., Swanson, J.J., Field, L.M., Devonshire, A.L., Siegfried, B.D., 1999. Amplification and methylation of an esterase gene associated with insecticide-resistance in greenbugs, *Schizaphis graminum* (Rondani) (Homoptera: Aphididae). Insect Biochem. Mol. Biol. 29, 1065–73.

Oppold, A., Kreß, A., Vanden Bussche, J., Diogo, J.B., Kuch, U., Oehlmann, J., Vandegehuchte, M.B., Müller, R., 2015. Epigenetic alterations and decreasing insecticide sensitivity of the Asian tiger mosquito *Aedes albopictus*. Ecotoxicol. Environ. Saf. 122, 45–53. doi:10.1016/j.ecoenv.2015.06.036

Owusu, H.F., Jančáryová, D., Malone, D., Müller, P., 2015. Comparability between insecticide resistance bioassays for mosquito vectors: time to review current methodology? Parasit. Vectors 8, 357. doi:10.1186/s13071-015-0971-6

Paluch, G., Grodnitzky, J., Bartholomay, L., Coats, J., 2009. Quantitative structure-activity relationship of botanical sesquiterpenes: Spatial and contact repellency to the yellow fever mosquito, *Aedes aegypti*. J. Agric. Food Chem. 57, 7618–7625. doi:10.1021/jf900964e

Paupy, C., Delatte, H., Bagny, L., Corbel, V., Fontenille, D., 2009. *Aedes albopictus*, an arbovirus vector: From the darkness to the light. Microbes Infect. 11, 1177–1185. doi:10.1016/j.micinf.2009.05.005

Pavlidi, N., Monastirioti, M., Daborn, P., Livadaras, I., Van Leeuwen, T., Vontas, J., 2012. Transgenic expression of the *Aedes aegypti* CYP9J28 confers pyrethroid resistance in *Drosophila melanogaster*. Pestic. Biochem. Physiol. 104, 132–135. doi:10.1016/j.pestbp.2012.07.003

PCST, 2016. Report from physicians in the crop-sprayed villages regarding Dengue-Zika, microcephaly, and mass-spraying with chemical poisons. Physicians in the Crop-Sprayed Towns. [WWW Document]. URL http://www.reduas.com.ar/wp-content/uploads/downloads/2016/02/Informe-Zika-de-Reduas_TRAD.pdf (accessed 3.24.16).

Pegoraro, M., Bafna, A., Davies, N.J., Shuker, D.M., Tauber, E., 2016. DNA methylation changes induced by long and short photoperiods in *Nasonia*. Genome Res. 26, 203–10.

doi:10.1101/gr.196204.115

Pluskota, B., Jöst, A., Augsten, X., Stelzner, L., Ferstl, I., Becker, N., 2016. Successful overwintering of *Aedes albopictus* in Germany. Parasitol. Res. doi:10.1007/s00436-016-5078-2

Pluskota, B., Storch, V., Braunbeck, T., Beck, M., Becker, N., 2008. First record of *Stegomyia albopicta* (Skuse) (Diptera : Culicidae) in Germany. J. Eur. Mosq. Control Assoc. 26, 1–5.

Poelchau, M.F., Reynolds, J.A., Elsik, C.G., Denlinger, D.L., Armbruster, P., 2013. Deep sequencing reveals complex mechanisms of diapause preparation in the invasive mosquito, *Aedes albopictus*. Proc. Biol. Sci. 280, 20130143. doi:10.1098/rspb.2013.0143

Poinar, G., Zavortinik, T., 2000. *Paleoculicis minututs* (Diptera: Culicidae) n. Gen., n. Sp., from Cretaceous Canadian amber, with a summary of described fossil mosquitoes. Acta Geol. ….

Porter, J.H., Parry, M.L., Carter, T.R., 1991. The potential effects of climatic change on agricultural insect pests. Agric. For. Meteorol. 57, 221–240. doi:10.1016/0168-1923(91)90088-8

Posadas, D.M., Carthew, R.W., 2014. MicroRNAs and their roles in developmental canalization. Curr. Opin. Genet. Dev. 27, 1–6. doi:10.1016/j.gde.2014.03.005

Pothikasikorn, J., Bangs, M.J., Boonplueang, R., Chareonviriyaphap, T., 2008. Susceptibility of various mosquitoes of Thailand to nocturnal subperiodic *Wuchereria bancrofti*. J. Vector Ecol. 33, 313–20.

Poulin, B., Lefebvre, G., Paz, L., 2010. Red flag for green spray: adverse trophic effects of Bti on breeding birds. J. Appl. Ecol. 47, 884–889. doi:10.1111/j.1365-2664.2010.01821.x

Pumpuni, C.B., Knepler, J., Craig, G.B., 1992. Influence of temperature and larval nutrition on the diapause inducing photoperiod of *Aedes albopictus*. J. Am. Mosq. Control Assoc. 8, 223–7.

Puthiyakunnon, S., Yao, Y., Li, Y., Gu, J., Peng, H., Chen, X., 2013. Functional characterization of three MicroRNAs of the Asian tiger mosquito, *Aedes albopictus*. Parasit. Vectors 6, 230. doi:10.1186/1756-3305-6-230

Qiao, C.-L., Raymond, M., 1995. The same esterase B1 haplotype is amplified in insecticide-resistant mosquitoes of the *Culex pipiens* complex from the Americas and China. Heredity (Edinb). 74, 339–345. doi:10.1038/hdy.1995.51

Ranson, H., Jensen, B., Vulule, J.M., Wang, X., Hemingway, J., Collins, F.H., 2000. Identification of a point mutation in the voltage-gated sodium channel gene of Kenyan *Anopheles gambiae* associated with resistance to DDT and pyrethroids. Insect Mol. Biol. 9, 491–497. doi:10.1046/j.1365-2583.2000.00209.x

Rasmussen, S.A., Jamieson, D.J., Honein, M.A., Petersen, L.R., 2016. Zika Virus and Birth Defects — Reviewing the Evidence for Causality. N. Engl. J. Med. NEJMsr1604338. doi:10.1056/NEJMsr1604338

Reamon-Buettner, S.M., Mutschler, V., Borlak, J., 2008. The next innovation cycle in toxicogenomics: environmental epigenetics. Mutat. Res. 659, 158–65. doi:10.1016/j.mrrev.2008.01.003

Reidenbach, K.R., Cook, S., Bertone, M.A., Harbach, R.E., Wiegmann, B.M., Besansky, N.J.,

2009. Phylogenetic analysis and temporal diversification of mosquitoes (Diptera: Culicidae) based on nuclear genes and morphology. BMC Evol. Biol. 9, 298. doi:10.1186/1471-2148-9-298

Reinert, J.F., 2000. New classification for the composite genus *Aedes* (Diptera: Culicidae: Aedini), elevation of subgenus *Ochlerotatus* to generic rank, reclassification of the other subgenera, and notes on certain subgenera and species. J. Am. Mosq. Control Assoc. 16, 175–88.

Reinert, J.F., Harbach, R.E., Kitching, I.J., 2009. Phylogeny and classification of tribe Aedini (Diptera: Culicidae). Zool. J. Linn. Soc. 157, 700–794. doi:10.1111/j.1096-3642.2009.00570.x

Reinert, J.F., Harbach, R.E., Kitsching, I.J., 2004. Phylogeny and classification of Aedini (Diptera: Culicidae), based on morphological characters of all life stages. Zool. J. Linn. Soc. 142, 289–368. doi:10.1111/j.1096-3642.2004.00144.x

Reisen, W.K., 2005. Journal policy on names of aedine mosquito genera and subgenera. J. Med. Entomol. 42, 511–511. doi:10.1093/jmedent/42.4.511

Reiter, P., 1998. *Aedes albopictus* and the world trade in used tires, 1988-1995: the shape of things to come? J. Am. Mosq. Control Assoc. 14, 83–94.

Reiter, P., Sprenger, D., 1987. The used tire trade: a mechanism for the worldwide dispersal of container breeding mosquitoes. J. Am. Mosq. Control Assoc. 3, 494–501.

Reynolds, J.A., Poelchau, M.F., Rahman, Z., Armbruster, P., Denlinger, D.L., 2012. Transcript profiling reveals mechanisms for lipid conservation during diapause in the mosquito, *Aedes albopictus*. J. Insect Physiol. 58, 966–973. doi:10.1016/j.jinsphys.2012.04.013

Rezende, G.L., Martins, A.J., Gentile, C., Farnesi, L.C., Pelajo-Machado, M., Peixoto, A.A., Valle, D., 2008. Embryonic desiccation resistance in *Aedes aegypti*: presumptive role of the chitinized serosal cuticle. BMC Dev. Biol. 8, 82. doi:10.1186/1471-213X-8-82

Rinker, D.C., Zhou, X., Pitts, R.J., Rokas, A., Zwiebel, L.J., 2013. Antennal transcriptome profiles of anopheline mosquitoes reveal human host olfactory specialization in *Anopheles gambiae*. BMC Genomics 14, 749. doi:10.1186/1471-2164-14-749

Roberts, S.B., Gavery, M.R., 2012. Is there a relationship between DNA methylation and phenotypic plasticity in invertebrates? Front. Physiol. 2 JAN, 116. doi:10.3389/fphys.2011.00116

Roessink, I., Arts, G.H.P., Belgers, J.D.M., Bransen, F., Maund, S.J., Brock, T.C.M., 2005. Effects of lambda-cyhalothrin in two ditch microcosm systems of different trophic status. Environ. Toxicol. Chem. 24, 1684–1696.

Rose, A., Kröckel, U., 2010. Prävention vektoriell übertragener Infektionen, in: Rieke, M.K. (Ed.), Moderne Reisemedizin: Handbuch Für Ärzte, Apotheker Und Reisenden. Gentner-Verlag, Stuttgart, p. 879.

Russo, V., Martienssen, R., Riggs, A., 1996. Epigenetic mechanisms of gene regulation. Cold Spring Harbor Laboratory Press, Long Island, New York.

Sabatini, A., Raineri, V., Trovato, G., Coluzzi, M., 1990. *Aedes albopictus* in Italy and possible diffusion of the species into the Mediterranean area. Parasitologia 32, 301–4.

Sabchareon, A., Wallace, D., Sirivichayakul, C., Limkittikul, K., Chanthavanich, P.,

Suvannadabba, S., Jiwariyavej, V., Dulyachai, W., Pengsaa, K., Wartel, T.A., Moureau, A., Saville, M., Bouckenooghe, A., Viviani, S., Tornieporth, N.G., Lang, J., 2012. Protective efficacy of the recombinant, live-attenuated, CYD tetravalent dengue vaccine in Thai schoolchildren: A randomised, controlled phase 2b trial. Lancet 380, 1559–1567. doi:10.1016/S0140-6736(12)61428-7

Sahlén, G., 1996. Eggshell ultrastructure in four mosquito genera (Diptera, culicidae). J. Am. Mosq. Control Assoc. 12, 263–270.

Savage, H., 2005. Classification of mosquitoes in tribe Aedini (Diptera: Culicidae): Paraphylyphobia, and classification versus cladistic analysis. J. Med. Entomol.

Schaefer, M., Lyko, F., 2007. DNA methylation with a sting: An active DNA methylation system in the honeybee. BioEssays 29, 208–211. doi:10.1002/bies.20548

Schäfer, B., 2014. Artemisinin. Chemie unserer Zeit 48, 134–145. doi:10.1002/ciuz.201400645

Schaffner, F., Mathis, A., 2014. Dengue and dengue vectors in the WHO European region: past, present, and scenarios for the future. Lancet. Infect. Dis. 14, 1271–80. doi:10.1016/S1473-3099(14)70834-5

Schmidt-Chanasit, J., Haditsch, M., Schöneberg, I., Günther, S., Stark, K., Frank, C., 2010. Dengue virus infection in a traveller returning from Croatia to Germany. Eurosurveillance 15, 2–3. doi:19677

Scholte, E.J., Dijkstra, E., Blok, H., De Vries, A., Takken, W., Hofhuis, A., Koopmans, M., De Boer, A., Reusken, C.B.E.M., 2008. Accidental importation of the mosquito *Aedes albopictus* into the Netherlands: A survey of mosquito distribution and the presence of dengue virus. Med. Vet. Entomol. 22, 352–358. doi:10.1111/j.1365-2915.2008.00763.x

Scholte, E.-J., Jacobs, F., Linton, Y.-M., Dijkstra, E., Takken, J.F.& W., 2007. First record of *Aedes* (*Stegomyia*) *albopictus* in the Netherlands. Eur. Mosq. Bull. 22, 201–203.

Scholte, E.J., Schaffner, F., 2007. Waiting for the tiger: Establishment and spread of the Asian tiger mosquito in Europe, in: Takken, W., Knols, B.G.J. (Eds.), Emerging Pests and Vector-Borne Diseases in Europe. Wagening Academics, pp. 241–261. doi:10.3201/eid1411.080945

Schroer, A.F.W., Belgers, J.D.M., Brock, C.M., Matser, A.M., Maund, S.J., Brink, P.J., Brock, T.C.M., Matser, A.M., Maund, S.J., Van den Brink, P.J., 2004. Comparison of laboratory single species and field population-level effects of the pyrethroid insecticide lambda-cyhalothrin on freshwater invertebrates. Arch. Environ. Contam. Toxicol. 46, 324–335. doi:10.1007/s00244-003-2315-3

Selck, F.W., Adalja, A.A., Boddie, C.R., 2014. An estimate of the global health care and lost productivity costs of dengue. Vector Borne Zoonotic Dis. 14, 824–6. doi:10.1089/vbz.2013.1528

Shroyer, D., 1986. *Aedes albopictus* and arboviruses: a concise review of the literature. J. Am. Mosq. Control Assoc. 2, 424–428.

Siler, J., Hall, M., Hitchens, A., 1926. Dengue: its history, epidemiology, mechanism of transmission, etiology, clinical manifestations, immunity, and prevention. The Government of the Philippine Islands Department of Agriculture and Natural Resources Bureau of Science Manila.

Skuse, F.A.A., 1894. The banded mosquito of Bengal. Indian Museum Not. 3, 20.

Sota, T., Mogi, M., 1992a. Interspecific variation in desiccation survival time of *Aedes* (*Stegomyia*) mosquito eggs is correlated with habitat and egg size. Oecologia 90, 353–358. doi:10.1007/BF00317691

Sota, T., Mogi, M., 1992b. Survival time and resistance to dessication of diapause and non-diapause eggs of temperate *Aedes* (*Stegomyia*) mosquitoes. Entomol. Exp. Appl. 63, 155–161. doi:10.1111/j.1570-7458.1992.tb01570.x

Sota, T., Mogi, M., 2006. Origin of pitcher plant mosquitoes in *Aedes* (*Stegomyia*): a molecular phylogenetic analysis using mitochondrial and nuclear gene sequences. J. Med. Entomol. 43, 795–800.

Sota, T., Mogi, M., Miyagi, I., Sembel, D.T., 1993. Desiccation Survival Time of Two *Aedes* (*Stegomyia*) Mosquito Eggs from North Sulawesi. Japanese J. Entomol. 61, 121–124.

Stark, K., Niedrig, M., Biederbick, W., Merkert, H., Hacker, J., 2009. Climate changes and emerging diseases. What new infectious diseases and health problem can be expected? Bundesgesundheitsblatt. Gesundheitsforschung. Gesundheitsschutz 52, 699–714. doi:10.1007/s00103-009-0874-9

Stolz, M., 2008. Drucksache 14/2628: Antrag der Abg. Dieter Ehret u. a. FDP/DVP und Stellungnahme des Ministeriums für Arbeit und Soziales: Gesundheitliche Folgen des Klimawandels, in: Zo1125-14,27. Landtag von Baden-Württemberg, pp. 1–6. doi:Drucksache 14 / 2628

Suk, J., Semenz, J., 2013. Environmental risk mapping: *Aedes albopictus* in Europe. University of Oxford, United Kingdom.

Sulaiman, S., Omar, B., Jeffery, J., Busparani, V., 1991. Evaluation of pyrethroids lambda-cyhalothrin, deltamethrin and permethrin against *Aedes albopictus* in the laboratory. J. Am. Mosq. Control Assoc. 7, 322–323.

Suwansirisilp, K., Visetson, S., Prabaripai, A., Tanasinchayakul, S., Grieco, J.P., Bangs, M.J., Chareonviriyaphap, T., 2012. Behavioral responses of *Aedes aegypti* and *Culex quinquefasciatus* (Diptera: Culicidae) to four essential oils in Thailand. J. Pest Sci. (2004). 86, 309–320. doi:10.1007/s10340-012-0464-8

Telford, A.D., 1957. The pasture *Aedes* of central and northern California. The egg stage: gross embryology and resistance to desiccation. Ann. Entomol. Soc. Am. 50, 537–543. doi:10.1093/aesa/50.6.537

Thomas, S.M., Obermayr, U., Fischer, D., Kreyling, J., Beierkuhnlein, C., 2012. Low-temperature threshold for egg survival of a post-diapause and non-diapause European aedine strain, *Aedes albopictus* (Diptera: Culicidae). Parasit. Vectors 5, 100. doi:10.1186/1756-3305-5-100

Tweedie, S., Ng, H.H., Barlow, a L., Turner, B.M., Hendrich, B., Bird, A., 1999. Vestiges of a DNA methylation system in *Drosophila melanogaster*? Nat. Genet. 23, 389–90. doi:10.1038/70490

Tzschentke, B., Basta, D., 2002. Early development of neuronal hypothalamic thermosensitivity in birds: influence of epigenetic temperature adaptation. Comp. Biochem. Physiol. Part A Mol. Integr. Physiol. 131, 825–832. doi:10.1016/S1095-6433(02)00020-X

Urbanski, J.M., Aruda, A., Armbruster, P., 2010a. A transcriptional element of the diapause program in the Asian tiger mosquito, *Aedes albopictus*, identified by suppressive

subtractive hybridization. J. Insect Physiol. 56, 1147–1154. doi:10.1016/j.jinsphys.2010.03.008

Urbanski, J.M., Benoit, J.B., Michaud, M.R., Denlinger, D.L., Armbruster, P., 2010b. The molecular physiology of increased egg desiccation resistance during diapause in the invasive mosquito, *Aedes albopictus*. Proc. Biol. Sci. 277, 2683–92. doi:10.1098/rspb.2010.0362

Urbanski, J.M., Mogi, M., O'Donnell, D., DeCotiis, M., Toma, T., Armbruster, P., O'Donnell, D., DeCotiis, M., Toma, T., Armbruster, P., O'Donnell, D., DeCotiis, M., Toma, T., Armbruster, P., 2012. Rapid adaptive evolution of photoperiodic response during invasion and range expansion across a climatic gradient. Am. Nat. 179, 490–500. doi:10.1086/664709

Valle, D., Monnerat, A.T.T., Soares, M.J.J., Rosa-Freitas, M.G.G., Pelajo-Machado, M., Vale, B.S.S., Lenzi, H.L.L., Galler, R., Lima, J.B.P.B.P., 1999. Mosquito embryos and eggs: Polarity and terminology of chorionic layers. J. Insect Physiol. 45, 701–708. doi:10.1016/S0022-1910(98)00154-1

Vandegehuchte, M.B., Janssen, C.R., 2011. Epigenetics and its implications for ecotoxicology. Ecotoxicology 20, 607–624. doi:10.1007/s10646-011-0634-0

Vega-Rua, A., Zouache, K., Caro, V., Diancourt, L., Delaunay, P., Grandadam, M., Failloux, A.-B., 2013. High efficiency of temperate *Aedes albopictus* to transmit chikungunya and dengue viruses in the southeast of France. PLoS One 8, e59716. doi:10.1371/journal.pone.0059716

Vinnersten, T.Z.P., Lundström, J.O., Schäfer, M.L., Petersson, E., Landin, J., 2010. A six-year study of insect emergence from temporary flooded wetlands in central Sweden, with and without Bti-based mosquito control. Bull. Entomol. Res. 100, 715–25. doi:10.1017/S0007485310000076

Vinogradova, E.B., 2000. Mosquitoes *Culex pipiens pipiens*: taxonomy, distribution, ecology, physiology, genetics, applied importance and control., PenSoft, Sofia. PenSoft, Sofia.

Vontas, J., Kioulos, E., Pavlidi, N., Morou, E., della Torre, A., Ranson, H., 2012. Insecticide resistance in the major dengue vectors *Aedes albopictus* and *Aedes aegypti*. Pestic. Biochem. Physiol. 104, 126–131. doi:10.1016/j.pestbp.2012.05.008

Walton, W.E., 2007. Larvivorous fish including *Gambusia*. J. Am. Mosq. Control Assoc. 23, 184–220. doi:10.2987/8756-971X(2007)23[184:LFIG]2.0.CO;2

Wang, R.-L., 1966. Observations on the influence of photoperiod on egg diapause in *Aedes albopictus* (Skuse). Acta Enomol. Sin. 15, 3.

Wang, X., Wheeler, D., Avery, A., Rago, A., Choi, J.-H., Colbourne, J.K., Clark, A.G., Werren, J.H., 2013. Function and evolution of DNA methylation in *Nasonia vitripennis*. PLoS Genet. 9, e1003872. doi:10.1371/journal.pgen.1003872

Wang, Y., Jorda, M., Jones, P.L., Maleszka, R., Ling, X., Robertson, H.M., Mizzen, C.A., Peinado, M.A., Robinson, G.E., 2006. Functional CpG methylation system in a social insect. Science 314, 645–7. doi:10.1126/science.1135213

Weaver, S., 2005. Journal policy on names of aedine mosquito genera and subgenera. Am. J. Trop. Med. Hyg. 73, 481. doi:10.1089/vbz.2005.5.93

Weill, M., Fort, P., Berthomieu, A., Dubois, M.P., Pasteur, N., Raymond, M., 2002. A novel acetylcholinesterase gene in mosquitoes codes for the insecticide target and is non-homologous to the ace gene in *Drosophila*. Proc. Biol. Sci. 269, 2007–16. doi:10.1098/rspb.2002.2122

Werner, D., Kampen, H., 2014. *Aedes albopictus* breeding in southern Germany, 2014. Parasitol. Res. 114, 831–834. doi:10.1007/s00436-014-4244-7

Werner, D., Kronefeld, M., Schaffner, F., Kampen, H., 2012. Two invasive mosquito species, *Aedes albopictus* and *Aedes japonicus japonicus*, trapped in south-west Germany, July to August 2011. Euro Surveill. 17, 1–4.

Weston, D.P., You, J., Harwood, A.D., Lydy, M.J., 2009. Whole sediment toxicity identification evaluation tools for pyrethroid insecticides: III. Temperature manipulation. Environ. Toxicol. Chem. 28, 173–80. doi:10.1897/08-143.1

Whelan, P., Kulbac, M., Bowbridge, D., Krause, V., 2009. The eradication of *Aedes aegypti* from Groote Eylandt NT Australia 2006-2008. Arbovirus Res. Aust. 10, 188–199.

WHO, 1998. Test procedures for insecticide resistance monitoring in malaria vectors, bio-efficacy and persistence of insecticides on treated surfaces. WHO/CDS/CPC/MAL/98.12 1–43. doi:WHO/CDS/CPC/MAL/98.12

WHO, 2009. Dengue: guidelines for diagnosis, treatment, prevention, and control. Spec. Program. Res. Train. Trop. Dis. 147. doi:WHO/HTM/NTD/DEN/2009.1

WHO, 2013. Yellow fever vaccination booster not needed [WWW Document]. URL http://www.who.int/mediacentre/news/releases/2013/yellow_fever_20130517/en/ (accessed 4.4.16).

WHO, 2016a. Zika virus and potential complications: Questions and answers. World Health Organization. [WWW Document]. URL http://www.who.int/features/qa/zika/en/ (accessed 4.4.16).

WHO, 2016b. WHO statement on the first meeting of the International Health Regulations (2005) (IHR 2005) Emergency Committee on Zika virus and observed increase in neurological disorders and neonatal malformations [WWW Document]. URL http://www.who.int/mediacentre/news/statements/2016/1st-emergency-committee-zika/en/ (accessed 4.4.16).

WHO, 2016c. Meeting of the Emergency Committee under the International Health Regulations (2005) concerning Yellow Fever [WWW Document]. URL http://www.who.int/mediacentre/news/statements/2016/ec-yellow-fever/en/ (accessed 5.15.16).

Williamson, M., 1996. Biological invasions. Springer Science & Business Media, Berlin.

Wong, P.-S.J., Li, M.I., Chong, C.-S., Ng, L.-C., Tan, C.-H., 2013. *Aedes* (*Stegomyia*) *albopictus* (Skuse): a potential vector of Zika virus in Singapore. PLoS Negl. Trop. Dis. 7, e2348. doi:10.1371/journal.pntd.0002348

Ye, Y.H., Woolfit, M., Huttley, G.A., Rancès, E., Caragata, E.P., Popovici, J., O'Neill, S.L., McGraw, E.A., 2013. Infection with a virulent strain of *Wolbachia* disrupts genome wide-patterns of cytosine methylation in the mosquito *Aedes aegypti*. PLoS One 8, e66482. doi:10.1371/journal.pone.0066482

# 4 Abkürzungsverzeichnis

| | |
|---|---|
| ADE | Infektionsverstärkende Antikörpern (Antibody-dependent enhancement) |
| B.t.i. | Bacillus thuringiensis israelensis |
| CDC | Centers for Disease Control and Prevention |
| $CO_2$ | Kohlenstoffdioxid |
| DDT | Dichlordiphenyltrichlorethan |
| DDV | Deutsche Vereinigung zur Bekämpfung der Viruskrankheiten |
| DE | Dunkles Endochorion (dark endochorion) |
| DEET | N,N-Diethyl-Toluamide |
| DENV | Dengue-Virus |
| DHF | Hämorrhagisches Denguefieber (Dengue Haemorrhagic Fever) |
| DKFZ | Deutsches Krebsforschungszentrum |
| DNA | Desoxyribonukleinsäure (deoxyribonucleic acid) |
| DNMT | DNA methyltransferase |
| DOI | Digitaler Objektbezeichner (Digital Object Identifier) |
| DSS | Dengue-Schock-Syndrom |
| et al. | et alia |
| EX | Exochorion |
| FLI | Friedrich-Löffler-Institut |
| ga | Galea |
| GVO | Gentechnisch veränderter Organismus |
| H | Habitat |
| $HC_{10}$ | Hazardous Concentration = schädliche Konzentration für 10% der getesteten Arten |
| $HC_5$ | Siehe $HC_{10}$ |
| iAChE | insektizin-insensitiven Acetylcholinesterase |
| ICZN | Internationale Komission für die Zoologische Nomenklatur (International Commission on Zoological Nomenclature) |
| ID | Idendifikations Nummer |
| IfD | Institut für Dipterologie |
| IPCC | Weltklimarat der Vereinten Nationen (Intergovernmental Panel on Climate Change) |
| IR3535 | Ethylbutylacetylaminopropionat |
| iSC | Innere Serosa-Cuticula |
| ISSN | Internationale Standardnummer für fortlaufende Sammelwerke |
| ITN's | Insecticide Treated Nets |
| JME | Journal of Medical Entomology |
| KABS e.V | Kommunale Aktionsgemeinschaft zur Bekämpfung der Schnakenplage |
| kdr | Knock Down Resistance Allele |
| KI | Konfidenzintervall |
| lb | Labellae |
| $LC_{10}$ | Lethale Konzentration bei der 10% im Vergleich zur Kontrolle sterben |
| $LC_{50}$ | Siehe $LC_{10}$ |
| $LC_{90}$ | Siehe $LC_{10}$ |
| LE | Helles Endochorion (light endochorion) |
| lg | Ligula |
| lr | Labrum |
| $LT_{50}$ | Halbwertszeit (Lethal Time, tötliche Zeit für 50% der Individuen) |
| microRNA | Kleine Nichtcodierende Ribonukleinsäure |
| mSC | Mittlere Serosa-Cuticula |
| n.d. | Nicht determiniert |
| OECD | Organisation für wirtschaftliche Zusammenarbeit und Entwicklung (Organisation for Economic Co-operation and Development) |
| oSC | äußere Serosa-Cuticula (outer serosal cuticle) |
| PCNA | Proliferating-Cell-Nuclear-Antigen |
| PCST | Physicians in the Crop-Sprayed Town |
| Pepck | Phosphoenolpyruvat-Carboxykinase |
| PMD | Para-Menthan-3,8-diol |
| qRT-PCR | Quantitative-Echtzeit-Polymerase-Kettenreaktion |
| RIDL | Release of Insects carrying a Dominant Lethal |
| RNI | Reaktionsnorm des Individuums |
| RNP | Reaktionsnorm der Population |
| RNS | Reaktionsnorm der Spezies |
| SD | Standardabweichung (standard deviation) |
| siRNA | Small interfering RNA |
| SIT | Sterile insect technique |
| SM | Serosa-Membran |
| SSD | Species-Sensitivity-Distribution-Modell |
| Swiss TPH | Swiss Tropical and Public Health Institute |
| US | Vereinigte Staaten |
| USA | Vereinigte Staaten von Amerika |
| WHO | Weltgesundheitsorganisation (World Health Organization) |
| WRAIR | Walter Reed Army Institute of Research |
| ZALF | Leibniz-Zentrum für Agrarlandschaftsforschung |